\documentclass[fleqn]{2022AMSE}
\usepackage{bm}

\usepackage{times}
\usepackage{algorithmic}
\usepackage{algorithm}
\usepackage{amssymb}
\usepackage{amsmath}
\usepackage{amsthm}
\usepackage{mathrsfs}
\DeclareMathAlphabet{\mathcal}{OMS}{cmsy}{m}{n}
\DeclareSymbolFont{largesymbols}{OMX}{cmex}{m}{n}
\usepackage{etoolbox}
\usepackage{subfig}
\usepackage{graphicx}
\usepackage{dcolumn}
\usepackage{bm}

\begin{document}

\ensubject{Fluid Dynamics}

\ArticleType{RESEARCH PAPER}

\title{An Implicit Adaptive Fourier Neural Operator for Long-term Predictions of Three-dimensional Turbulence}

\author[1,2]{Yuchi Jiang}{}%
\author[3]{Zhijie Li}{}%
\author[1,2]{Yunpeng Wang}{}
\author[1,2]{Huiyu Yang}{}%
\author[1,2]{Jianchun Wang}{wangjc@sustech.edu.cn}



\address[1]{Department of Mechanics and Aerospace Engineering, Southern University of Science and Technology, Shenzhen 518055, China}
\address[2]{Guangdong Provincial Key Laboratory of Turbulence Research and Applications, Southern University of Science and Technology, Shenzhen 518055, China}
\address[3]{Department of Biomedical Engineering Applications, National University of Singapore, Singapore 117583, Singapore}

\contributions{Executive Editor: ???}

\abstract{Long-term prediction of three-dimensional (3D) turbulent flows is one of the most challenging problems for machine learning approaches. Although some existing machine learning approaches such as implicit U-net enhanced Fourier neural operator (IUFNO) have been proven to be capable of achieving stable long-term predictions for turbulent flows, their computational costs are usually high. In this paper, we use the adaptive Fourier neural operator (AFNO) as the backbone to construct a model that can predict 3D turbulence. Furthermore, we employ the implicit iteration to our constructed AFNO and propose the implicit adaptive Fourier neural operator (IAFNO). IAFNO is systematically tested in three types of 3D turbulence, including forced homogeneous isotropic turbulence (HIT), temporally evolving turbulent mixing layer and turbulent channel flow. The numerical results demonstrate that IAFNO is more accurate and stable than IUFNO and the traditional large-eddy simulation using dynamic Smagorinsky model (DSM). Meanwhile, the AFNO model exhibits instability in numerical simulations. Moreover, the training efficiency of IAFNO is 4 times higher than that of IUFNO, and the number of parameters and GPU memory occupation of IAFNO are only 1/80 and 1/3 of IUFNO, respectively in HIT. In other tests, the improvements are slightly lower but still considerable. These improvements mainly come from patching and self-attention in 3D space. Besides, the well-trained IAFNO is significantly faster than the DSM.}

\keywords{Fourier Neural Operator, Self-Attention, Turbulence, Large-Eddy Simulation}

\setlength{\textheight}{23.6cm}
\thispagestyle{empty}

\maketitle
\setlength{\parindent}{1em}

\vspace{-1mm}
\begin{multicols}{2}

\section{Introduction}
Turbulence is ubiquitous in various environments and has attracted great attention in aerospace, marine and meteorological studies, making the modeling and prediction of turbulence a cutting-edge scientific issue \cite{pope2000turbulent,cook2019optimization,rippeth2022turbulent,he2022observations}. With the rapid advances of computer science and numerical method, computational fluid dynamics becomes an important, efficient, and transferable approach to studying turbulence \cite{bauweraerts2021reconstruction,abdulkadir2015comparison}. Direct numerical simulations solve all scales of turbulent flows and become impractical in the situation of high Reynolds numbers \cite{moin1998direct,ishihara2009study,yang2021grid,meneveau2000scale}. Therefore, Reynolds-averaged Navier–Stokes (RANS) simulation and \Authorfootnote large-eddy simulation (LES) methods, which achieve high efficiency by using coarser grids, have been widely used for industrial applications \cite{duraisamy2019turbulence,templeton2005eddy,bhushan2012dynamic,wu2019reynolds,morgan2022large,li2022synchronizing}.

In recent years, machine learning methods are widely applied in turbulence research \cite{brunton2020machine,chen2021theory,beck2019deep}. Applications include, but are not limited to, the use of neural networks to enhance or develop RANS and LES models \cite{duraisamy2021perspectives,yuan2021dynamic,wang2021artificial,zhao2020rans,xie2019artificial,tabe2023priori,kochkov2021machine}; optimizing input boundary conditions \cite{veras2023reconstruction}, wall modeling \cite{lozano2023machine,bae2022scientific}, and grid generation using machine learning methods \cite{huang2021machine,tingfan2022mesh}; leveraging machine learning models to solve nonlinear partial differential equations related to fluid dynamics \cite{kovachki2023neural,font2021deep,sirignano2018dgm,lusch2018deep}, to predict the evolution of flow fields over time \cite{peng2023linear,srinivasan2019predictions}, or to generate the initial state and temporal evolution of turbulent flows \cite{shu2023physics,gao2024bayesian,oommen2024integrating}. Among these applications, surrogate models that predict time evolution of the flow field have great potential to achieve both high efficiency and high accuracy. Therefore, how to make these models efficient, accurate and numerically stable for long-term predictions of turbulent flows has received a lot of attention \cite{li2022fourier,yang2024implicit,liu2022analysis}.

Guo et al. proposed a general and flexible model for real-time prediction of non-uniform steady laminar flow based on convolutional neural networks (CNNs) \cite{guo2016convolutional}. Nakamura et al. investigated the applicability of the machine learning based reduced order model (ML-ROM) by combining a three-dimensional convolutional neural network autoencoder (CNN-AE) and a LSTM neural network in a three-dimensional (3D) turbulent channel flow at friction Reynolds number of $Re_{\tau}=110$ \cite{nakamura2021convolutional}. Physics-informed neural networks (PINNs) was developed by Raissi et al. to predict the solutions of general nonlinear partial differential equations \cite{raissi2019physics}. Jin et al. proposed the Navier-Stokes flow nets (NSFnets) based on PINN framework to simulate turbulent channel flow at friction Reynolds number $Re_{\tau}=999$ \cite{jin2021nsfnets}.

Although many neural networks (NNs) are good at approximating mappings between finite-dimensional Euclidean spaces for specific datasets, it is difficult for these models to generalize to different flow conditions or boundary conditions \cite{wang2024prediction,raissi2019physics,li2020fourier}. In 2020, Li et al. proposed the Fourier neural operator (FNO) by parameterizing the integral kernel in Fourier space, and it enables reconstructing information in infinite-dimensional spaces while achieving superior accuracy compared to previous NN-based solvers \cite{li2020fourier}. Wen et al. proposed a U-FNO model using the U-net structure to enhance FNO, which is more accurate than FNO in solving multiphase flow problems \cite{wen2022u}. Moreover, You et al. \cite{you2022learning} pointed out that as the network gets deeper, the number of trainable parameters increases linearly in the FNOs, causing the model to be difficult to train. To address this problem, they proposed an implicit Fourier neural operator (IFNO), which significantly reduces the number of trainable parameters and maintains stability in deep networks \cite{you2022learning}. Recently, Li et al. proposed an implicit U-Net enhanced FNO (IUFNO) for long-term prediction of 3D turbulence at high Reynolds numbers, using the coarse-grid filtered DNS (fDNS) data of 3D isotropic turbulence and turbulent mixing layer \cite{li2023long}. Wang et al. further employed the IUFNO network to predict the 3D turbulent channel flows at different friction Reynolds numbers $Re_{\tau}$ from 180 to 590 \cite{wang2024prediction}. 

Vaswani et al. proposed the Transformer in 2017, which relies on attention mechanisms to draw global dependencies between inputs and outputs, and outperforms other machine learning methods in various benchmark tests \cite{vaswani2017attention}. In 2021, Cao et al. conceptualized a learnable Petrov-Galerkin projection and proposed the Galerkin transformer to deal with data-driven operator learning problems related to partial differential equations (PDEs) \cite{cao2021choose}. In 2022, Li et al. applied an attention-based encoder-decoder structure to Transformer for tasks governed by PDEs \cite{li2022transformer}. Moreover, Hao et al. proposed a general neural operator transformer (GNOT) capable of flexibly encoding multiple input functions and irregular meshes in practical PDEs problems \cite{hao2023gnot}.

Although Transformer neural operators have achieved competitive results in various benchmark tests of PDEs, their substantial memory requirements often make their application in high-dimensional PDEs impractical. To overcome this issue, Li et al. introduced a factorized transformer (FactFormer) built on an axial factorized kernel integral, providing an efficient low-rank alternative surrogate modeling \cite{li2024scalable}. Wu et al. proposed the Transolver, which achieves consistent state-of-the-art in large-scale industrial simulations \cite{wu2024transolver}. Moreover, Yang et al. proposed an implicit factorized transformer (IFactFormer) model which enables stable training at greater depths through implicit iteration over factorized attention for 3D isotropic turbulence \cite{yang2024implicit}. Yang et al. further introduced a modified implicit factorized transformer (IFactFormer-m) model that replaces the original chained factorized attention with parallel factorized attention, which gives a better prediction compared to the original IFactFormer in 3D turbulent channel flows \cite{yang2024implicit2}.

Another method to reduce the computational cost of Transformer-based models was developed by Guibas et al., i.e., the adaptive Fourier neural operator (AFNO) as an efficient token mixer that learns to mix in the Fourier domain \cite{guibas2021adaptive}. Pathak et al. proposed a Fourier forecasting neural network (FourCastNet) for weather forecast \cite{pathak2022fourcastnet} which combined the Fourier transform-based token-mixing scheme \cite{guibas2021adaptive} with a vision transformer (ViT) backbone \cite{dosovitskiy2020image}. The FourCastNet can accurately forecast high-resolution, fast-timescale variables including the surface wind speed, precipitation, and atmospheric water vapor \cite{pathak2022fourcastnet}.

Fast simulations of three-dimensional nonlinear partial differential equations (PDEs) is of great importance in engineering applications. While many data-driven approaches have been widely successful in solving one-dimensional (1D) and two-dimensional (2D) PDEs, the relevant works on data-driven fast simulations of 3D PDEs are relatively rare \cite{li2023long}. Moreover, among many practical problems governed by three-dimensional partial differential equations, long-term prediction of turbulent flow is one of the most challenging problems. In recent years, the IUFNO model performed well in long-term prediction of turbulence. However, the IUFNO model has some drawbacks including large number of parameters, high GPU memory usage and long training time.

In this paper, we use the AFNO model as the backbone to construct a model adapted for learning complex turbulent flows in 3D space. Furthermore, we propose an implicit adaptive Fourier neural operator (IAFNO) model for the fast prediction of turbulence. We used fDNS data to train both AFNO and IAFNO models in a similar manner as in Li et al. \cite{li2022fourier}, and compare the results with those of IUFNO models \cite{li2023long,wang2024prediction}. In our work, the IAFNO model achieves a more accurate long-term prediction of various turbulence with higher computational efficiency compared to traditional dynamic Smagorinsky model (DSM), the original AFNO model and IUFNO model.

The rest of the paper is organized as follows. In Section \ref{method}, governing equations of the large-eddy simulation, and the architecture of AFNO and IAFNO are presented. We then present the results of DSM and several machine learning models for homogeneous isotropic turbulence, free-shear turbulent mixing layer and turbulent channel flow in Section \ref{numerical}. Moreover, in Section \ref{numerical}, we also compare the computational cost and efficiency of the IAFNO model with DSM and other data-driven models. In Section \ref{conclusion}, conclusions are drawn.


\section{Methodology}
\label{method}
\subsection{Governing equations}

The governing equations of the three-dimensional incompressible turbulence are given by \cite{pope2000turbulent,ishihara2009study}:

\begin{equation}
\frac{\partial u_i}{\partial x_i}=0 ~, \label{eq:NS1}
\end{equation}
\begin{equation}
\frac{\partial u_i}{\partial t}+\frac{\partial u_iu_j}{\partial x_j}=-\frac{\partial p}{\partial x_i}+\nu\frac{\partial^2 u_i}{\partial x_j\partial x_j}+\mathcal{F}_i ~, \label{eq:NS2}
\end{equation} where $u_i$ denotes the $i$th component of velocity, $p$ is the pressure divided by the constant density, $\nu$ represents the kinematic viscosity, and $\mathcal{F}_i$ stands for a large-scale forcing to the  momentum of the fluid in the $i$th coordinate direction. Throughout this paper, the summation convention is used unless otherwise specified.

The energy spectrum $E(k)$ of turbulence is one of the most fundamental quantities characterizing the multi-scale energy statistics of turbulent flows \cite{ishihara2016energy}. The kinetic energy $E_k$ per unit mass is defined as \cite{pope2000turbulent}:

\begin{equation}
E_k=\int_{0}^{\infty}E(k)\mathrm{d}k=\frac{1}{2}(u^{\textrm{rms}})^2 ~, \label{eq:Ek}
\end{equation} where $E(k)$ is the energy spectrum, and $u^{\textrm{rms}} = \sqrt{\langle u_iu_i\rangle}$ is the root mean square (RMS) of the velocity, with $\langle \cdot \rangle$ denoting a spatial average over the homogeneous directions \cite{pope2000turbulent,li2023long}. In addition, the Kolmogorov length scale $\eta$, the Taylor length scale $\lambda$, and the Taylor-scale Reynolds number $Re_{\lambda}$ are defined, respectively, as \cite{pope2000turbulent,wang2022constant}:

\begin{equation}
\eta=\left ( \frac{\nu^3}{\epsilon}\right )^{\frac{1}{4}},~~\lambda=\sqrt{\frac{5\nu}{\epsilon}}u^{\textrm{rms}},~~Re_{\lambda}=\frac{u^{\textrm{rms}}\lambda}{\sqrt{3}\nu} ~, \label{eq:variousPhysq}
\end{equation} where $\epsilon=2\nu\langle S_{ij}S_{ij} \rangle$ denotes the average kinetic energy dissipation rate and $S_{ij}=(\partial u_i/\partial x_j+\partial u_j/\partial x_i)/2$ represents the strain rate tensor. Furthermore, the integral length scale $L_I$ and the large-eddy turnover time $\tau$ are respectively given by \cite{pope2000turbulent}:

\begin{equation}
L_I=\frac{3\pi}{2(u^{\textrm{rms}})^2}\int_{0}^{\infty}\frac{E(k)}{k}\mathrm{d}k,~~\tau=\frac{L_I}{u^{\textrm{rms}}} ~. \label{eq:turnoverT}
\end{equation}

For wall-bounded turbulence, the friction Reynolds number is defined as \cite{pope2000turbulent}:

\begin{equation}
Re_{\tau}=\frac{u_{\tau}\delta_{\nu}}{\nu} ~, \label{eq:REtau}
\end{equation} where $\delta_{\nu}=\nu/u_{\tau}$ is viscous length scale and $u_{\tau}=\sqrt{\tau_{\omega}/\rho}$ is the wall shear velocity. Here, the wall-shear stress is calculated as $\tau_{\omega}=\mu\partial\langle u\rangle/\partial y$ at the wall ($y = 0$), with $\langle \cdot \rangle$ denoting a spatial average over the homogeneous streamwise and spanwise directions \cite{wang2024prediction}.

Although the Navier-Stokes (NS) equations have been discovered for more than a century, seeking full-scale solutions of these equations using DNS is yet impractical at high Reynolds numbers mainly due to the numerous degrees of freedom \cite{moser2021statistical,meneveau2000scale,ishihara2009study,yang2021grid}. Unlike DNS, LES only solves the major energy-containing large-scale motions using a coarse grid, leaving the effects of subgrid-scale (SGS) motions handled by the SGS models \cite{smagorinsky1963general,deardorff1970numerical,germano1992turbulence}. A filtering method can be implemented to decompose the physical variables of turbulence into distinct large-scale and sub-filter small-scale components, which is defined as \cite{lesieur1996new,meneveau2000scale}:

\begin{equation}
\bar{f}(\textbf{x})=\int_{D}f(\textbf{x}-\textbf{r})G(\textbf{r};\Delta)\mathrm{d}\textbf{r} ~, \label{eq:filter}
\end{equation} where $f$ represents any physical quantity of interest in physical space associated with vector $\textbf{x}$, and $D$ is the entire domain. $G$ and $\Delta$ are the filter kernel and filter width, respectively. As shown in Eq.~\ref{eq:filter}, filtering is essentially a convolution calculation, hence in Fourier space a filtered quantity is given by $\bar{f}(\textbf{k})=\hat{G}(\textbf{k})f(\textbf{k})$, where $\hat{G}$ is $G$ after Fourier transform: $\hat{G}(\textbf{k})=\int_{-\infty}^{\infty}G(\textbf{x})e^{-i\textbf{k}\textbf{x}}\mathrm{d}\textbf{x}$. In the present study, a sharp spectral filter $\hat{G}(\textbf{k})=H(k_c-\textbf{k})$ is utilized in Fourier space for homogeneous isotropic turbulence \cite{pope2000turbulent}, where the Heaviside function $H(x)=1$ if $x\geq0$; otherwise $H(x)=0$. Here, the cutoff wavenumber $k_c=\pi/\Delta$, and $\Delta$ denotes the filter width.

Applying filtering to Eqs.~\ref{eq:NS1} and \ref{eq:NS2} yields:

\begin{equation}
\frac{\partial \bar{u}_i}{\partial x_i}=0 ~, \label{eq:fNS1}
\end{equation}
\begin{equation}
\frac{\partial \bar{u}_i}{\partial t}+\frac{\partial \bar{u}_i\bar{u}_j}{\partial x_j}=-\frac{\partial \bar{p}}{\partial x_i}-\frac{\partial \tau_{ij}}{\partial x_j}+\nu\frac{\partial^2 \bar{u}_i}{\partial x_j\partial x_j}+\bar{\mathcal{F}}_i ~, \label{eq:fNS2}
\end{equation} where the unclosed SGS stress $\tau_{ij}$ is defined by:

\begin{equation}
\tau_{ij}=\overline{u_iu_j}-\bar{u}_i\bar{u}_j ~, \label{eq:sgstress}
\end{equation} and it represents the nonlinear interactions between the resolved flow structures and SGS flow motions.

\subsection{The adaptive Fourier neural operator}

In recent research, self-attention-based architectures, in particular Transformers \cite{vaswani2017attention}, have become the de-facto standard in natural language processing (NLP) \cite{dosovitskiy2020image}.  Inspired by the success of the Transformer in NLP, Dosovitskiy et al. modified the Transformer and proposed the Vision Transformer (ViT) in 2020 \cite{dosovitskiy2020image}, which achieved excellent results for image recognition tasks. However, although Transformer and ViT are parameter efficient and exhibit excellent performance, they suffer from quadratic complexity in sequence size \cite{guibas2021adaptive}. To address this shortcoming, Guibas et al. proposed the adaptive Fourier neural operator (AFNO) \cite{li2020fourier,guibas2021adaptive}, which embedded the Fourier neural operator (FNO) with the self-attention mechanism.

\begin{figure*}[ht]
    \centering
    \includegraphics[width=14cm,height=7.875cm]{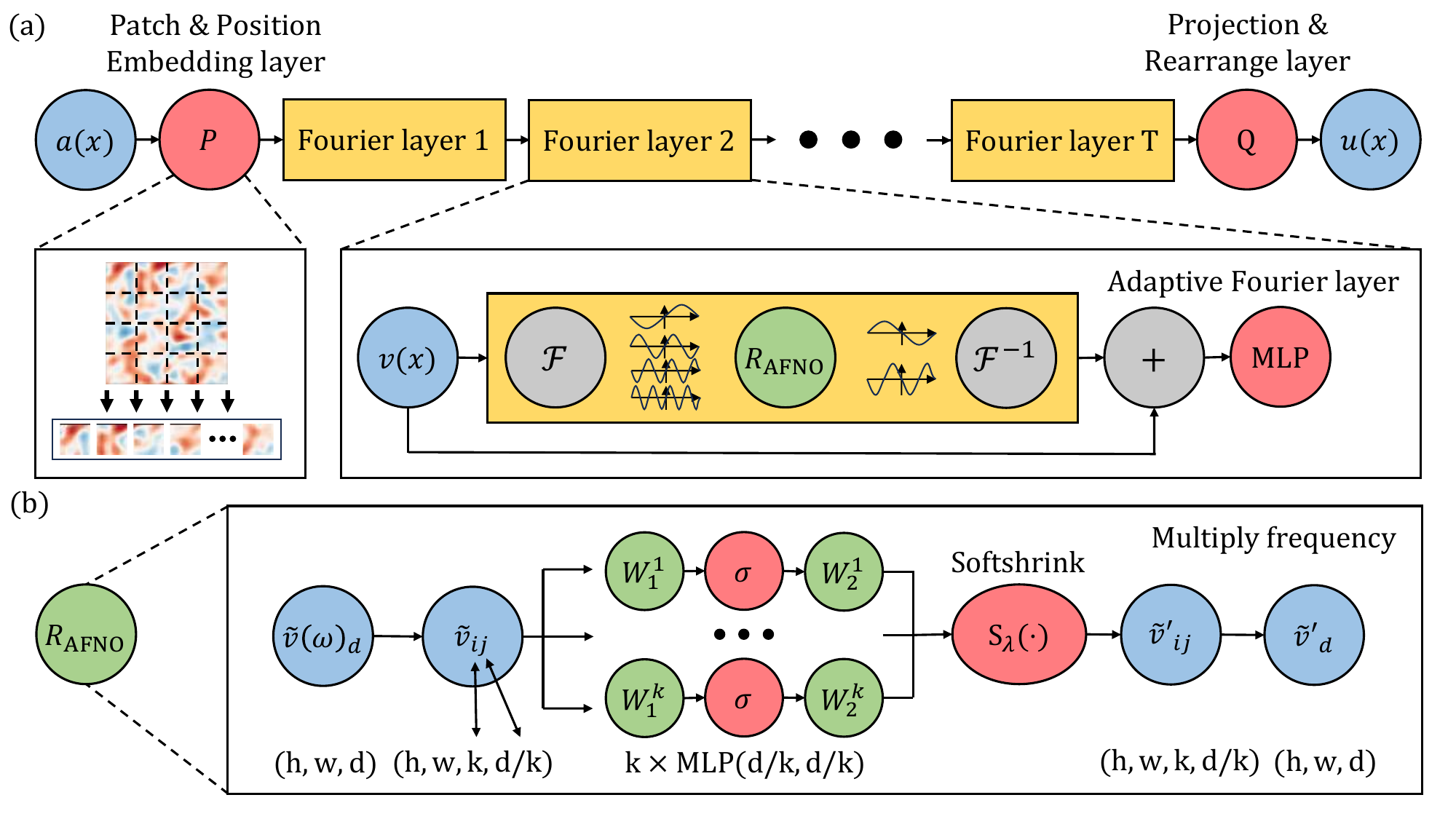}
    \caption{\label{fig:ModelStructureAFNO} (a) The model constructed with AFNO as the backbone; (b) the architecture of AFNO proposed by Guibas et al. \cite{guibas2021adaptive}.}
\end{figure*}

Hence, the key difference between the AFNO model and the FNO (shown in \ref{app1}) or the implicit U-net enhanced Fourier neural operator (IUFNO) (shown in \ref{app2}) is that AFNO introduces the self-attention mechanism, which is defined by \cite{tsai2019transformer,kovachki2023neural,guibas2021adaptive}:

\begin{equation}
\textbf{q}=XW_q,~\textbf{k}=XW_k,~\textbf{v}=XW_v ~, \label{eq:selfattention1}
\end{equation}
\begin{equation}
\mathrm{Att}(X):=\mathrm{softmax}\left( \frac{\textbf{q}\textbf{k}^{\mathrm{T}}}{\sqrt{d}} \right)\textbf{v} ~, \label{eq:selfattention2}
\end{equation} where $\textbf{q}, \textbf{k}, \textbf{v} \in \mathbb{R}^{N\times d}$ are the query, key and value vectors, respectively. $W_q, W_k, W_v \in \mathbb{R}^{d\times d}$ are corresponding weight matrices. Here, as shown in Fig.~\ref{fig:ModelStructureAFNO}(b), the input tensor $X$ can also be denoted as $\tilde{v}(\omega)$ with a tensor size of $[h,w,d]$. $h$ is the height and $w$ is the width of an image, and $d$ is the channel width. For notation convenience, we define $N:=hw$ and index the second order array $X$ as a token sequence which has the form of $X[s]=X[n_s,m_s]$ for some discrete index $s$ and $t$ where $s,t\in[hw]$. Therefore $X\in \mathbb{R}^{N\times d}$. Hence, the Eq.~\ref{eq:selfattention2} represents a kernel integration of $\mathbb{R}^{N\times d}\rightarrow\mathbb{R}^{N\times d}$ \cite{guibas2021adaptive}. In fact, combining this index approach with the convolutional neural network for data dimensionality reduction gives the patch layer shown in Fig.~\ref{fig:ModelStructureAFNO}.


Moreover, define $K:=\mathrm{softmax}(\langle XW_q, XW_k \rangle/\sqrt{d})$ as the $N\times N$ score array with $\langle \cdot,\cdot\rangle$ being inner product in $\mathbb{R}^d$. We then treat self-attention as an asymmetric matrix-valued kernel $\kappa:[N]\times[N]\rightarrow\mathbb{R}^{d\times d}$ parameterized as $\kappa[s,t]=K[s,t]\cdot W_v$ (where $K[s,t]$ is scalar valued and ``$\cdot$'' is scalar-matrix multiplication). Therefore, the alternative kernel summation form of the self-attention can be written as \cite{guibas2021adaptive}:

\begin{equation}
\mathrm{Att}(X)[s]:=\sum_{t=1}^{N}X[t]\kappa[s,t],~~\forall s\in [N] ~, \label{eq:SAsumform}
\end{equation} where $t$ as a discrete index can be extended into a continuous small change $\mathrm{d}t$ in the integral calculation. With this extension from discrete to continuous, the input tensor $X$ is no longer a finite-dimensional vector in the Euclidean space $X\in \mathbb{R}^{N\times d}$, but rather a spatial function in the function space $X\in (D,\mathbb{R}^{d})$ defined on domain $D\subset \mathbb{R}^2$ which is the physical space of the images. In this continuum formulation, the neural network becomes an operator that acts on the input functions, thus the kernel integral operator $\mathcal{K}:(D,\mathbb{R}^{d})\rightarrow(D,\mathbb{R}^{d})$ can be defined as \cite{guibas2021adaptive}:

\begin{equation}
\mathcal{K}(X)(s)=\int_{D}\kappa(s,t)X(t)\mathrm{d}t,~~\forall s\in D ~, \label{eq:SAintegralform}
\end{equation} with a continuous kernel function $\kappa:D\times D\rightarrow\mathbb{R}^{d\times d}$ \cite{li2020neural}. For the special case of the Green’s kernel $\kappa(s,t)=\kappa(s-t)$, the integral leads to global convolution defined \cite{guibas2021adaptive}:

\begin{equation}
\mathcal{K}(X)(s)=\int_{D}\kappa(s-t)X(t)\mathrm{d}t,~~\forall s\in D ~. \label{eq:SAconvolutionform}
\end{equation} With FFT and inverse FFT, Eq.~\ref{eq:SAconvolutionform} becomes:

\begin{equation}
\mathcal{K}(X)(s)=\mathcal{F}^{-1}\left(\mathcal{F}(\kappa)\cdot\mathcal{F}(X)\right)(s),~~\forall s\in D ~. \label{eq:SAfftform}
\end{equation}

Next we introduced the Block-Diagonal structure on $W$ and formulated the multiply frequency process shown in Fig.~\ref{fig:ModelStructureAFNO}(b) as:

\begin{equation}
\tilde{v}'_{d}=\tilde{v}_{ij}^{'(l)}:=S_{\lambda}\left[W_2^{(l)}\sigma\left(W_1^{(l)}\tilde{v}_{ij}^{(l)}\right)\right],~~l=1,...,k ~, \label{eq:SAmf}
\end{equation} where the weight matrix $W$ is divided into $k$ weight blocks of size $d/k\times d/k$. This block-diagonal-weight's methodology can be computationally parallelizable, in which each block can be interpreted as a head in multi-head self-attention which projects into a subspace of the data \cite{guibas2021adaptive}. Furthermore, in order to preserve the ability of increasing the parameters, a hyperparameter ``hidden-size-factor'' (HSF) is introduced. Therefore, the shape of $W_1$ can be scaled from $(d/k,d/k)$ to $(d/k,f*d/k)$, and $W_2$ can be scaled from $(d/k,d/k)$ to $(f*d/k,d/k)$. To sparsify the tokens, soft-thresholding and shrinkage operation: $S_{\lambda}[x]=\mathrm{sign}(x)\mathrm{max}\{|x|-\lambda,0\}$ is introduced, where $\lambda$ is a tuning parameter that controls the sparsity \cite{tibshirani1996regression,guibas2021adaptive}.

Hence, the AFNO architecture can be described as:

\begin{equation}
v_{t+1}=\mathrm{MLP}\left[ v_{t}+\mathcal{F}^{-1}\left( 
R_{\mathrm{AFNO}}\cdot\mathcal{F}(v_t) \right)(s) \right] ,~~\forall s\in D ~. \label{eq:fullAFNO}
\end{equation}

\subsection{The implicit adaptive Fourier neural operator}

\begin{figure*}[t]
    \centering
    \includegraphics[width=14cm,height=7.875cm]{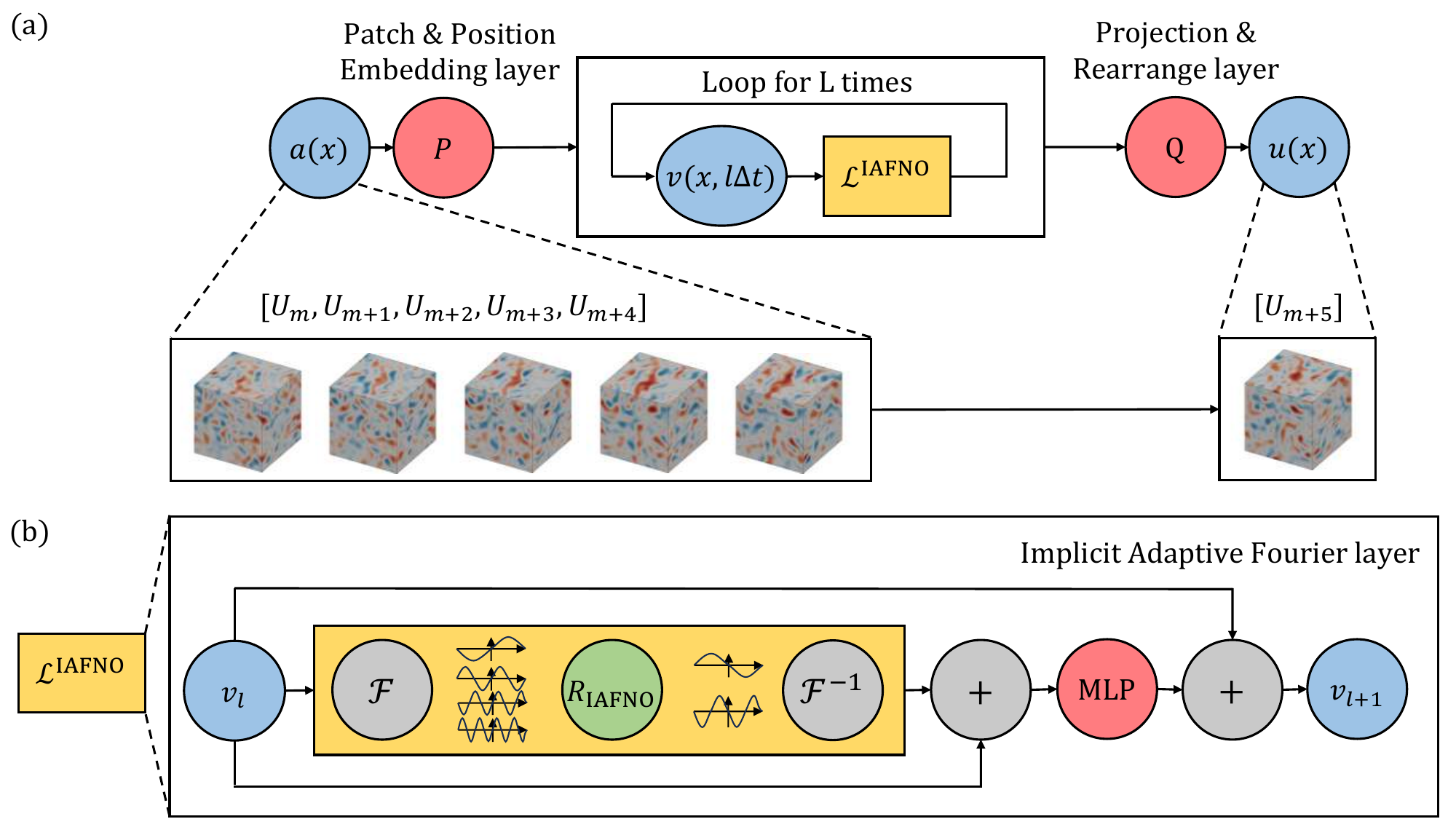}
    \caption{\label{fig:ModelStructureIAFNO} The architecture of IAFNO.}
\end{figure*}

However, in our tests, the model that uses only AFNO as the backbone without additional modifications is unstable. Therefore, based on the inspiration given by the IFNO model, we introduce the implicit iteration approach and propose the implicit adaptive Fourier neural operator (IAFNO) model. The architecture of IAFNO is shown in Fig.~\ref{fig:ModelStructureIAFNO}. The corresponding pseudocode of IAFNO is shown in Algorithm~\ref{alg:IAFNO}. Here, the IAFNO model consists of three main steps:

(1) The velocity field from the first $5$ time nodes is utilized as the input to the model, where the contained information is extracted via the patch and position embedding layer $P$. The patch and position embedding layer $P$ is specially annotated on lines $26\sim27$, making use of the first defined function PatchEmbed(x) in Algorithm~\ref{alg:IAFNO}.

(2) Then the processed velocity field is iteratively updated through the implicit adaptive Fourier layers. The formulation of iterative implicit adaptive Fourier layer is given by:

\end{multicols}

\begin{align}
v(x,(l+1)\Delta t)=\mathcal{L}^{\mathrm{IAFNO}}[v(x,l\Delta t)]:=v(x,l\Delta t)+\Delta tc(x,l\Delta t),~~\forall x\in D ~, \label{eq:iter1IAFNO}
\end{align}
\begin{align}
c(x,l\Delta t):=\mathrm{MLP}\left[ v(x,l\Delta t)+\mathcal{F}^{-1}\left( 
R_{\mathrm{IAFNO}}\cdot\mathcal{F}(v(x,l\Delta t)) \right)(x) \right],~~\forall x\in D ~, \label{eq:iter2IAFNO}
\end{align}
\begin{align}
R_{\mathrm{IAFNO}}\cdot\mathcal{F}(v(x,l\Delta t)):=S_{\lambda}\left[W_2\sigma\left(W_1\mathcal{F}(v(x,l\Delta t))+b_1\right)+b_2\right],~~\forall x\in D ~. \label{eq:iter3IAFNO}
\end{align}

\begin{multicols}{2}
Here, $\Delta t=1/L$, and $L$ represents the total number of iterations. The corresponding pseudocode for $R_{\mathrm{IAFNO}}\cdot\mathcal{F}(v(x,l\Delta t))$, which is defined in Eq.~\ref{eq:iter3IAFNO}, is the second defined function IAFNO(x) in Algorithm~\ref{alg:IAFNO}. Altogether, Eq.~\ref{eq:iter1IAFNO} and Eq.~\ref{eq:iter2IAFNO} state the function IAFNOnet(x) and Block(x) in Algorithm~\ref{alg:IAFNO}, and the graphical representations are shown in Fig.~\ref{fig:ModelStructureIAFNO}(a) and (b), respectively.

(3) After $L$ times of iterations, the output is obtained through the projection and rearrange of $Q$, which is the velocity field of the next time node. Here, the projection and rearrange layer $Q$ is specially annotated on lines $32\sim33$ in Algorithm~\ref{alg:IAFNO}.

\end{multicols}
\begin{algorithm}[!b]
    \caption{Pseudocode of IAFNO}
    \label{alg:IAFNO}
    \renewcommand{\algorithmicrequire}{\textbf{Input:}}
    \renewcommand{\algorithmicensure}{\textbf{Output:}}
    
    \begin{algorithmic}[1]
        \REQUIRE $\{a_m\}_{m=0}^{N-5}=[U_m,U_{m+1},U_{m+2},U_{m+3},U_{m+4}]\leftarrow(bs,x,y,z,c,5)$~~~~~~\# $bs$:batchsize, $c$:channel width  
        \ENSURE $\{u_m\}_{m=0}^{N-5}=U_{m+5}\leftarrow(bs,x,y,z,c)$    
        
        \STATE  \textbf{def} PatchEmbed(x)~~~~~~\# $p$:patchsize, $d$:embedded dimension (EmbedDim)
            \STATE~~~~x = x.flatten(4) $\leftarrow(bs,x,y,z,5c)$
            \STATE~~~~x = Conv3d(x)
            \STATE~~~~\textbf{return}    x $\leftarrow(bs,x/p,y/p,z/p,d):=(bs,x',y',z',d)$
        \STATE  \textbf{def} IAFNO(x)~~~~~~\# $k$:number of blocks (\#ofBlocks)
            \STATE~~~~bias = x
            \STATE~~~~x = RFFTn(x)
            \STATE~~~~x = x.reshape($bs,x',y',z'//2+1,k,d/k$)
            \STATE~~~~x = MatMul(x, $W_1$) $+~b_1$ $\leftarrow W_1:(2,k,d/k,f*d/k)$~~~~~~\# $f$:hidden-size-factor (HSF)
            \STATE~~~~x = ReLU(x)
            \STATE~~~~x = MatMul(x, $W_2$) $+~b_2$ $\leftarrow W_2:(2,k,f*d/k,d/k)$
            \STATE~~~~x = SoftShrink(x)
            \STATE~~~~x = ViewAsComplex(x)
            \STATE~~~~x = x.reshape($bs,x',y',z'//2+1,d$)
            \STATE~~~~x = IRFFTn(x)
            \STATE~~~~\textbf{return}    x + bias $\leftarrow(bs,x',y',z',d)$
        \STATE  \textbf{def} Block(x)
            \STATE~~~~residual = x
            \STATE~~~~x = norm(x)
            \STATE~~~~x = IAFNO(x)
            \STATE~~~~x = x + residual
            \STATE~~~~residual = x
            \STATE~~~~x = norm(x)
            \STATE~~~~x = MLP(x)
            \STATE~~~~\textbf{return}    x + residual $\leftarrow(bs,x',y',z',d)$
        \STATE  \textbf{def} IAFNOnet(x)
            \STATE~~~~x = PatchEmbed(x)~~~~~~\# patch
            \STATE~~~~x = x + nn.Parameter(zeros($1,x',y',z',d$))~~~~~~\# position embedding
            \STATE  ~~~~\textbf{for} each $i \in [1,L]$~~~~~~\# $L$:number of implicit iterations
                \STATE~~~~~~~~coef = $1/L$
                \STATE~~~~~~~~x = x + Block[0](x)*coef
            \STATE  ~~~~\textbf{end for}
            \STATE~~~~x = nn.Linear(x) $\leftarrow(bs,x',y',z',p^3c)$~~~~~~\# projection
            \STATE~~~~x = rearrange(x) $\leftarrow(bs,x,y,z,c)$~~~~~~\# rearrange
            \STATE~~~~\textbf{return}    x~~~~~~\# Output
    \end{algorithmic}
\end{algorithm}

\begin{multicols}{2}

\section{Numerical Results}
\label{numerical}

In this section, the flow fields of the filtered direct numerical simulation of three types of turbulence are used for the evaluations of three FNO-based models including IUFNO, AFNO and IAFNO, by comparing them against traditional LES with dynamic Smagorinsky model \cite{yuan2023adjoint}. The three types of turbulent flows includes forced homogeneous isotropic turbulence (HIT), temporally evolving turbulent mixing layer and turbulent channel flow.

In order to objectively evaluate the ability of previous models to simulate the three different turbulent flows, we reproduced the results from references \cite{li2023long,wang2024prediction}, and used the same dataset for training and validating the IAFNO model. For the \textit{a posteriori} analysis, we perform the numerical simulations for both IUFNO and IAFNO model with five different random initializations in forced HIT, five different initializations in temporally evolving turbulent mixing layer and one initialization in the turbulent channel flow. We report the ensemble average of the statistical results of different random initializations in the \textit{a posteriori} analysis.
\subsection{Forced homogeneous isotropic turbulence}
The homogeneous isotropic turbulence datasets are generated by applying a periodic boundary condition to a cube of size $(2\pi)^3$, performing a pseudo-spectral spatial discretization of the NS equations \cite{ku1987pseudospectral}, and processing the time advance using a second-order two-step explicit Adams-Bashforth scheme \cite{chen1993statistical,he2007stability}. It should be noted that the use of periodic boundary conditions here can efficiently capture the information of the global flow field \cite{yuan2020deconvolutional,xie2020modeling,munters2016shifted}, and can conveniently satisfy the requirement of the Fourier transform for periodic boundaries. However, the Fourier neural operator can conduct on non-periodic boundary conditions which benefits from the bias term \cite{li2020fourier}. The aliasing errors from nonlinear convective terms are eliminated by truncating the high wavenumbers of Fourier modes by the two-thirds rule \cite{hussaini1986spectral}. The flow field has a kinematic viscosity $\nu \approx 0.00625$, so the Taylor Reynolds number $Re_{\lambda}\approx100$. In order to ensure that the physical quantities of the flow field remain statistically steady, the data is collected after 10 large-eddy turnover times. Here, the large-eddy turnover time $\tau$ is defined as $\tau \equiv L_I/u^{\textrm{rms}} \approx 1.0$.

\begin{table*}[!ht]
\centering
\caption{\label{tab:hitfdns}Parameters and statistics for DNS and fDNS of forced HIT.}
\renewcommand{\arraystretch}{1.5}
\begin{tabular}{cccccccc}
\hline\hline
\mbox{Reso.(DNS)}&\mbox{Reso.(fDNS)}&\mbox{Domain}&\mbox{$Re_{\lambda}$}&\mbox{$\nu$}&\mbox{$\mathrm{d} t$}&\mbox{$k_c$}&\mbox{$\tau$}\\
\hline
\mbox{$256^3$}&\mbox{$32^3$}&\mbox{$(2\pi)^3$}&\mbox{100}&\mbox{0.00625}&\mbox{0.001}&\mbox{10}&\mbox{1.00}\\
\hline\hline
\end{tabular}
\end{table*}

Before the dataset enters the machine learning network, it will be filtered by a sharp spectral filter into an input tensor with a resolution of $32^3$ \cite{pope2000turbulent}. The truncation frequency $k_c = 10$ is used in the filtering process \cite{pope2000turbulent}. In addition, every 200 time steps is taken as a time node, and the time interval is equivalent to one-fifth of the large-eddy turnover time ($\Delta t=200\mathrm{d}t=0.2\tau$). The flow field data will be saved every time node as defined above, for a total of 600 time nodes. To ensure that the model training results are generalizable to different HITs, the dataset used for training contains 50 different sets of HIT flow fields generated from 50 different independent initial fields using the above method, of which 45 sets are used for training ($80\%$) and testing ($20\%$), and the last five sets are used to validate the models. The size of the tensor used for training and testing is $[45\times600\times32\times32\times32\times3]$, where the last index ``3'' represents that this tensor contains the velocity in all three axes.

For training and testing, we choose 5 neighboring flow field data and stack them in the second dimension to form the input of the neural operator, which can be denoted as $(U_m,U_{m+1},U_{m+2},U_{m+3},U_{m+4})$, with a size of $[45\times5\times32\times32\times32\times3]$. Meanwhile, the flow field of the output is given by: $\Delta U_{m+5}=U_{m+6}-U_{m+5}$. With the above methodology, we can generate $45\times(600-5)=26775$ input-output pairs, among which 80\% will be used to train the model and the remaining 20\% will be used to test the model \cite{li2023long,li2022fourier}.

In the process of training and testing, all data-driven models output the predicted increment of the flow field at the next moment ($\Delta U_{m+5}^{\textrm{pre}}$), and calculate the loss function defined as:
\begin{equation}
\mathrm{Loss} =\frac{\vert\vert u^*-u\vert\vert_2}{\vert\vert u \vert\vert_2},~~\mathrm{where}~~\vert\vert \textbf{A} \vert\vert = \frac{1}{n}\sqrt{\sum_{\textbf{k}=1}^{n}\vert \textbf{A}_{\textbf{k}} \vert^2} ~, \label{eq:lossfunction}
\end{equation} where $u^*$ denotes the prediction of velocity increment fields and $u$ is the ground truth in the output sub-dataset ($\Delta U_{m+5}$) \cite{li2023long,li2022fourier}. Furthermore, the autoregressive approach aimed at achieving long-term predictions of turbulent flows is given as follows:
\end{multicols}

\begin{align}
[U_1,U_2,U_3,U_4,U_5]&\rightarrow U_6^{\textrm{pre}} = \Delta U_5^{\textrm{pre}} + U_5 ,\nonumber \\
[U_2,U_3,U_4,U_5,U_6^{\textrm{pre}}]&\rightarrow U_7^{\textrm{pre}} = \Delta U_6^{\textrm{pre}} + U_6^{\textrm{pre}} ,\nonumber \\
[U_3,U_4,U_5,U_6^{\textrm{pre}},U_7^{\textrm{pre}}]&\rightarrow U_8^{\textrm{pre}} = \Delta U_7^{\textrm{pre}} + U_7^{\textrm{pre}} ,\nonumber \\
&\dots ,\nonumber \\
[U_{m}^{\textrm{pre}},U_{m+1}^{\textrm{pre}},U_{m+2}^{\textrm{pre}},U_{m+3}^{\textrm{pre}},U_{m+4}^{\textrm{pre}}]&\rightarrow U_{m+5}^{\textrm{pre}} = \Delta U_{m+4}^{\textrm{pre}} + U_{m+4}^{\textrm{pre}} ~. \label{eq:predict}
\end{align}

\begin{multicols}{2}

For the numerical simulations of HIT, the hyperparameters used by the IUFNO model are shown in Tab.~\ref{tab:hitiufnohyps}. The source code of IUFNO does not use a scheduler, so the learning rate is kept constant \cite{li2023long}. This setting is intentionally kept the same in AFNO and IAFNO. The optimizer is Adam optimizer with weight-decay-value of $10^{-11}$. In Tab.~\ref{tab:hitiufnohyps}, ``Width'' refers to the dimension of features of the tensor output by the lifting layer $P$. In general, the IUFNO model is carried out with 30 epochs, batchsize of 5, learning rate of 0.001 and width of 36.

\begin{table*}[htb]
\centering
\caption{\label{tab:hitiufnohyps}The hyperparameters settings for IUFNO in the forced HIT.}
\renewcommand{\arraystretch}{1.5}
\begin{tabular}{cccc}
\hline\hline
\mbox{Epochs}&\mbox{Batchsize}&\mbox{LearningRate(LR)}&\mbox{Width}\\
\hline
\mbox{30}&\mbox{5}&\mbox{0.001}&\mbox{36}\\
\hline\hline
\end{tabular}
\end{table*}

The hyperparameters in the AFNO and IAFNO model are shown in Tab.~\ref{tab:hitafnohyps}. The optimizer is Adam optimizer with weight-decay-value of $10^{-11}$. In Tab.~\ref{tab:hitafnohyps}, ``Patchsize'' refers to the size of every patches in $(x,y,z)$ directions. ``EmbedDim'' refers to the dimension of features of the tensor output by the patch and position embedding layer $P$. ``HSF'' (hidden-size-factor) refers to the scaling factor of the dimension of features in a hidden layer. ``\#ofBlocks'' (number of blocks) refers to the number of blocks in a block diagonal matrix. In general, AFNO and IAFNO are carried out with 30 epochs, batchsize of 5, learning rate of 0.001, patchsize of $(2, 2, 2)$, embedded dimension of 162, hidden-size-factor of 3 and 1 blocks after the block-diagonalize method.

\begin{table*}[htb]
\centering
\caption{\label{tab:hitafnohyps}The hyperparameters settings for AFNO and IAFNO in the forced HIT.}
\renewcommand{\arraystretch}{1.5}
\begin{tabular}{ccccccc}
\hline\hline
\mbox{Epochs}&\mbox{Batchsize}&\mbox{LearningRate(LR)}&\mbox{Patchsize}&\mbox{EmbedDim}&\mbox{HSF}&\mbox{\#ofBlocks}\\
\hline
\mbox{30}&\mbox{5}&\mbox{0.001}&\mbox{(2,2,2)}&\mbox{162}&\mbox{3}&\mbox{1}\\
\hline\hline
\end{tabular}
\end{table*}



However, in scenarios of long-term predictions of turbulence, a low test loss does not necessarily mean that predictions made based on the validation set will yield accurate results. Therefore, in order to judge the performance of the model, we must post-process the predictions produced on the validation set and compare them with the benchmark of the fDNS data. Thus, we divide the \textit{a posteriori} study into two parts, the first part for verifying the stability of all three models and the second part for comparing the model performance in detail. The models used for the different \textit{a posteriori} analyses with their training and testing losses are presented in the Tab.~\ref{tab:hitLoss}. To be more specific, the models with 10 layers are used to demonstrate the stability of each neural operator. The rest are used to compare the performance between the IAFNO model and the IUFNO model.

\begin{table*}[!t]
\centering
\caption{\label{tab:hitLoss}Comparison of minimum training and testing loss of different data-driven models in forced HIT.}
\renewcommand{\arraystretch}{1.5}
\begin{tabular}{cccc}
\hline\hline
\multicolumn{4}{c}{(Training Loss, Testing Loss)}
\\
\hline
\mbox{Model}&\mbox{$L=10$}&\mbox{$L=20$}&\mbox{$L=40$}\\
\hline
\mbox{IUFNO}&\mbox{(0.1393, 0.1857)}&\mbox{(\textbf{0.1261}, 0.1633)}&\mbox{(\textbf{0.1252}, 0.1614)}\\
\mbox{AFNO}&\mbox{(\textbf{0.1333}, \textbf{0.1338})}&\mbox{N/A}&\mbox{N/A}\\
\mbox{IAFNO}&\mbox{(0.1680, 0.1730)}&\mbox{(0.1600, 0.1639)}&\mbox{(0.1576, 0.1631)}\\
\mbox{IAFNO(Ep50)}&\mbox{N/A}&\mbox{(0.1531, \textbf{0.1538})}&\mbox{(0.1526, \textbf{0.1532})}\\
\hline\hline
\end{tabular}
\end{table*}

    \subsubsection{The a posteriori study of numerical stability}

\begin{figure*}[t]
    \centering
    \subfloat{
    \includegraphics[width=6cm,height = 4.5cm]{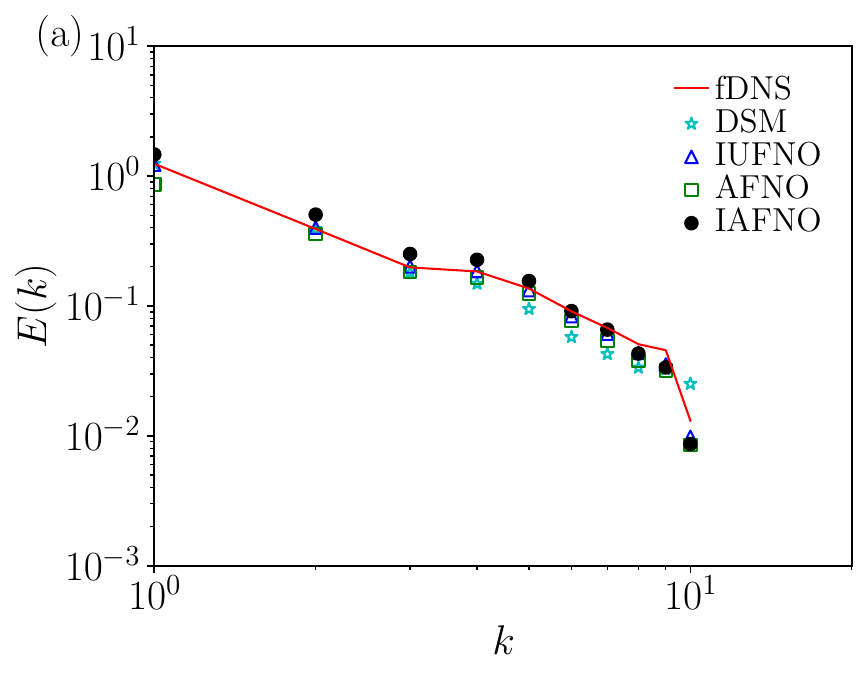}}
    \subfloat{
    \includegraphics[width=6cm,height = 4.5cm]{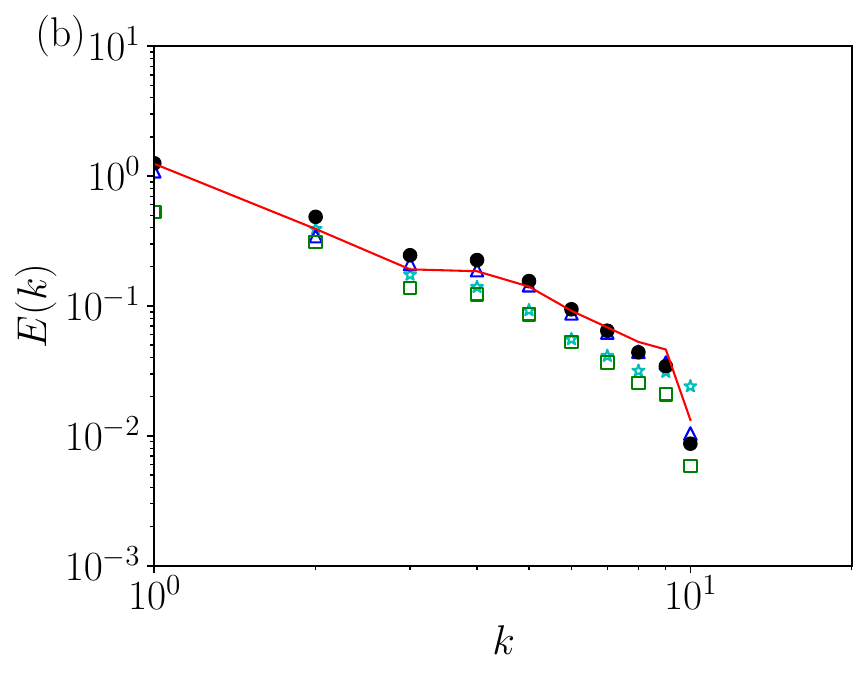}}
    \\
    \subfloat{
    \includegraphics[width=6cm,height = 4.5cm]{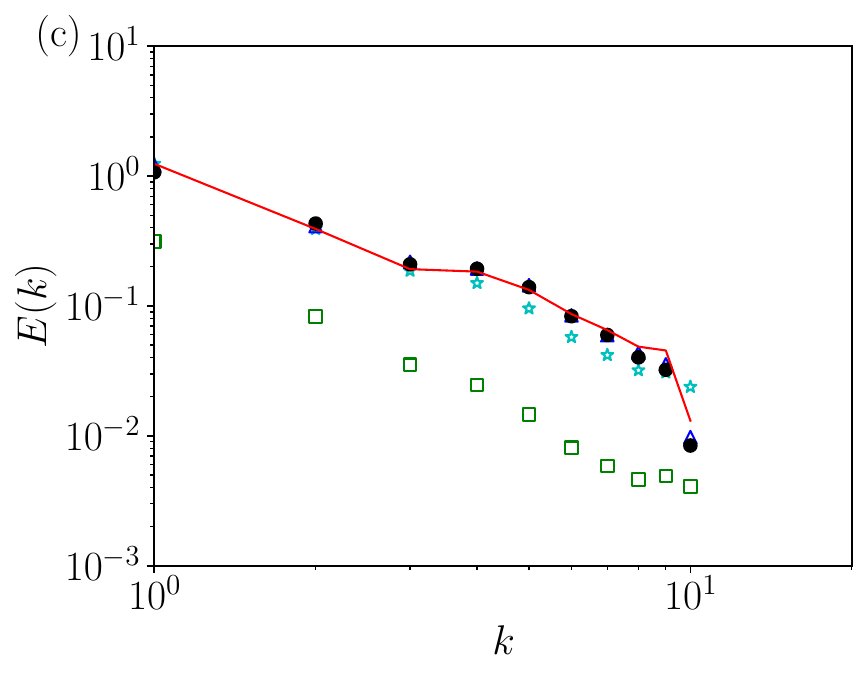}}
    \subfloat{
    \includegraphics[width=6cm,height = 4.5cm]{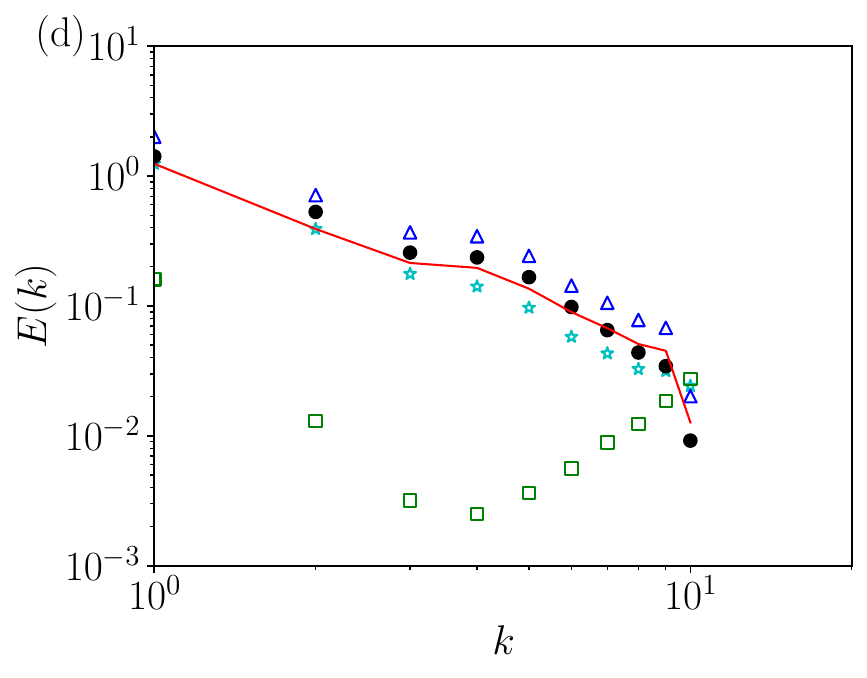}}
    \caption{\label{fig:HIT10spec} The velocity spectra in the forced HIT at different time instants: (a) $t/\tau\approx 4.0$; (b) $t/\tau\approx 8.0$; (c) $t/\tau\approx 20.0$; (d) $t/\tau\approx 50.0$.}
\end{figure*}

The velocity spectra of various models in the forced HIT at different time instants are shown in Fig.~\ref{fig:HIT10spec}. Here, each data-driven model contains a network depth equivalent to ten Fourier layers, with IUFNO and IAFNO modeled as one Fourier layer for ten implicit iterations, and AFNO as ten separate Fourier layers in series. It is observed that all three data-driven models perform well in the short-term prediction of the velocity spectrum ($t/\tau\leq4$), with the IUFNO having the best results, the IAFNO model being slightly inferior, and the AFNO model being the worst. For the traditional LES model DSM, an obvious shift from the benchmark can be observed in the interval of $5\leq k\leq9$. At the 8th large-eddy turnover time, AFNO shows a noticeable shift with respect to the benchmark, and the predictions of the other two data-driven models and the DSM are basically the same. At $t/\tau \approx 20$, the spectrum predicted by the AFNO model is much smaller than that of fDNS, while the IUFNO and IAFNO models give a reasonable prediction. Moreover, when $t/\tau\approx50$, the spectrum given by the AFNO model still has a very large deviation from fDNS result, and the spectrum predicted by IUFNO is larger than fDNS result. It is important to note that the IAFNO model gives stable and accurate results at all predicted time nodes. With all time nodes considered, the DSM is always worse than IAFNO and IUFNO.

\begin{figure*}[!t]
    \centering
    \subfloat{
    \includegraphics[width=6cm,height = 4.5cm]{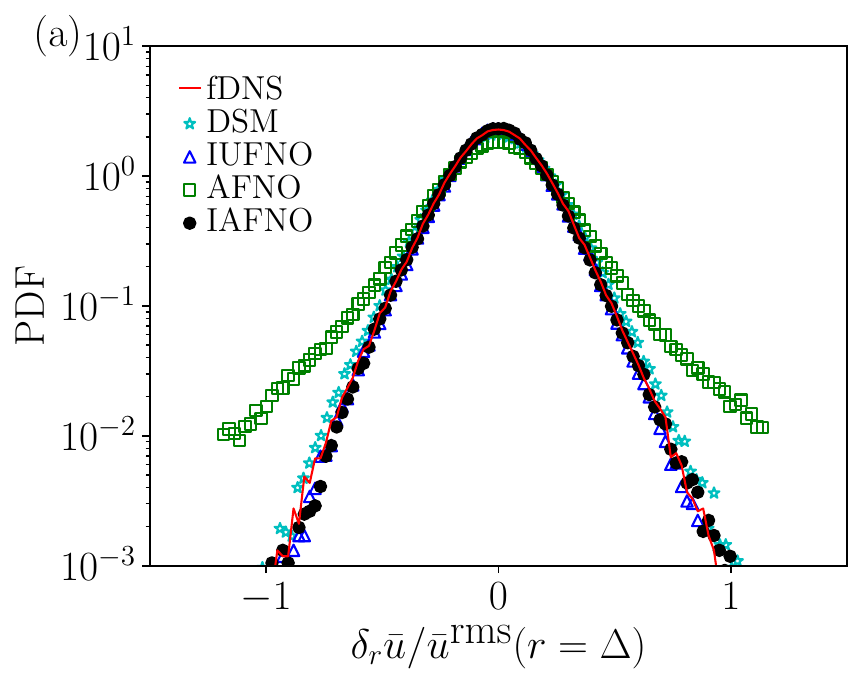}}
    \subfloat{
    \includegraphics[width=6cm,height = 4.5cm]{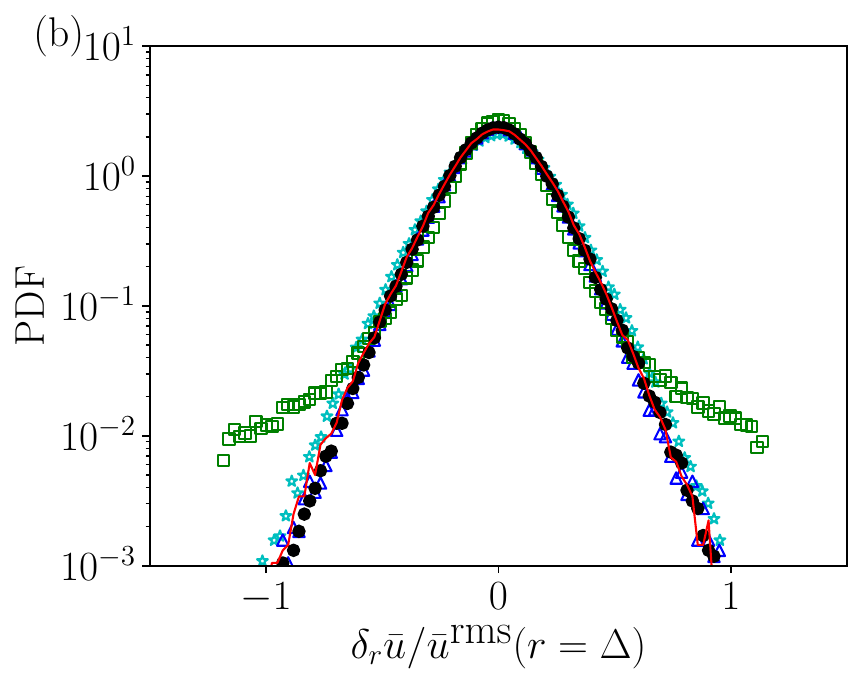}}
    \\
    \subfloat{\hspace{4mm}
    \includegraphics[width=6cm,height = 4.5cm]{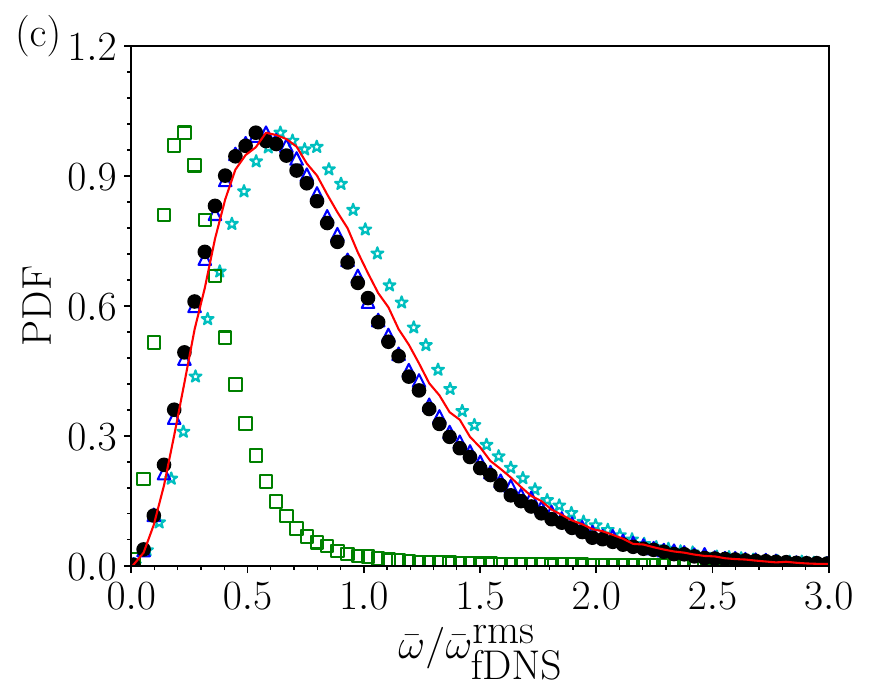}}
    \subfloat{
    \includegraphics[width=6cm,height = 4.5cm]{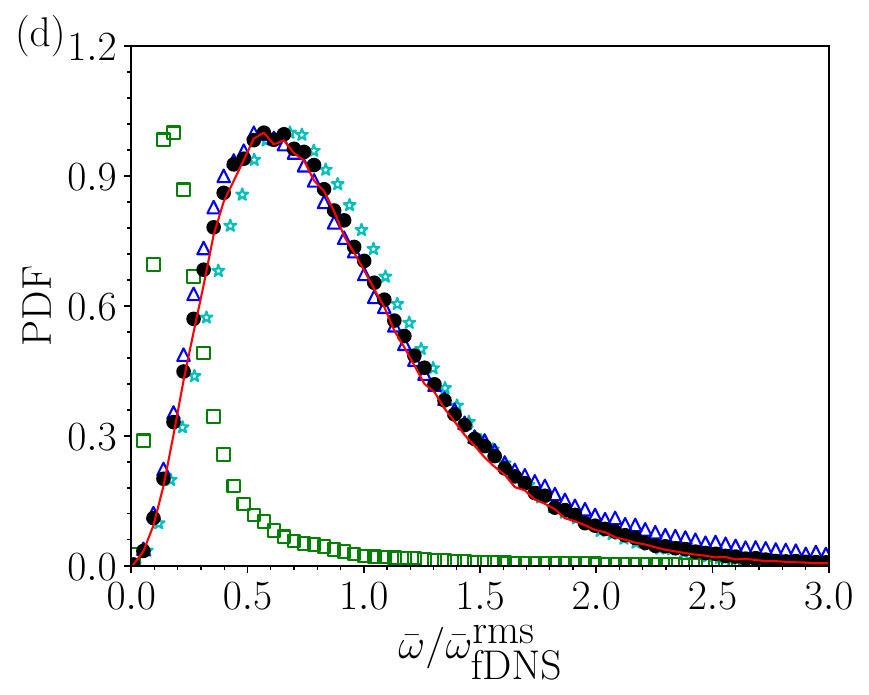}}
    \caption{\label{fig:HIT10inc-vort} The PDFs of the normalized velocity increments $\delta_{r} \bar{u} / \bar{u}^{\textrm{rms}}$ in the forced HIT at  at (a) $t/\tau\approx 20.0$; (b) $t/\tau\approx 50.0$. The PDFs of the normalized vorticity $\bar{\omega} / \bar{\omega}^{\textrm{rms}}_{\textrm{fDNS}}$ at (c) $t/\tau\approx 20.0$; (d) $t/\tau\approx 50.0$.}
\end{figure*}

Through previous validation of velocity spectra, it is indicated that the long term dynamics of flow field at $t/\tau \geq 20$ are harder to predict than the short term dynamics, and we will focus on the results at $t/\tau \approx 20$ and $t/\tau \approx 50$. Fig.~\ref{fig:HIT10inc-vort} plots the probability density functions (PDFs) of the normalized velocity increments $\delta_{r} \bar{u} / \bar{u}^{\textrm{rms}}$ and normalized vorticity $\bar{\omega} / \bar{\omega}^{\textrm{rms}}_{\textrm{fDNS}}$ at $t/\tau \approx 20$ and $t/\tau \approx 50$ predicted by different models in the forced HIT. Here, $\delta_{r} \bar{u}=[\bar{\textbf{u}}(\textbf{x}+\textbf{r})-\bar{\textbf{u}}(\textbf{x})]\cdot \hat{\textbf{r}}$, denotes the velocity increment between two points at a distance of $\vert\textbf{r}\vert=\Delta$ ($\Delta$ is the filtering width), and $\hat{\textbf{r}}=\textbf{r}/\vert\textbf{r}\vert$.

As shown in Fig.~\ref{fig:HIT10inc-vort}(a)(b), IAFNO model, IUFNO model and DSM give a satisfying result on the prediction of the PDFs of normalized velocity increment. However, the PDF tails given by AFNO model show a significant upward tendency at the 20th large-eddy turnover time. In Fig.~\ref{fig:HIT10inc-vort}(c)(d), it can be seen that the AFNO model shows a very large deviation for the PDFs of the normalized vorticity at $t/\tau \approx 20$, and this deviation becomes larger with time. As for the IUFNO and IAFNO models, the predictions of PDFs of normalized vorticity fit well with the fDNS data. The results of the DSM slightly shift to the right compared to the benchmark, which in turn makes the DSM inferior to the IAFNO model.

\begin{figure*}[!t]
    \centering
    \subfloat{\hspace{3mm}
    \includegraphics[width=6cm,height = 4.5cm]{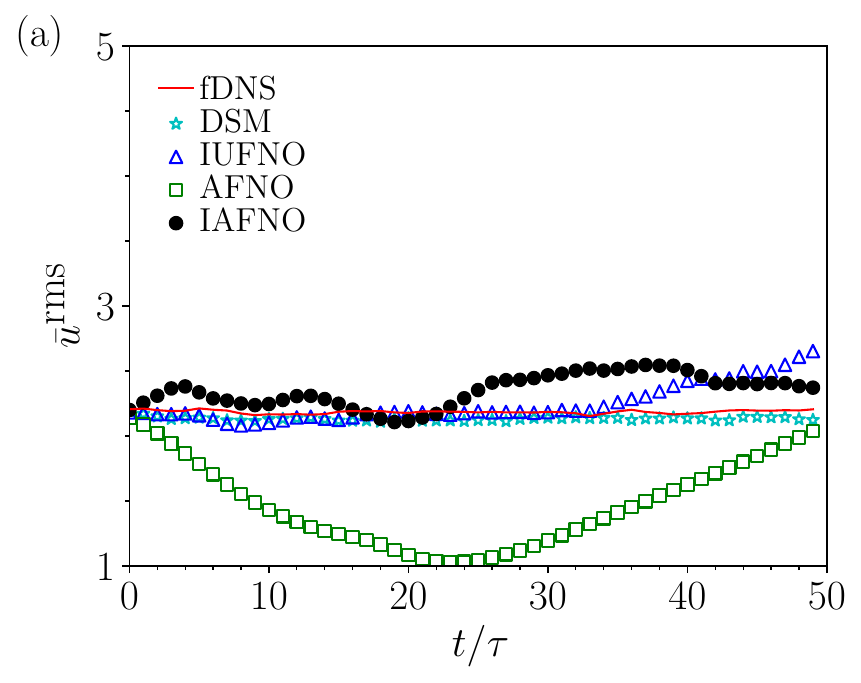}}
    \subfloat{
    \includegraphics[width=6cm,height = 4.5cm]{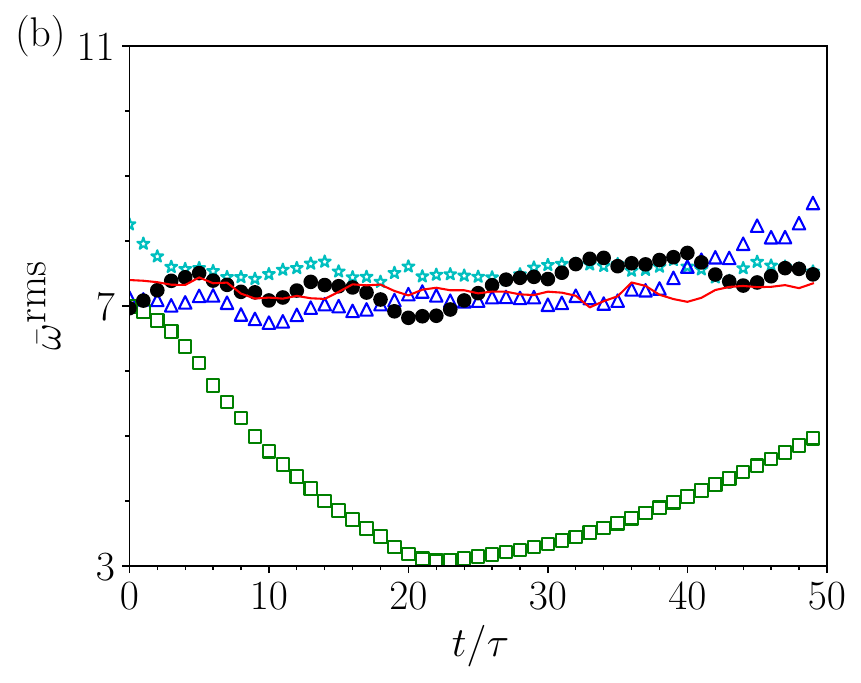}}
    \caption{\label{fig:HIT10rms} Temporal evolutions of the velocity rms value and vorticity rms value in the forced HIT.}
\end{figure*}

After the comparison of the three physical statistics above, it is observed that the AFNO model is unstable in long-term prediction of turbulent flow. Moreover, tracking some physical quantities over time can better demonstrate a model's overall prediction accuracy and stability. We show the temporal evolutions of the rms values of velocity and vorticity predicted by different models in the forced HIT in Fig.~\ref{fig:HIT10rms}.

The rms values of both velocity and vorticity predicted by the AFNO model have a tendency to deviate from the benchmark starting from the beginning, and reach the maximum deviation around $t/\tau \approx22$, thereafter it will regress to the benchmark at a slower rate compared to the deviation process. This result further confirms that the AFNO model is not stable, so the AFNO model will not be discussed in subsequent more complex numerical examples.

Now we focus on the other three results given by DSM, IUFNO and IAFNO. Through the demonstration in Fig.~\ref{fig:HIT10rms}, the IUFNO model is more accurate than the IAFNO model and DSM at $t/\tau \leq 30$. However, the IUFNO model will experience a deviation around $t/\tau \approx 36$, which gradually increases with time, while the IAFNO model can maintain relatively stable values throughout the entire prediction duration. At this point, we recall that the shifting observed in the velocity spectra for IUFNO model at the time of $t/\tau \approx 50$ is consistent with the behavior of the root-mean-square value of the velocity.



    \subsubsection{A complete a posteriori study}

In this subsection we will further compare in detail the performance of the two models, IUFNO and IAFNO, in various aspects for different numbers of implicit layers and epochs. Tab.~\ref{tab:hitinfo} shows the number of implicit layers and training epochs used in all data-driven models of this subsection. 

\begin{table*}[!t]
\centering
\caption{\label{tab:hitinfo}The number of implicit layers and training epochs used in all data-driven models in the forced HIT.}
\renewcommand{\arraystretch}{1.5}
\begin{tabular}{ccc}
\hline\hline
\mbox{Model}&\mbox{Number of Implicit Layers}&\mbox{Epochs}\\
\hline
\mbox{IUFNO$_{(1)}$}&\mbox{20}&\mbox{30}\\
\mbox{IUFNO$_{(2)}$}&\mbox{40}&\mbox{30}\\
\mbox{IAFNO$_{(1)}$}&\mbox{40}&\mbox{30}\\
\mbox{IAFNO$_{(2)}$}&\mbox{20}&\mbox{30}\\
\mbox{IAFNO$_{(3)}$}&\mbox{20}&\mbox{50}\\
\hline\hline
\end{tabular}
\end{table*}

\begin{figure*}[!t]
    \centering
    \subfloat{
    \includegraphics[width=5.5cm,height = 4.31cm]{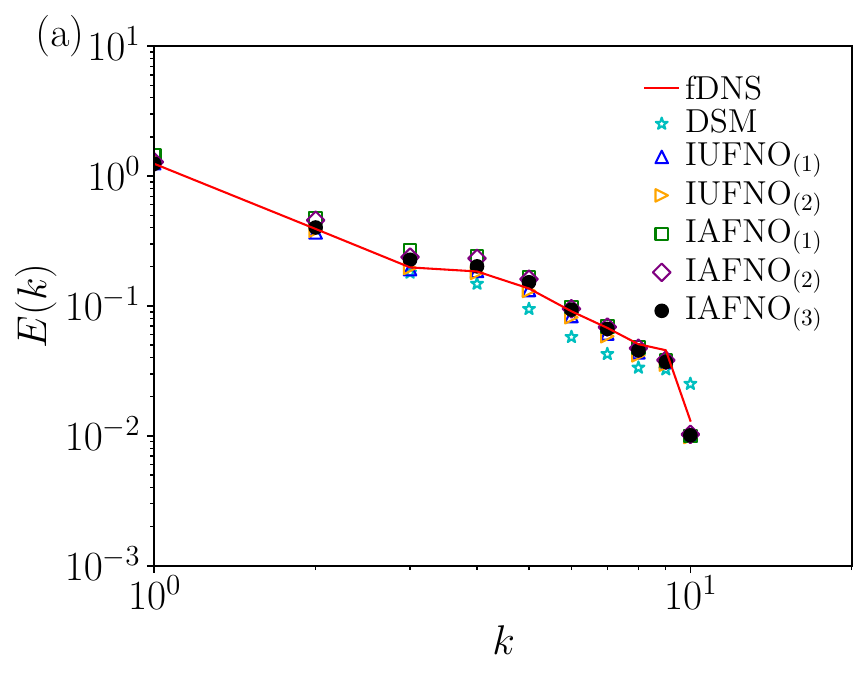}}
    \subfloat{
    \includegraphics[width=5.5cm,height = 4.31cm]{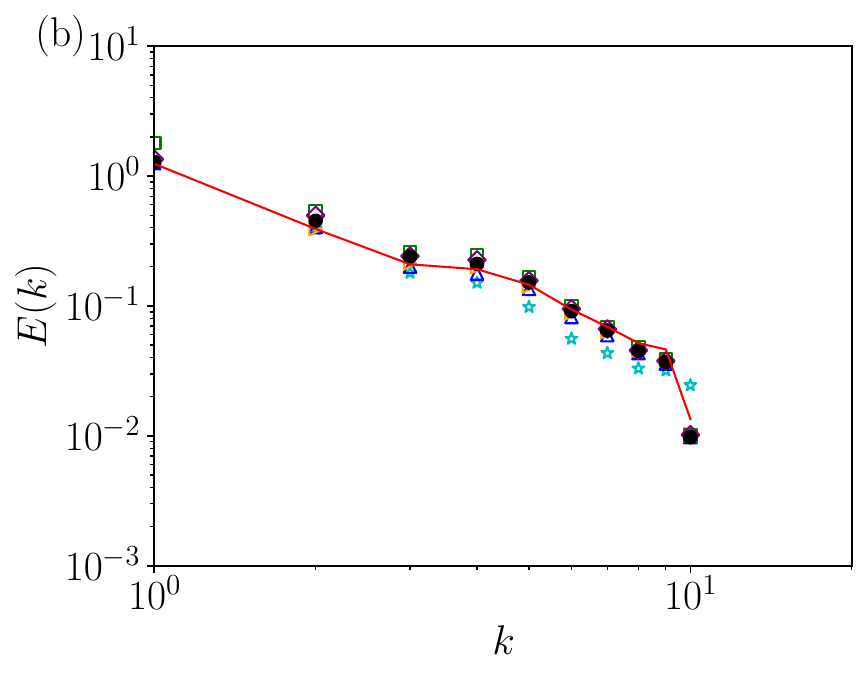}}
    \\
    \subfloat{
    \includegraphics[width=5.5cm,height = 4.31cm]{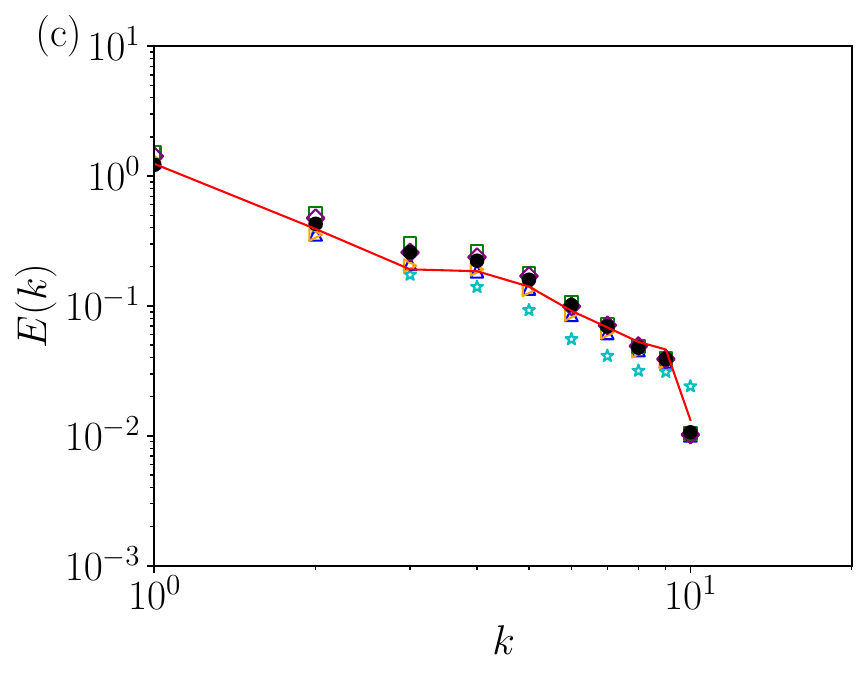}}
    \subfloat{
    \includegraphics[width=5.5cm,height = 4.31cm]{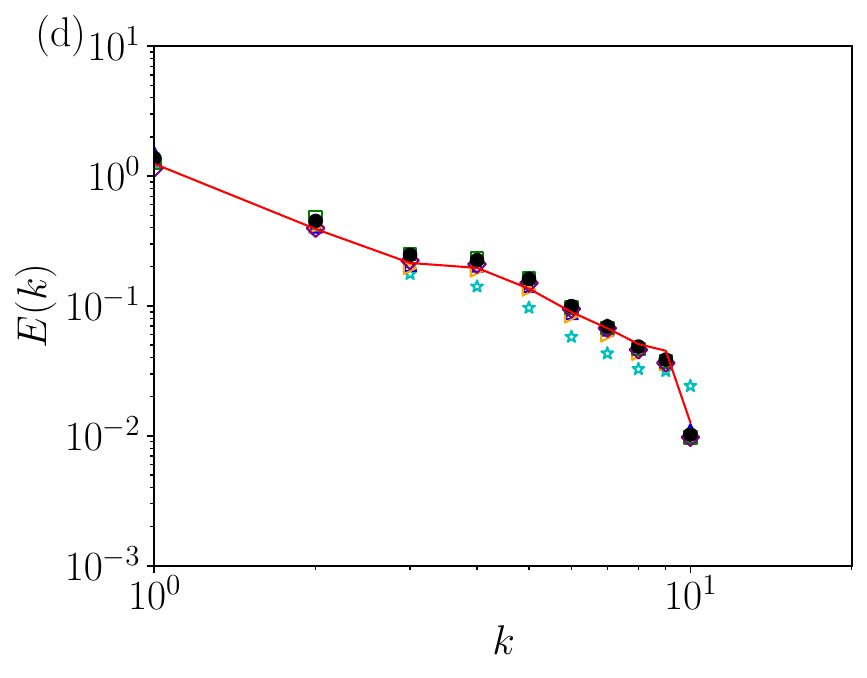}}
    \caption{\label{fig:HIT20spec} The velocity spectra of various models in the forced HIT at different time instants: (a) $t/\tau\approx 4.0$; (b) $t/\tau\approx 6.0$; (c) $t/\tau\approx 8.0$; (d) $t/\tau\approx 50.0$. Here, we magnify the area where $3\leq k\leq7$ to provide a clearer comparison.}
\end{figure*}

To start with, the velocity spectra predicted by IUFNO and IAFNO models with different hyperparameters and the DSM in the forced HIT at different time instants are shown in Fig.~\ref{fig:HIT20spec}. Overall, each data-driven model can accurately construct the velocity spectrum of the flow field at all time, while the DSM model is not satisfactory. A detailed comparison with the benchmark reveals that the predictions of the IUFNO model are more accurate compared to the IAFNO model for $3\leq k\leq5$, and the IAFNO model is more accurate for other values of $k$. In the comparison of the same model with different numbers of implicit layers, for the IUFNO model, the improvement brought by increasing the number of implicit layers from 10 to 20 is obvious, but the benefit is not significant when the number of implicit layers increasing from 20 to 40 layers, implying that the IUFNO model with 20 layers is already performing well enough. A similar phenomenon is observed in the IAFNO model. By comparing IAFNO$_{(1)}$ with IAFNO$_{(2)}$ presented in Fig.~\ref{fig:HIT20spec}, it can be shown that when the number of implicit layers increases from 20 to 40, the performance of the IAFNO model decreases slightly. The comparison between IAFNO$_{(2)}$ and IAFNO$_{(3)}$ shows that the performance improvement brought about by increasing the number of epochs of the IAFNO model is not only reflected in the reduction of testing loss (as shown in Tab.~\ref{tab:hitLoss}), but also in the increased accuracy of the predicted physical statistics.

\begin{figure*}[!t]
    \centering
    \subfloat{\hspace{15mm}
    \includegraphics[width=5.5cm,height = 4.31cm]{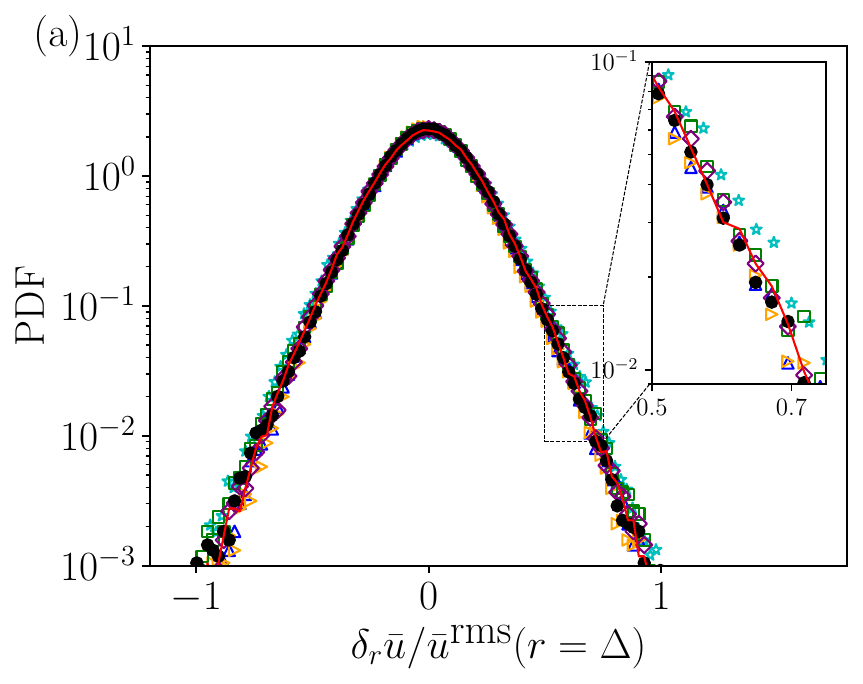}}
    \subfloat{
    \includegraphics[width=6.95cm,height = 4.31cm]{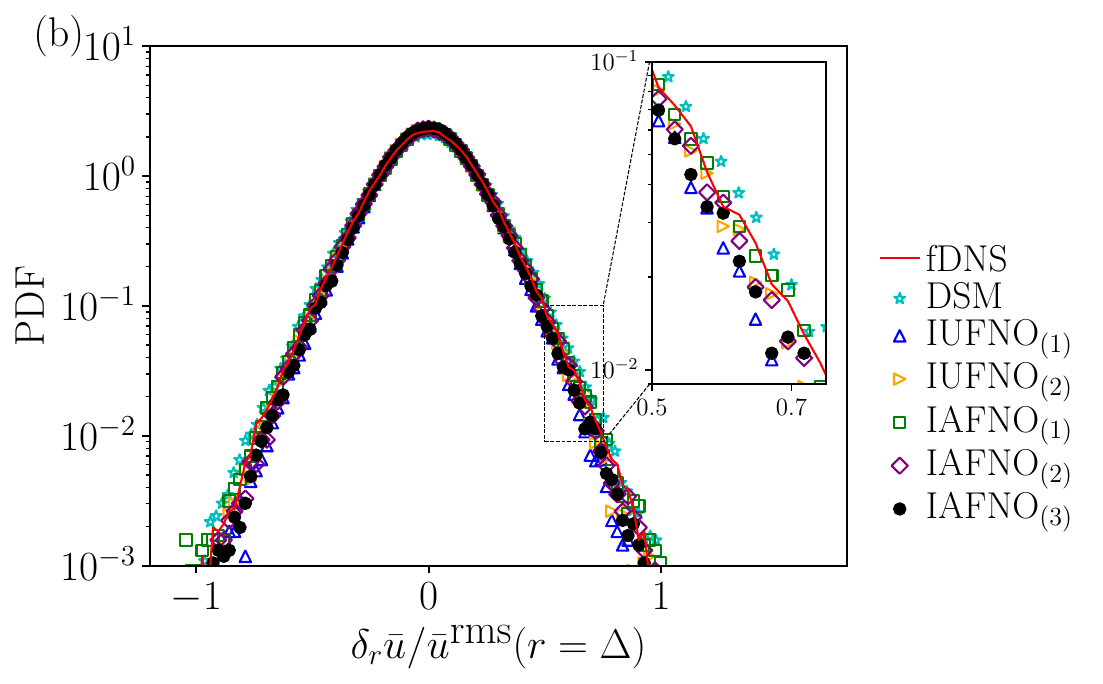}}
    \\
    \subfloat{
    \includegraphics[width=5.5cm,height = 4.31cm]{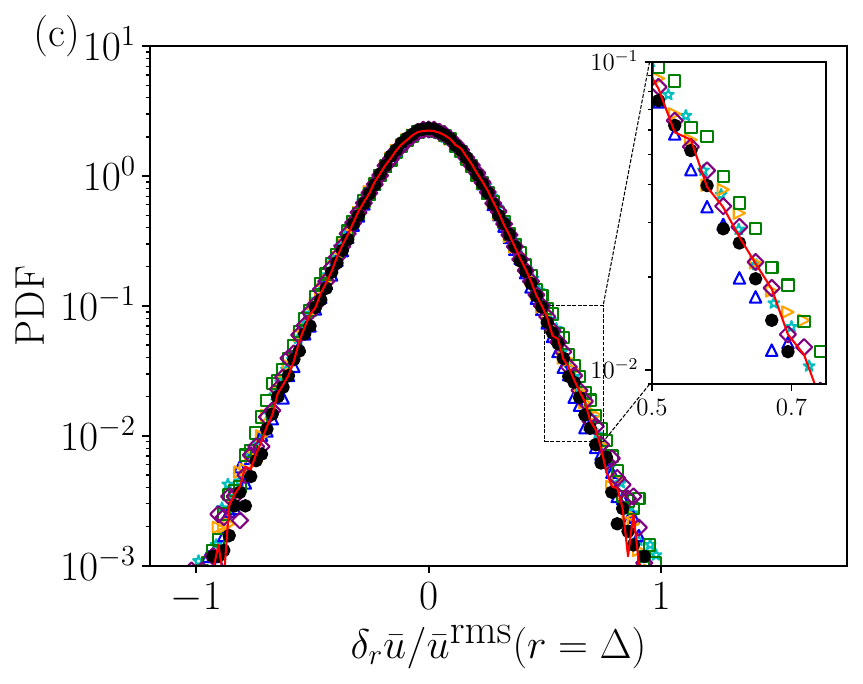}}
    \subfloat{
    \includegraphics[width=5.5cm,height = 4.31cm]{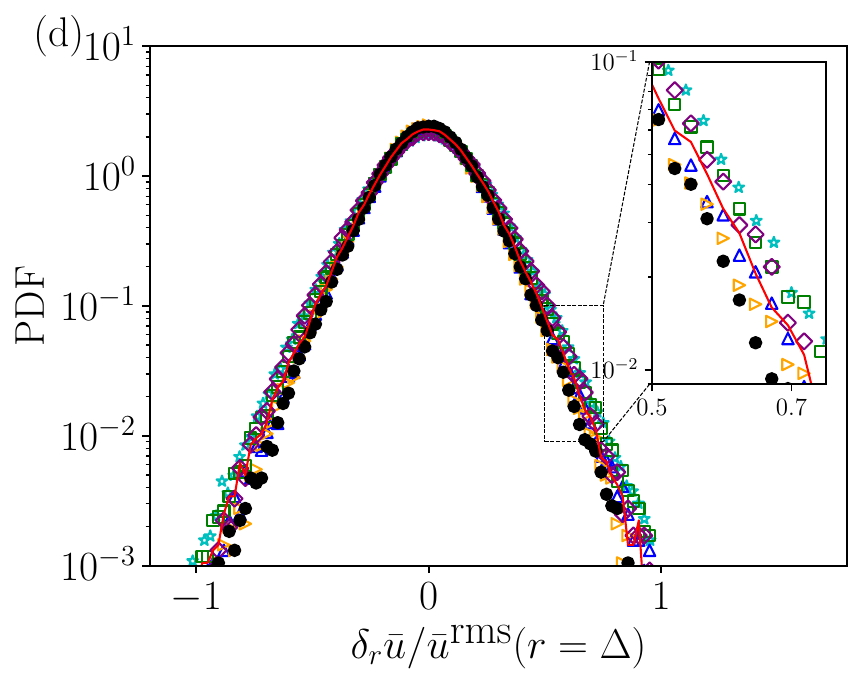}}
    \caption{\label{fig:HIT20vel} The PDFs of the normalized velocity increments $\delta_{r} \bar{u} / \bar{u}^{\textrm{rms}}$ of various models in the forced HIT at different time instants: (a) $t/\tau\approx 4.0$; (b) $t/\tau\approx 6.0$; (c) $t/\tau\approx 8.0$; (d) $t/\tau\approx 50.0$. Here, we magnify the area where $0.5\leq \delta_r\bar{u}/\bar{u}^{\textrm{rms}}\leq0.75$ to provide a clearer comparison.}
\end{figure*}

PDFs of the normalized velocity increments predicted by different models are plotted in Fig.~\ref{fig:HIT20vel}. It is shown that all models perform well enough at first glance. However, when we take a closer look at Fig.~\ref{fig:HIT20vel}, we can see that the overall performance of IAFNO$_{(2)}$ is better than other IAFNO models, and also slightly better than IUFNO$_{(2)}$.

\begin{figure*}[!t]
    \centering
    \subfloat{\hspace{15mm}
    \includegraphics[width=5.5cm,height = 4.31cm]{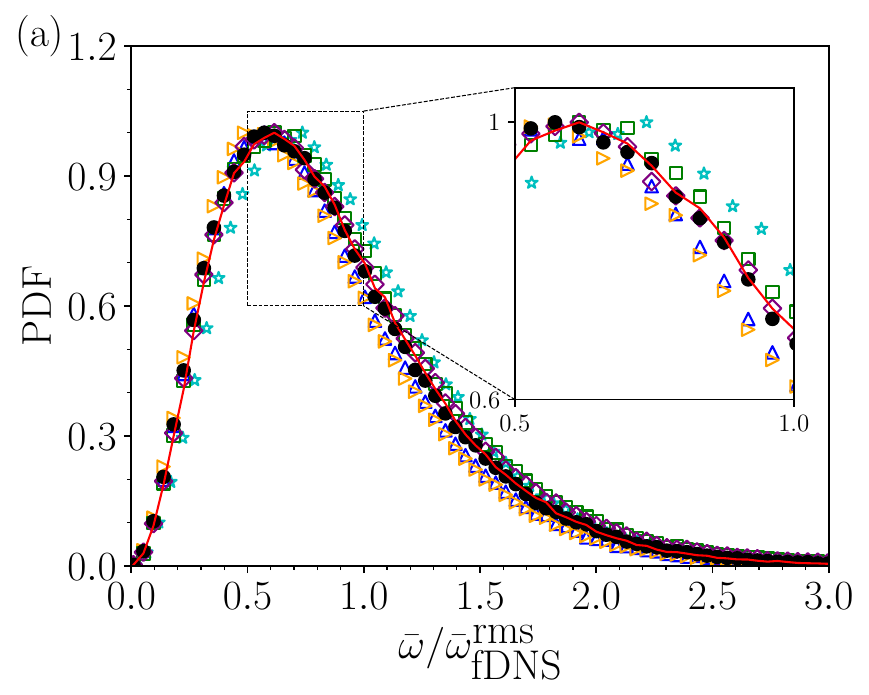}}
    \subfloat{
    \includegraphics[width=6.95cm,height = 4.31cm]{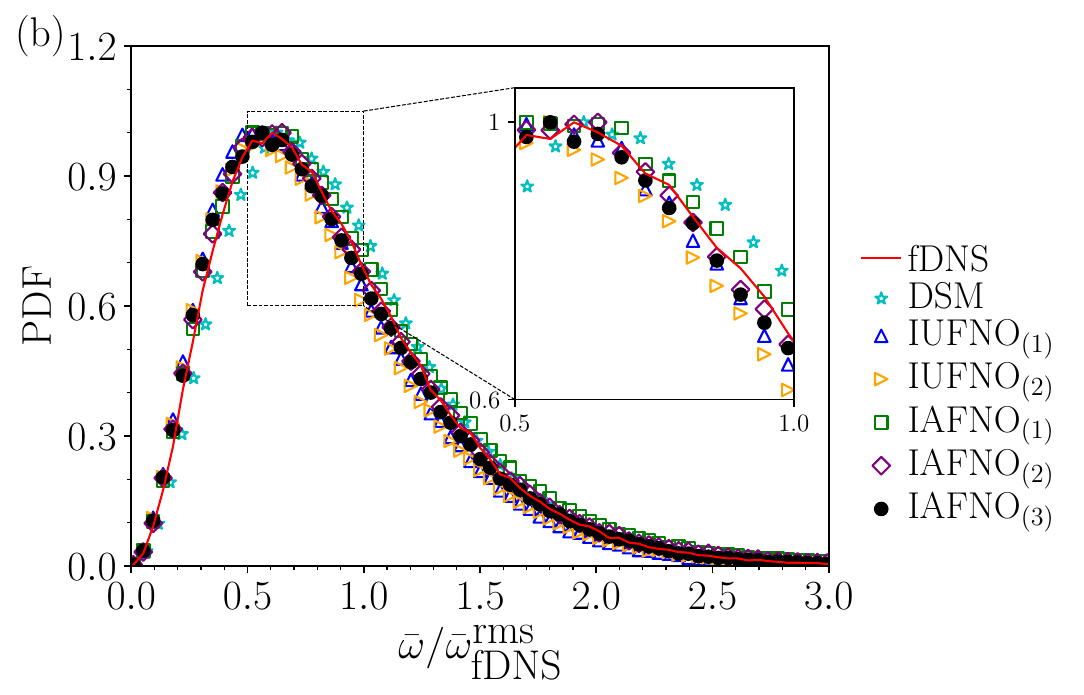}}
    \\
    \subfloat{
    \includegraphics[width=5.5cm,height = 4.31cm]{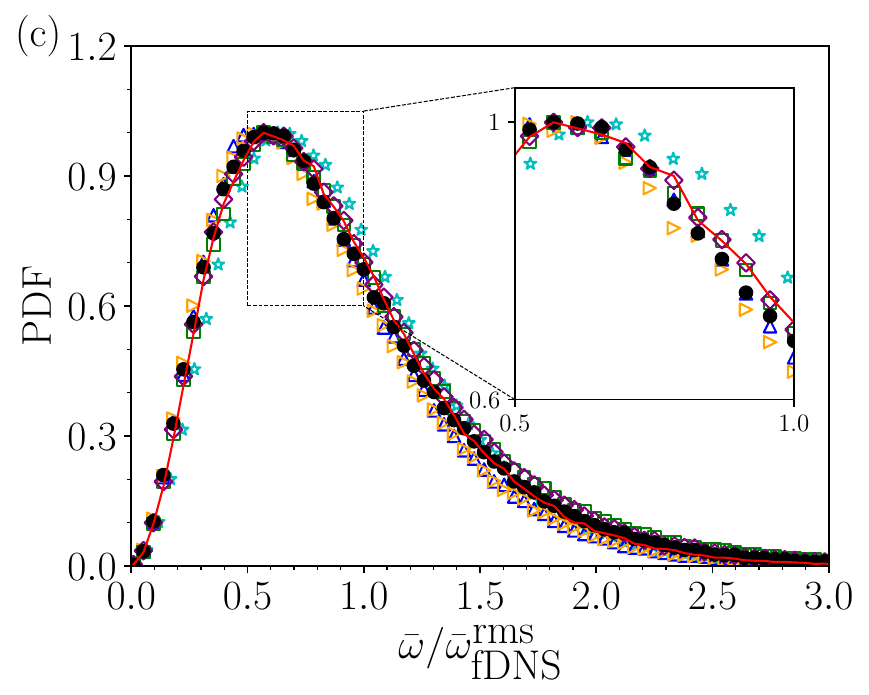}}
    \subfloat{
    \includegraphics[width=5.5cm,height = 4.31cm]{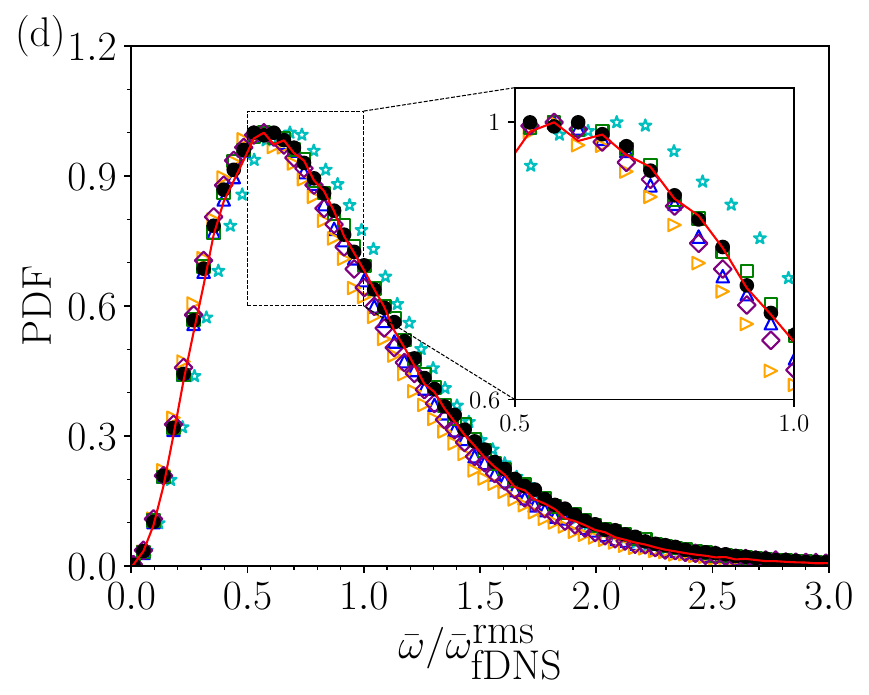}}
    \caption{\label{fig:HIT20vort} The PDFs of the normalized vorticity $\bar{\omega} / \bar{\omega}^{\textrm{rms}}_{\textrm{fDNS}}$ of various models in the forced HIT at different time instants: (a) $t/\tau\approx 4.0$; (b) $t/\tau\approx 6.0$; (c) $t/\tau\approx 8.0$; (d) $t/\tau\approx 50.0$. Here, we magnify the area where $0.5\leq \bar{\omega}/\bar{\omega}^{\textrm{rms}}_{\textrm{fDNS}}\leq1.0$ to provide a clearer comparison.}
\end{figure*}

Furthermore, we compare the PDFs of the normalized vorticity $\bar{\omega} / \bar{\omega}^{\textrm{rms}}_{\textrm{fDNS}}$ predicted by various data-driven models at different time instants in Fig.~\ref{fig:HIT20vort}. It can be seen that the IAFNO models outperform the IUFNO models and DSM, and IAFNO$_{(3)}$ has the best performance since most of the black points representing IAFNO$_{(3)}$ overlap the red line.

\begin{figure*}[!ht]
    \centering
    \subfloat{
    \includegraphics[width=5.5cm,height = 4.31cm]{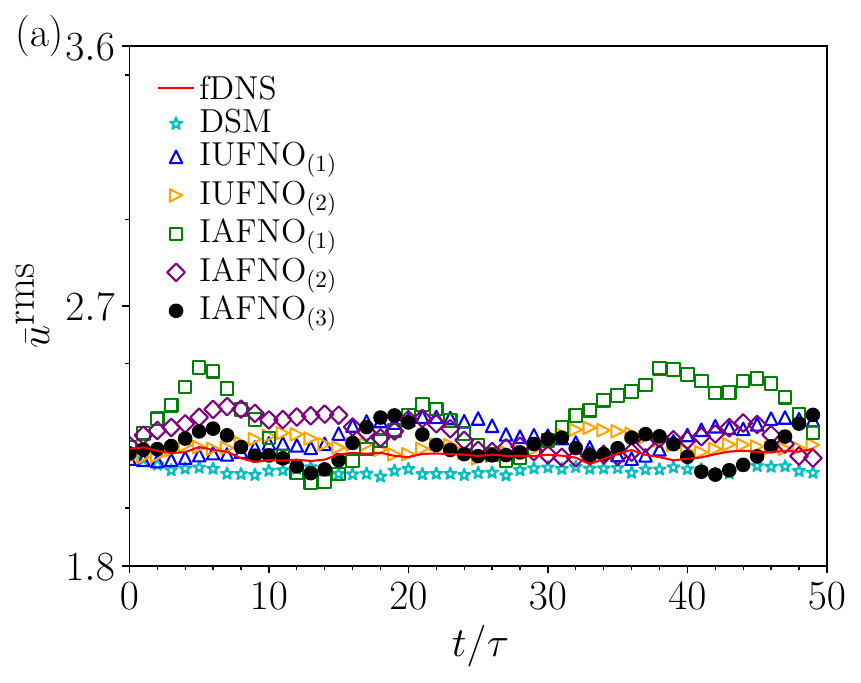}}
    \subfloat{
    \includegraphics[width=5.5cm,height = 4.31cm]{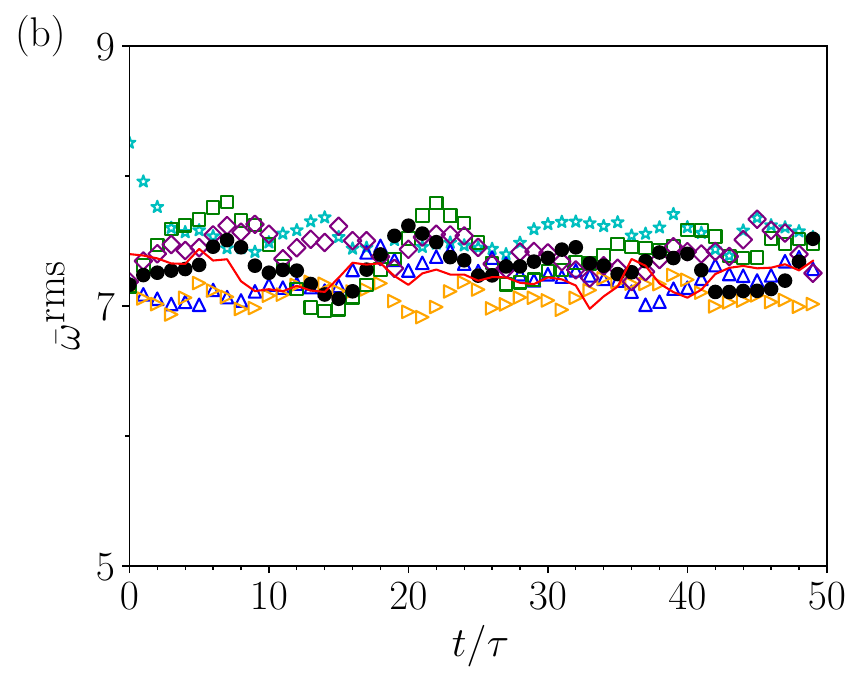}}
    \caption{\label{fig:HIT20rms} Temporal evolutions of the velocity rms value and vorticity rms value of various models in the forced HIT.}
\end{figure*}

The temporal evolutions of the rms values of velocity and vorticity predicted by the IAFNO and IUFNO models with different hyperparameters are shown in Fig.~\ref{fig:HIT20rms}. In Fig.~\ref{fig:HIT20rms}(a), it is observed that velocity rms value predicted by all data-driven models will oscillate to some degree relative to the benchmark with time, while the DSM gives a slightly downward-shifted result with no occurrence of oscillations. Among them, IAFNO$_{(1)}$ experiences a prolonged period of significant deviation relative to the other models at time $t/\tau \approx30$. For the other models, IAFNO$_{(3)}$'s performance is slightly inferior to the two IUFNO models. Overall, the results in Fig.~\ref{fig:HIT20rms}(a) are consistent with the results of the physical statistics shown in the previous figures (see Fig.~\ref{fig:HIT20spec} and Fig.~\ref{fig:HIT20vel}). In Fig.~\ref{fig:HIT20rms}(b), it is observed that IAFNO$_{(3)}$'s performance is slightly inferior to IUFNO$_{(2)}$. However, IAFNO$_{(3)}$ outperforms the other models including DSM.

\begin{figure*}[!t]
    \centering
    \includegraphics[width=12cm,height=13.16cm]{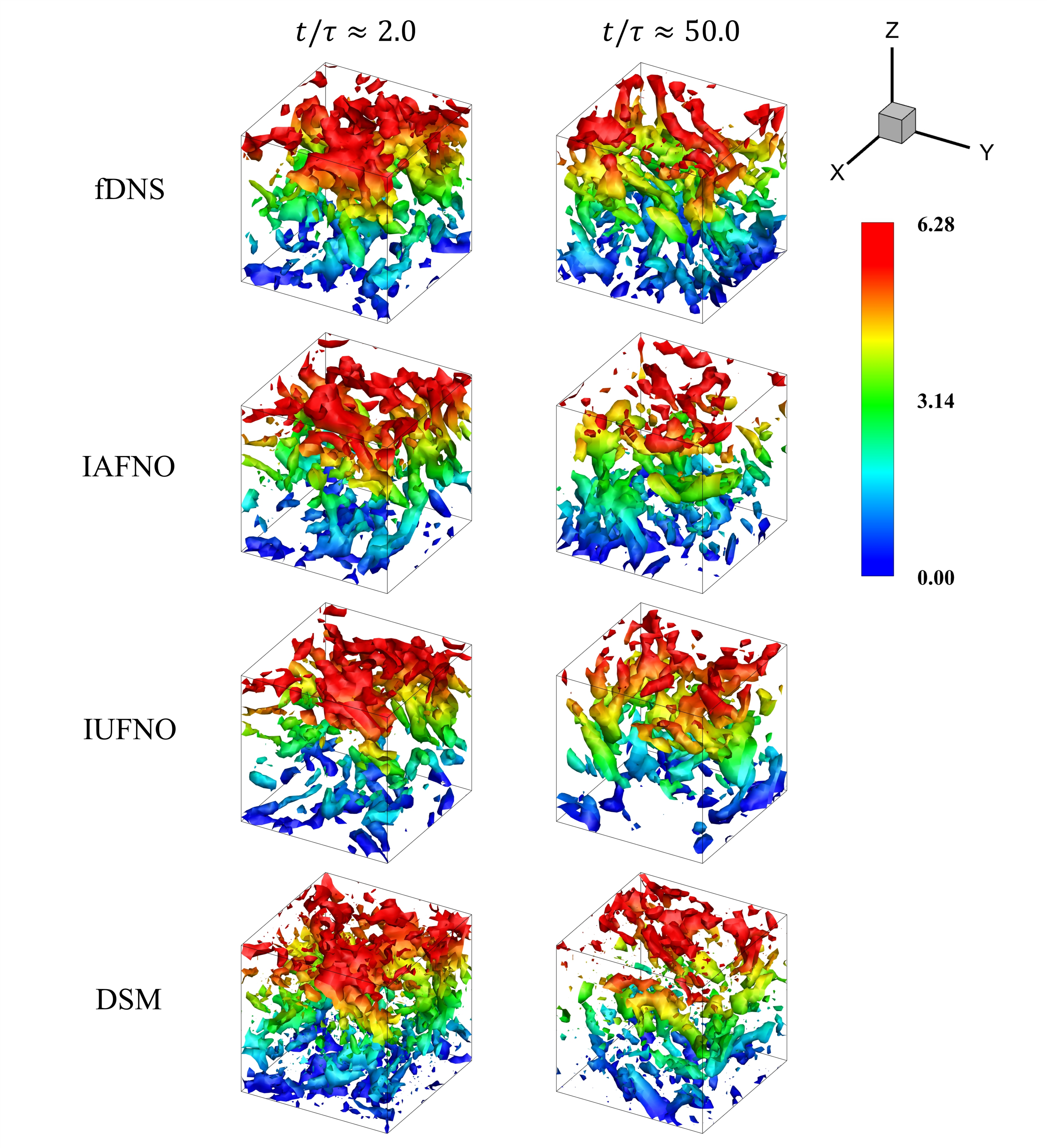}
    \caption{\label{fig:HITCoutour} Isosurface of the normalized vorticity $\bar{\omega} / \bar{\omega}^{\textrm{rms}}_{\textrm{fDNS}}=1.0$ (colored by altitude of z-direction) at $t/\tau\approx 2.0$ and $t/\tau\approx 50.0$ for HIT.}
\end{figure*}

We have shown that the IAFNO model works well in the prediction of several statistics of physical variables. Moreover, we demonstrate isosurfaces of the normalized vorticity $\bar{\omega} / \bar{\omega}^{\textrm{rms}}_{\textrm{fDNS}}=1.0$ (colored by altitude of z-direction) of the IAFNO, IUFNO model and DSM at $t/\tau\approx 2.0$ and $t/\tau\approx 50.0$ in Fig.~\ref{fig:HITCoutour}. IAFNO$_{(3)}$ and IUFNO$_{(2)}$ are chosen here. We can see from Fig.~\ref{fig:HITCoutour} that both IAFNO model and IUFNO model are capable of predicting the spatial structures of the vorticity field. However, the DSM has the least similarity to the benchmark.  

\subsection{Temporally evolving turbulent mixing layer}

In the previous section we have shown some advantages of the IAFNO model in predicting 3D forced homogeneous isotropic turbulence compared to the IUFNO model and the DSM. However, due to the relative simplicity of the HIT dataset, it is not sufficient to demonstrate the model performance adequately. Therefore, in this section we will further test the prediction ability of the IAFNO model in a 3D free-shear turbulent mixing layer, which is a more complex turbulent flow field.

The free-shear turbulent mixing layer is also governed by the Navier–Stokes equations as given in Eq.~\ref{eq:NS1} and Eq.~\ref{eq:NS2} without the forcing term $\mathcal{F}_i$. In Tab.~\ref{tab:fsfdns}, it is shown that the mixing layer is numerically simulated in a cuboid domain with lengths $L_1\times L_2\times L_3=8\pi\times8\pi\times4\pi$ using a uniform grid resolution of $N_1\times N_2\times N_3=256\times256\times128$. Here, $x_1\in[-4\pi, 4\pi], x_2\in[-4\pi, 4\pi]$ and $x_3\in[-2\pi, 2\pi]$ denote the streamwise, transverse, and spanwise coordinates, respectively \cite{yuan2023adjoint,li2023long}. In addition, $[x_1, x_2, x_3]$ is also denoted as $[x, y, z]$, which means that any physical quantity subscripted by 2 represents the component of its vector form in the $y$ direction.

\begin{table*}[!ht]
\centering
\caption{\label{tab:fsfdns}Parameters and statistics for DNS and fDNS of 3D free-shear turbulent mixing layer.}
\renewcommand{\arraystretch}{1.5}
\begin{tabular}{cccccccc}
\hline\hline
\mbox{Reso.(DNS:$N_1\times N_2\times N_3$)}&\mbox{Reso.(fDNS:$N_1\times N_2\times N_3$)}&\mbox{Domain}&\mbox{$Re^0_{\theta}$}&\mbox{$\nu$}&\mbox{$\mathrm{d} t$}&\mbox{$\delta^0_{\theta}$}&\mbox{$\Delta U$}\\
\hline
\mbox{$256\times256\times128$}&\mbox{$64\times64\times32$}&\mbox{$8\pi\times8\pi\times4\pi$}&\mbox{320}&\mbox{0.008}&\mbox{0.002}&\mbox{0.08}&\mbox{2}\\
\hline\hline
\end{tabular}
\end{table*}

The initial streamwise velocity is given by \cite{wang2022constant,wang2022compressibility,yuan2023adjoint}:
\end{multicols}
\begin{equation}
 u_1=\frac{\Delta U}{2}\left[\mathrm{tanh}\left(\frac{x_2}{2\delta^0_{\theta}}\right)-\mathrm{tanh}\left( \frac{x_2+L_2/2}{2\delta^0_{\theta}} \right)-\mathrm{tanh}\left( \frac{x_2-L_2/2}{2\delta^0_{\theta}} \right) \right]+\lambda_1 ~, \label{eq:fsvx}
\end{equation}
\begin{multicols}{2} where $\delta^0_{\theta}=0.08$ is the initial momentum thickness and $\Delta U=U_2-U_1$ is the velocity difference between two equal and opposite free streams across the shear layer \cite{yuan2023adjoint,wang2022constant}. The momentum thickness quantifies the range of turbulence region in the mixing layer, which is given by \cite{rogers1994direct}:
\begin{equation}
\delta_{\theta}=\int_{-L_2/2}^{L_2/2}\left[ \frac{1}{4}-\left( \frac{\langle\bar{u}_1\rangle}{\Delta U} \right)^2 \right]\mathrm{d}x_2~. \label{eq:fsmt}
\end{equation}

The initial normal and spanwise velocities are given as $u_2=\lambda_2,~u_3=\lambda_3$. Here, $\lambda_1,\lambda_2,\lambda_3\sim N(\mu,\sigma^2)$, i.e., $\lambda_1,\lambda_2,\lambda_3$ satisfy the Gaussian random distribution. The mean is $\mu=0$ and the variance is $\sigma^2=0.01$ \cite{li2023long}. The Reynolds number based on the momentum thickness $Re_{\theta}$ is defined as $Re_{\theta}=\Delta U\delta_{\theta}/\nu_{\infty}$. Here, the kinematic viscosity of shear layer is set to $\nu_{\infty}=5\times10^{-4}$, so the initial momentum thickness Reynolds number is $Re^0_{\theta}=320$ \cite{yuan2023adjoint}.

To mitigate the impact of the top and  bottom boundaries on the central mixing layer, two numerical diffusion buffer zones are implemented to the vertical edges of the computational domain \cite{yuan2023adjoint,wang2022constant,wang2022compressibility}. The periodic boundary conditions in all three directions are utilized and the pseudo-spectral method with the two-thirds dealiasing rule is employed for the spatial discretization. Moreover, an explicit two-step Adam-Bashforth scheme is chosen as the time-advancing scheme.

The DNS data are then explicitly filtered by the commonly-used Gaussian filter, which  is defined by \cite{pope2000turbulent,sagaut2005large}:
\begin{equation}
G(\textbf{r};\bar{\Delta})=\left( \frac{6}{\pi\bar{\Delta}^2} \right)^{1/2}\mathrm{exp}\left( -\frac{6\textbf{r}^2}{\bar{\Delta}^2} \right)~. \label{eq:fsgf}
\end{equation} Here, the filter scale $\bar{\Delta}=8h_{\mathrm{DNS}}$ is selected for the free-shear turbulent mixing layer, where $h_{\mathrm{DNS}}$ is the grid spacing of DNS. The filter-to-grid ratio $\mathrm{FGR}=\bar{\Delta}/h_{\mathrm{LES}}=2$ is utilized and then the corresponding grid resolution of fDNS: the grid number of $64\times64\times32$ can be obtained \cite{wang2022constant,chang2022effect}.

\begin{table}[!ht]
\centering
\caption{\label{tab:fsinfo}The number of implicit layers and training batchsize used in all the data-driven models in the free-shear turbulent mixing layer.}
\renewcommand{\arraystretch}{1.5}
\begin{tabular}{ccc}
\hline\hline
\mbox{Model}&\mbox{Number of Implicit Layers}&\mbox{Batchsize}\\
\hline
\mbox{IUFNO$_{(1)}$}&\mbox{20}&\mbox{2}\\
\mbox{IUFNO$_{(2)}$}&\mbox{40}&\mbox{2}\\
\mbox{IAFNO$_{(1)}$}&\mbox{20}&\mbox{2}\\
\mbox{IAFNO$_{(2)}$}&\mbox{20}&\mbox{5}\\
\hline\hline
\end{tabular}
\end{table}

By the method described above we generate a total of 150 sets of flow field data with different initial conditions, and we save the results for 90 temporal snapshots in each set, of which we have kept 5 sets as validation sets and the remaining 145 sets will be used as training and testing sets. The time interval for each snapshot is $200dt$,  where $dt = 0.002$ is the time step of DNS. Therefore, a datasets of size $[145\times90\times64\times64\times32\times3]$ is used as training ($80\%$) and testing ($20\%$), which is similar to the situation of HIT. The inference (prediction) process has been described in Eq.~\ref{eq:predict}, and the loss function has been defined in Eq. ~\ref{eq:lossfunction}.

We have obtained the trained raw data (IUFNO model with $L=40$) used in the previous paper \cite{li2023long}, and with the permission of the authors, we now compare it with the IAFNO model. The number of implicit layers and training batchsize settings for all data-driven models in the free-shear turbulent mixing layer are shown in Tab.~\ref{tab:fsinfo}. The other unchanged hyperparameters for IUFNO and IAFNO are shown in Tab.~\ref{tab:fshypsContinuedIU} and Tab.~\ref{tab:fshypsContinuedIA}, respectively.

\begin{table*}[htb]
\centering
\caption{\label{tab:fshypsContinuedIU}The other hyperparameters settings for IUFNO in the free-shear turbulent mixing layer.}
\renewcommand{\arraystretch}{1.5}
\begin{tabular}{ccc}
\hline\hline
\mbox{Epochs}&\mbox{LearningRate(LR)}&\mbox{Width}\\
\hline
\mbox{30}&\mbox{0.001}&\mbox{36}\\
\hline\hline
\end{tabular}
\end{table*}

\begin{table*}[htb]
\centering
\caption{\label{tab:fshypsContinuedIA}The other hyperparameters settings for IAFNO in the free-shear turbulent mixing layer.}
\renewcommand{\arraystretch}{1.5}
\begin{tabular}{cccccc}
\hline\hline
\mbox{Epochs}&\mbox{LearningRate(LR)}&\mbox{Patchsize}&\mbox{EmbedDim}&\mbox{HSF}&\mbox{\#ofBlocks}\\
\hline
\mbox{30}&\mbox{0.001}&\mbox{(2,2,2)}&\mbox{162}&\mbox{3}&\mbox{1}\\
\hline\hline
\end{tabular}
\end{table*}

\begin{figure*}[!t]
    \centering
    \subfloat{\hspace{2.5mm}
    \includegraphics[width=5.81cm,height = 4.5cm]{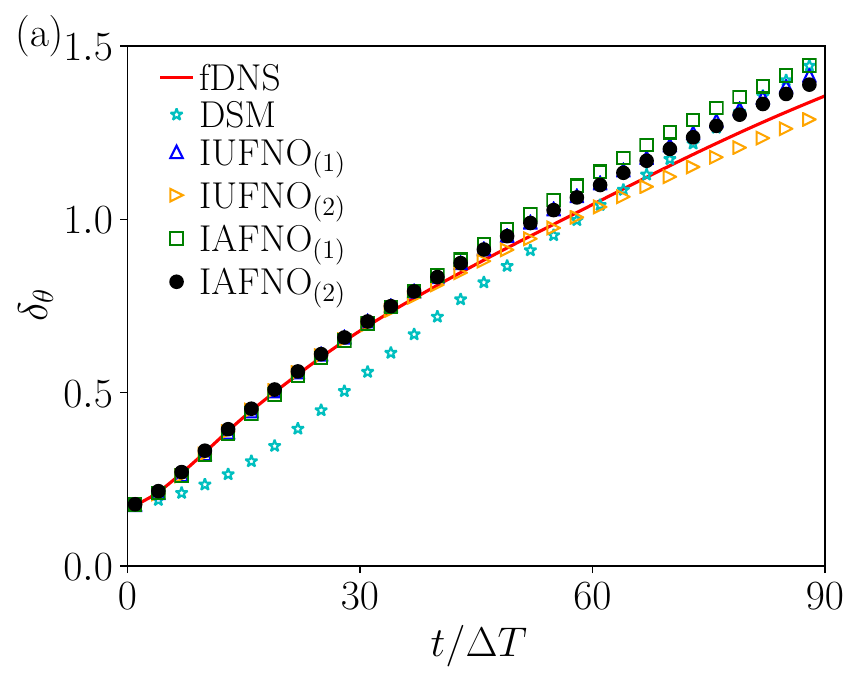}}
    \subfloat{
    \includegraphics[width=6cm,height = 4.5cm]{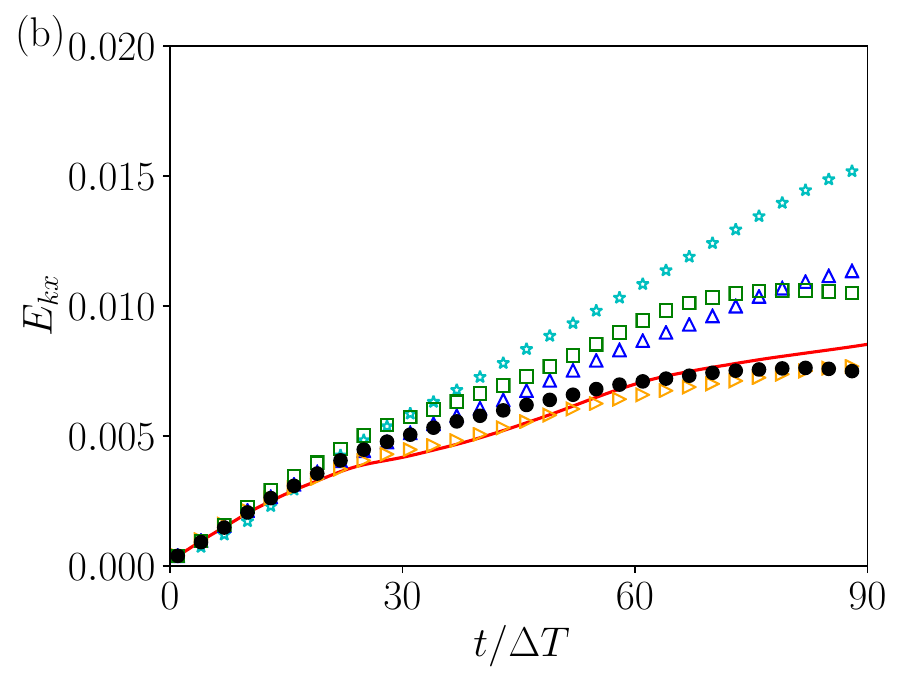}}
    \\
    \subfloat{
    \includegraphics[width=6cm,height = 4.5cm]{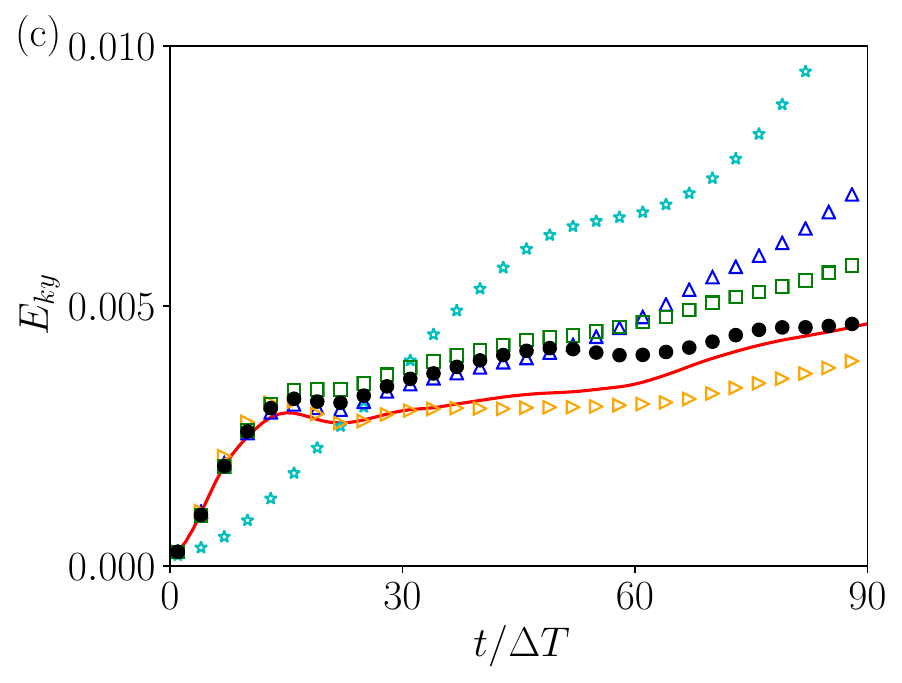}}
    \subfloat{
    \includegraphics[width=6cm,height = 4.5cm]{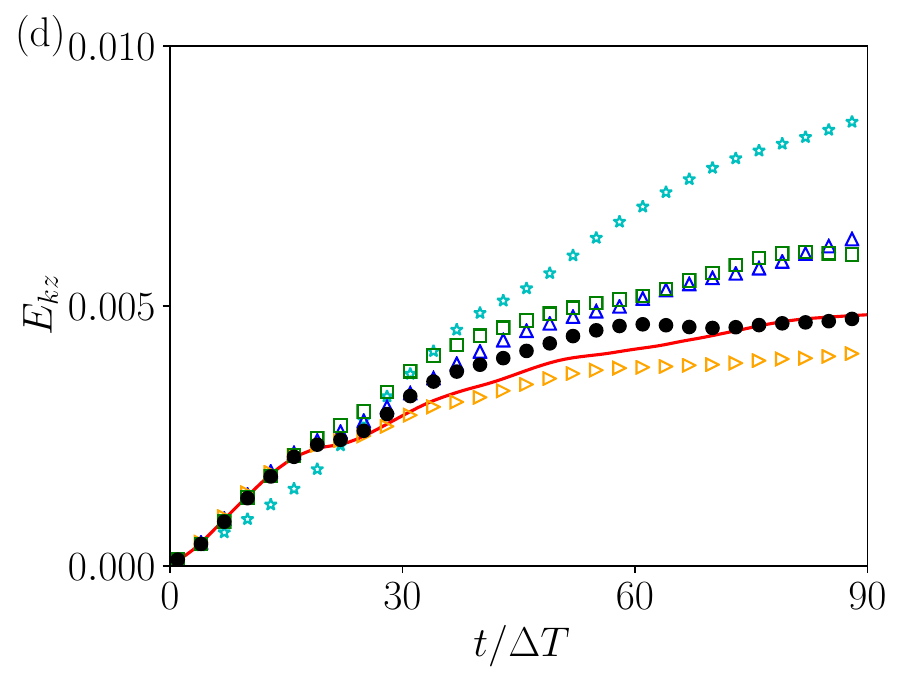}}
    \caption{\label{fig:FSEk} Temporal evolutions of different physical quantities in the free-shear turbulent mixing layer: (a) the momentum thickness $\delta_{\theta}$; (b) streamwise turbulent kinetic energy $E_{kx}$; (c) transverse turbulent kinetic energy $E_{ky}$; (d) spanwise turbulent kinetic energy $E_{kz}$.}
\end{figure*}

Fig.~\ref{fig:FSEk}(a) presents the temporal evolutions of the momentum thickness $\delta_{\theta}$ using different models. Here, dimensionless time interval: $\Delta T=200dt=0.4$. It can be seen that both IAFNO$_{(2)}$ and IUFNO$_{(2)}$ perform well since they are closer to the benchmark, while IAFNO$_{(1)}$ and IUFNO$_{(1)}$ perform slightly worse. The result of DSM deviates obviously when $9\leq t/\Delta T \leq 48$, and gives the worst prediction among all models.

The temporal evolutions of the turbulent kinetic energy in the streamwise direction $E_{kx}$ is shown in Fig.~\ref{fig:FSEk}(b), where $E_{kx}=E_{k1}=\frac{1}{2}(\sqrt{\langle u_1u_1 \rangle})^2$. It is revealed that IAFNO$_{(1)}$ and IUFNO$_{(1)}$ deviate from the benchmark when $t/\Delta T\geq 50$, and IAFNO$_{(2)}$ exhibits slightly inaccurate tendency when $t/\Delta T\geq 75$, while IUFNO$_{(2)}$ performs the best during the whole development of the shear layer. The DSM model has shown a very large up-tilt since $t/\Delta T\geq 42$, which is much larger than IUFNO$_{(1)}$ and IAFNO$_{(1)}$. Moreover, we compare the temporal evolutions of the turbulent kinetic energy in the transverse and spanwise directions of different models in Fig.~\ref{fig:FSEk}(c)(d). It is shown that IAFNO$_{(2)}$ outperforms other models even though a deviation occurs when $30\leq t/\Delta T \leq 60$ in both transverse and spanwise directions.

\begin{figure*}[!t]
    \centering
    \subfloat{
    \includegraphics[width=6cm,height = 4.5cm]{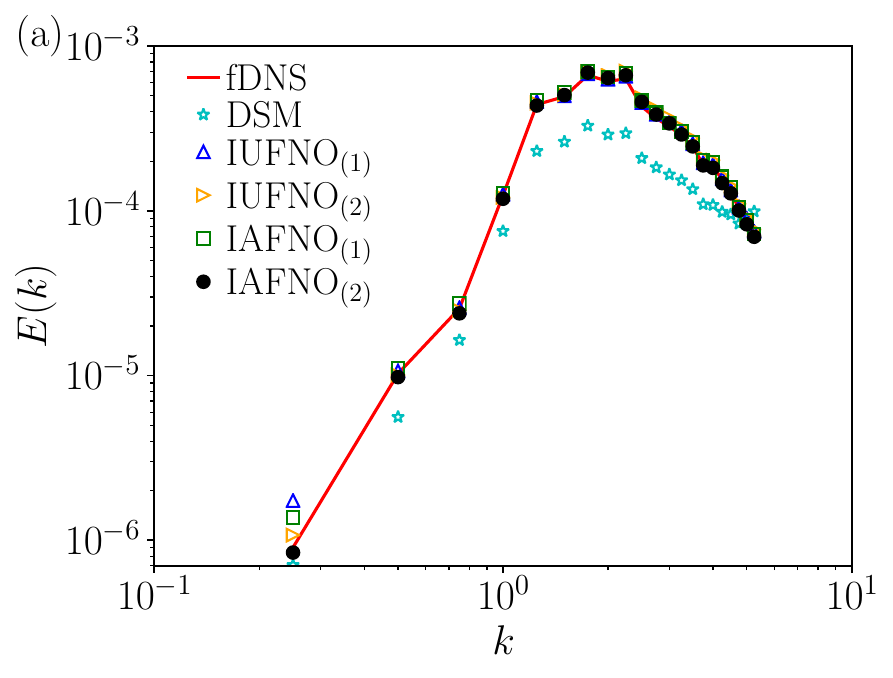}}
    \subfloat{
    \includegraphics[width=6cm,height = 4.5cm]{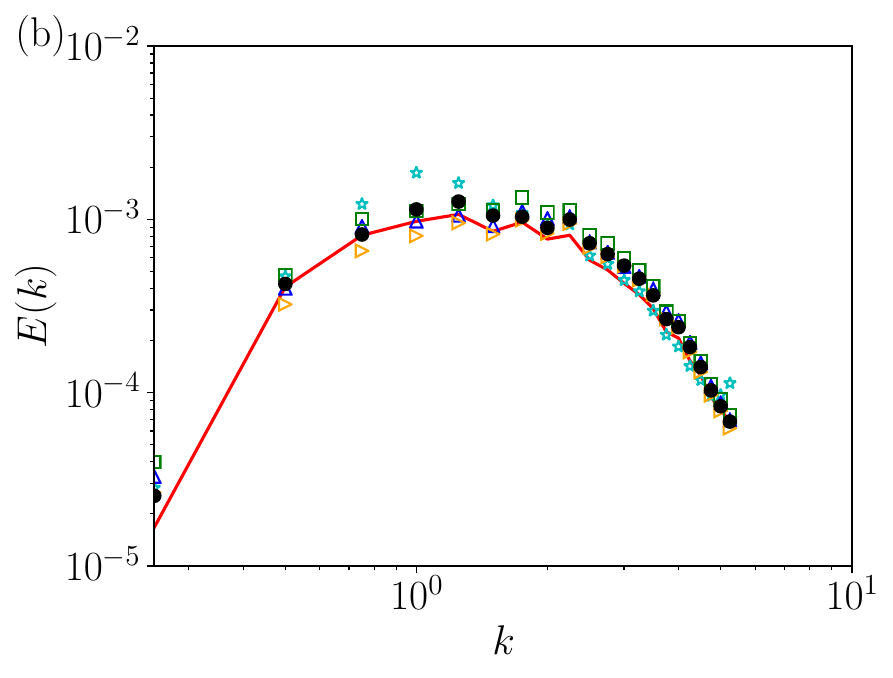}}
    \\
    \subfloat{
    \includegraphics[width=6cm,height = 4.5cm]{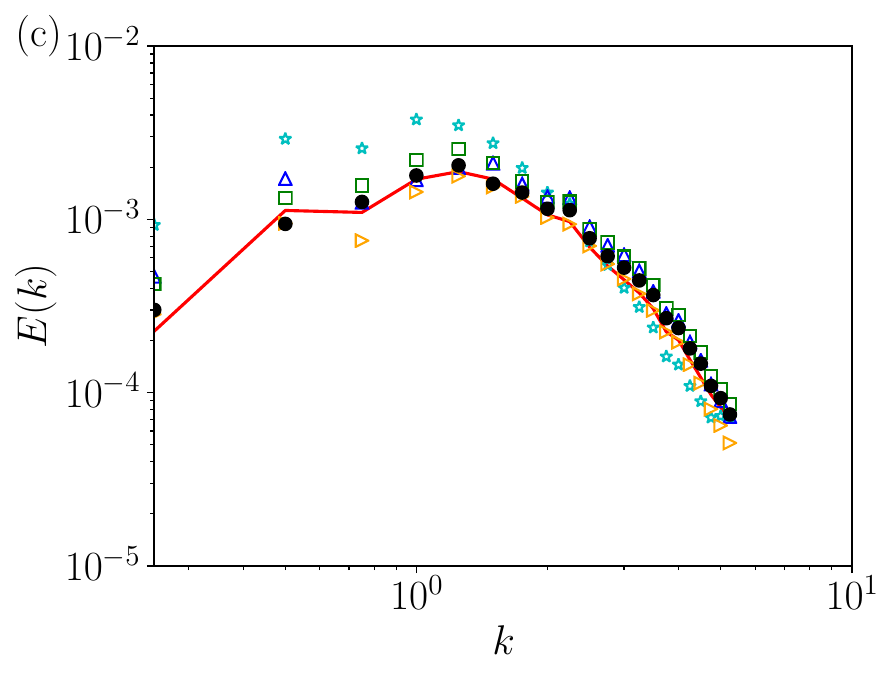}}
    \subfloat{
    \includegraphics[width=6cm,height = 4.5cm]{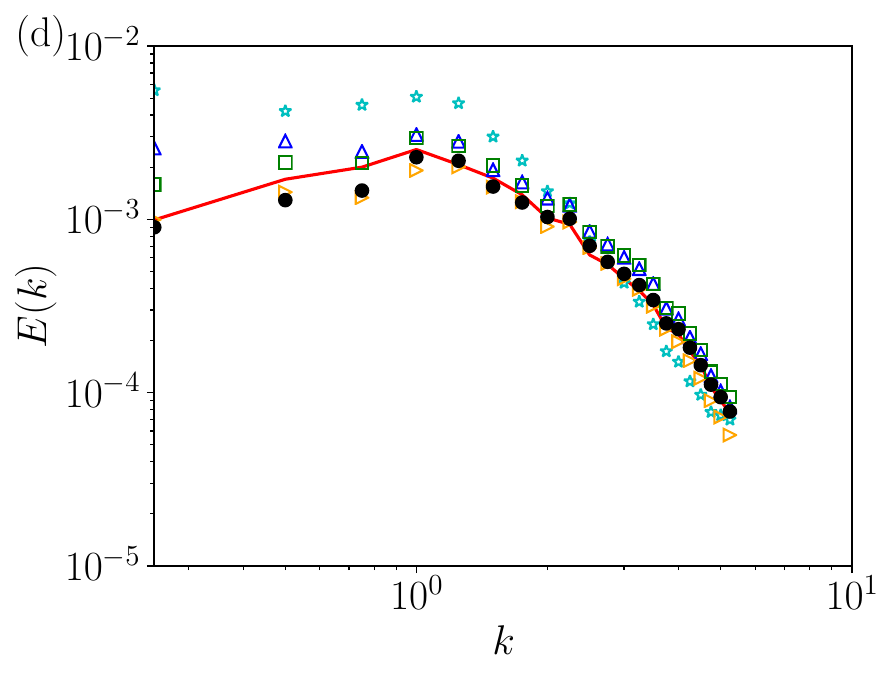}}
    \caption{\label{fig:FSspec} The energy spectra in the free-shear turbulent mixing layer at different time steps: (a) $t/\Delta T \approx 10$; (b) $t/\Delta T \approx 30$; (c) $t/\Delta T \approx 60$; (d) $t/\Delta T \approx 90$.}
\end{figure*}

Furthermore, the energy spectra predicted by various models in the free-shear turbulent mixing layer at four different time instants are shown in Fig.~\ref{fig:FSspec}. The DSM gives the worst result in general especially at $t/\Delta T \approx 10$ and $t/\Delta T \approx 90$. For the data-driven models, they all give satisfactory results at high wave numbers ($k\geq 2$), but at low wave numbers ($k\leq 1$), there are varying degrees of deviation. However, IAFNO$_{(2)}$ outperforms other models, giving the most accurate predictions of the velocity spectrum at any time instants.

\begin{figure*}[!t]
    \centering
    \includegraphics[width=14cm,height=8.425cm]{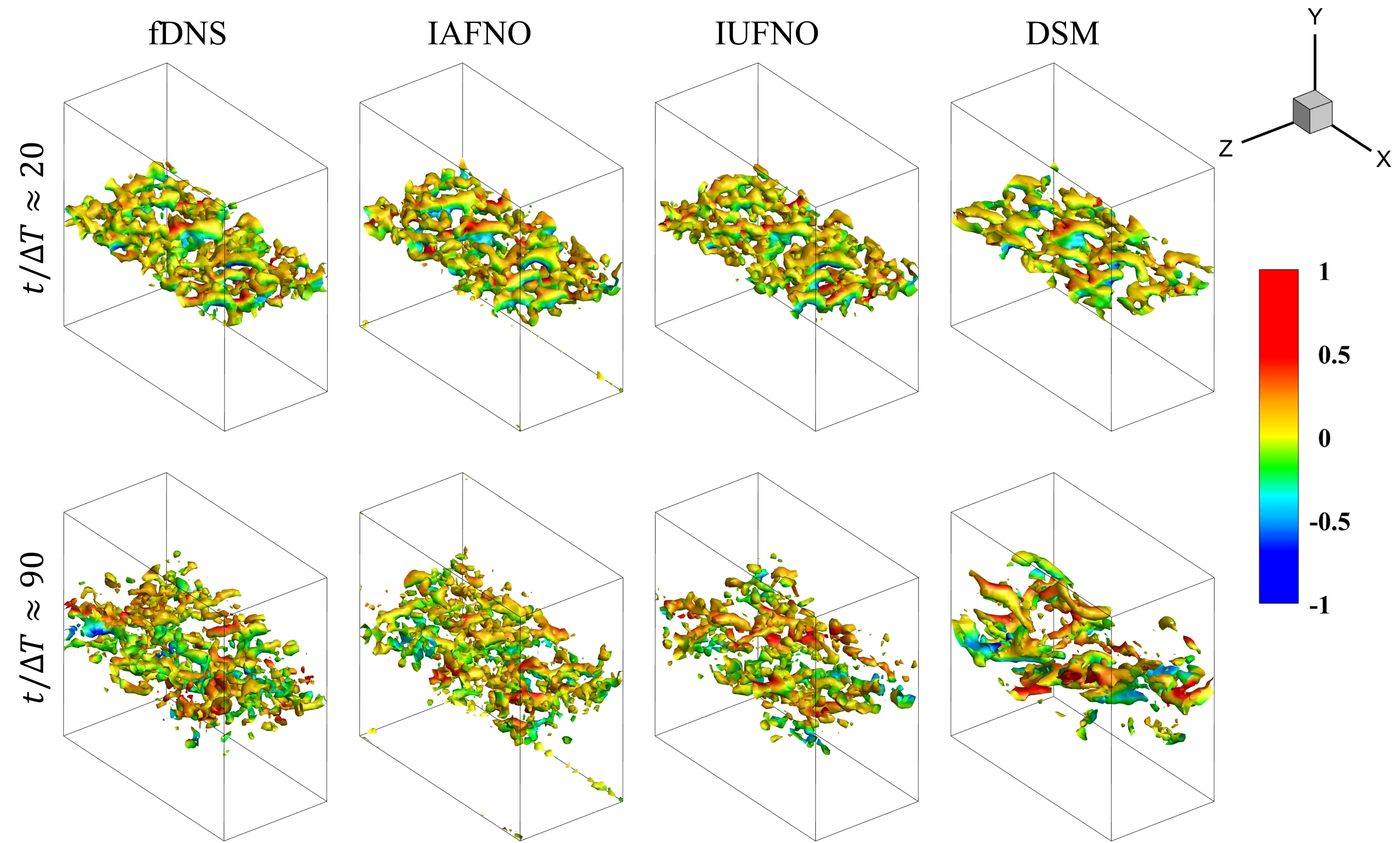}
    \caption{\label{fig:FSCoutour} The iso-surface of the Q-criterion at $Q = 0.05$ colored by the streamwise velocity at  $t/\Delta T \approx 20$ and $t/\Delta T \approx 90$ in the free-shear turbulent mixing layer.}
\end{figure*}

To visualize the vortex structure in the 3D free-shear turbulent mixing layer, we compare the Q-criterion predicted by the IAFNO, IUFNO model and DSM with fDNS data. The Q-criterion has been widely used for visualizing vortex structures in turbulent flows and is defined by \cite{dubief2000coherent}:

\begin{equation}
Q=\frac{1}{2}\left(\bar{\Omega}_{ij}\bar{\Omega}_{ij}-\bar{S}_{ij}\bar{S}_{ij}\right) ~, \label{eq:qcriterion}
\end{equation} where $\bar{\Omega}_{ij}=(\partial\bar{u}_i/\partial x_j-\partial\bar{u}_j/\partial x_i)/2$ is the filtered rotation-rate tensor and $\bar{S}_{ij}=(\partial\bar{u}_i/\partial x_j+\partial\bar{u}_j/\partial x_i)/2$ is the filtered strain-rate tensor. Fig.~\ref{fig:FSCoutour} displays the instantaneous iso-surfaces of $Q=0.05$ at $t/\Delta T \approx 20$ and $t/\Delta T \approx 90$ colored by the streamwise velocity. We chose the results of IAFNO$_{(2)}$ and IUFNO$_{(2)}$ which perform best in the prediction of physical statistics. As shown in the figures, both IAFNO and IUFNO results are close to that of fDNS, while DSM is the worst. Thus, the ability of the IAFNO model to predict vortex structure is confirmed.

\subsection{Turbulent channel flow}
In order to further validate the ability of the IAFNO model to predict complex turbulent flow fields, we test IAFNO model in turbulent channel flow at a friction Reynolds number $Re_{\tau}\approx590$. For the fDNS data, which serves as both the training dataset and the benchmark, we use the filtered direct numerical simulation data generated by Xcompact3D \cite{laizet2009high,bartholomew2020xcompact3d,wang2024prediction}.

\begin{table*}[!t]
\centering
\caption{\label{tab:cfdns}Parameters and statistics for DNS and fDNS of 3D turbulent channel flow.}
\renewcommand{\arraystretch}{1.5}
\begin{tabular}{cccccc}
\hline\hline
\mbox{Reso.(DNS:$N_1\times N_2\times N_3$)}&\mbox{Reso.(fDNS:$N_1\times N_2\times N_3$)}&\mbox{Domain}&\mbox{$Re_{\tau}$}&\mbox{$\nu$}&\mbox{$\mathrm{d} t$}\\
\hline
\mbox{$384\times257\times192$}&\mbox{$64\times65\times32$}&\mbox{$4\pi\times2\times4\pi/3$}&\mbox{590}&\mbox{1/16800}&\mbox{0.005}\\
\hline\hline
\end{tabular}
\end{table*}

\begin{table*}[!t]
\centering
\caption{\label{tab:cfhypsiufno}The hyperparameters settings for IUFNO in turbulent channel flow.}
\renewcommand{\arraystretch}{1.5}
\begin{tabular}{cccc}
\hline\hline
\mbox{Epochs}&\mbox{Batchsize}&\mbox{LearningRate(LR)}&\mbox{Width}\\
\hline
\mbox{100}&\mbox{4}&\mbox{0.001}&\mbox{50}\\
\hline\hline
\end{tabular}
\end{table*}

\begin{table*}[!t]
\centering
\caption{\label{tab:cfhypsiafno}The hyperparameters for the IAFNO model in turbulent channel flow.}
\renewcommand{\arraystretch}{1.5}
\begin{tabular}{ccccccc}
\hline\hline
\mbox{Epochs}&\mbox{Batchsize}&\mbox{LearningRate(LR)}&\mbox{Patchsize}&\mbox{EmbedDim}&\mbox{HSF}&\mbox{\#ofBlocks}\\
\hline
\mbox{100}&\mbox{5}&\mbox{0.001}&\mbox{(2,2,2)}&\mbox{200}&\mbox{3}&\mbox{1}\\
\hline\hline
\end{tabular}
\end{table*}

It is shown in Tab.~\ref{tab:cfdns} that the DNS of turbulent channel flow is performed in a three dimensional computational domain which has a streamwise, transverse and spanwise size of $(L_x,L_y,L_z)=(4\pi,2,4\pi/3)$, and the kinematic viscosity is $\nu=1/16800$. In addition, the grid points are uniformly distributed with a resolution of 384 and 192 in the x and z axes, and non-uniformly distributed with a resolution of 257 in the y-axis due to the need for the finer mesh near the wall. $(\Delta X^+,\Delta Z^+)=(19.3,12.9)$ are set to be the normalized distances between two neighboring grids in the streamwise and spanwise directions respectively, and $\Delta Y^+_\omega=1.6$ denotes the normalized distance between the grid point which is the nearest to the wall and the wall surface along normal direction. The superscript ``+'' indicates that the physical quantity has been normalized in viscous units, e.g. $y^+=y/\delta_{\nu}, u^+=u/u_{\tau}$ where $\delta_{\nu}$ and $u_{\tau}$ are the viscous length and wall-friction velocity, respectively \cite{wang2024prediction}. For turbulent channel flow, while the mesh is nonuniform in y direction, it is still a structured mesh and can be transformed into a uniform mesh \cite{li2023fourier}. Hence the FFT can still be conveniently applied.

The DNS is filtered in the streamwise and spanwise directions where the grid point distributions are homogeneous, meanwhile, there is no filter acting on the normal direction \cite{wang2024prediction}. For the coarsening operation, the linear interpolation method is introduced to calculate the values at each position of the coarsened grids. The obtained coarsened fDNS data has a resolution of $64\times65\times32$. In order to construct a sufficiently large dataset for model training and validation, a total of 21 groups of different fDNS data sets are generated by the above method through different initial flow fields. Of the 21 sets of data mentioned above, the last set will be kept as a validation set for all data-driven models. Each set of data contains 400 instants of flow field information with a dimensionless time interval $\Delta T=200\mathrm{d} t$. Here, the DNS time step $\mathrm{d} t=0.005$. Therefore, a datasets of size $[20\times400\times64\times65\times32\times3]$ can be used as training ($80\%$) and testing ($20\%$), which is the same as in the previous two sections. The inference (prediction) process has been described in Eq.~\ref{eq:predict}, and the loss function has been defined in Eq.~\ref{eq:lossfunction}.

We have obtained the trained raw data IUFNO model's prediction at $Re_{\tau}\approx590$ used in the previous paper \cite{wang2024prediction}, and compare the results with the IAFNO model. The hyperparameters used in the IUFNO and IAFNO model are shown in Tab.~\ref{tab:cfhypsiufno} and Tab.~\ref{tab:cfhypsiafno}, respectively. It can be observed that the number of implicit layers and the training epochs are kept the same for both data-driven models. Furthermore, the width of the IUFNO model is increased to 50 \cite{wang2024prediction}, while the embedded dimension of the IAFNO model is increased to 200. In order to demonstrate the advantages of the IAFNO model over the IUFNO model, we select the prediction results at the instant $t=400\Delta T$ for plotting the following figures. Moreover, since the IUFNO model seems to deviate farther away at longer prediction times, we will present the performance of the IAFNO model at $t=800\Delta T$ to show the stability of the IAFNO model over long time.

\begin{figure*}[!t]
    \centering
    \subfloat{
    \includegraphics[width=6cm,height = 4.5cm]{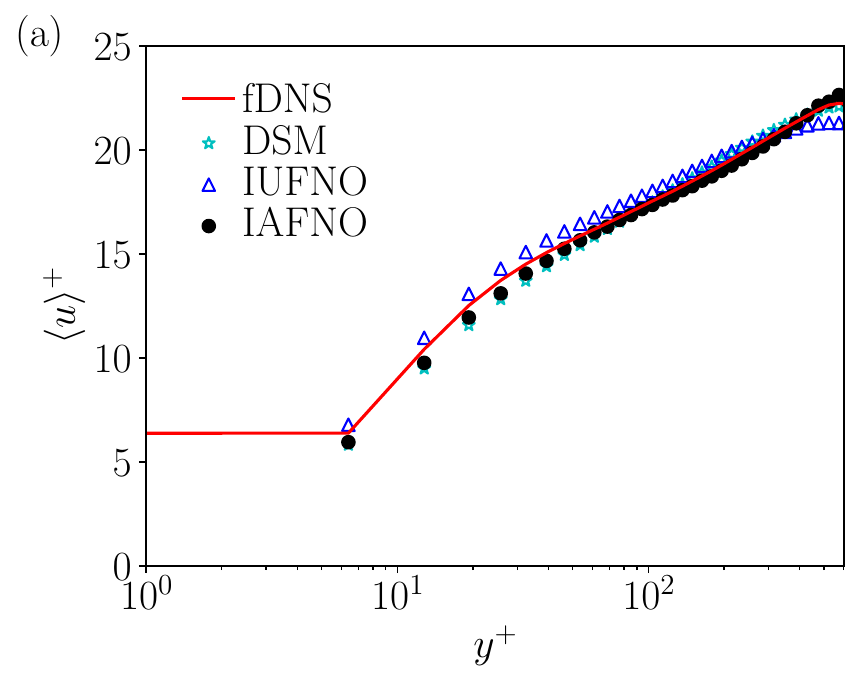}}
    \subfloat{
    \includegraphics[width=6cm,height = 4.5cm]{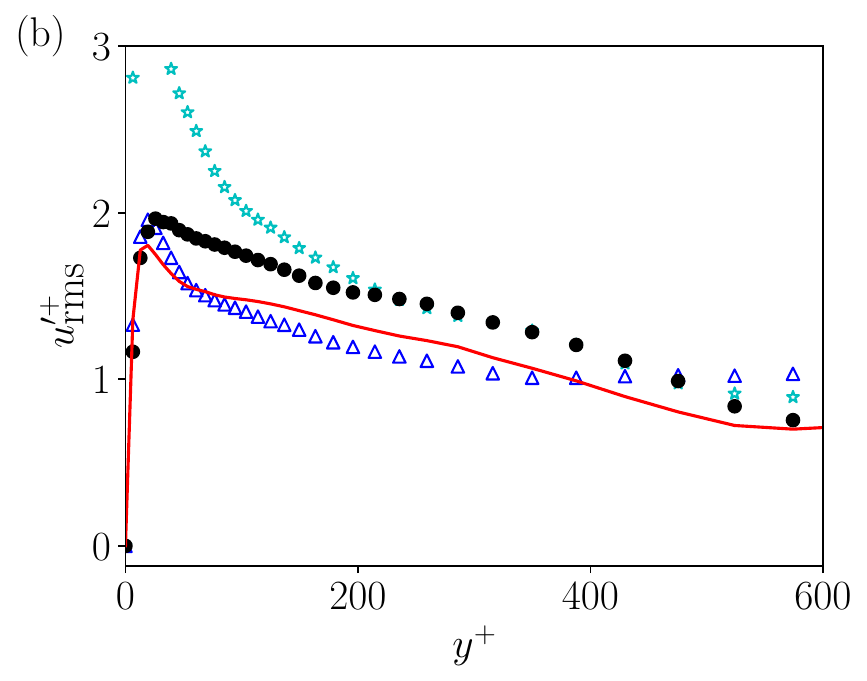}}
    \\
    \subfloat{\hspace{4mm}
    \includegraphics[width=6cm,height = 4.5cm]{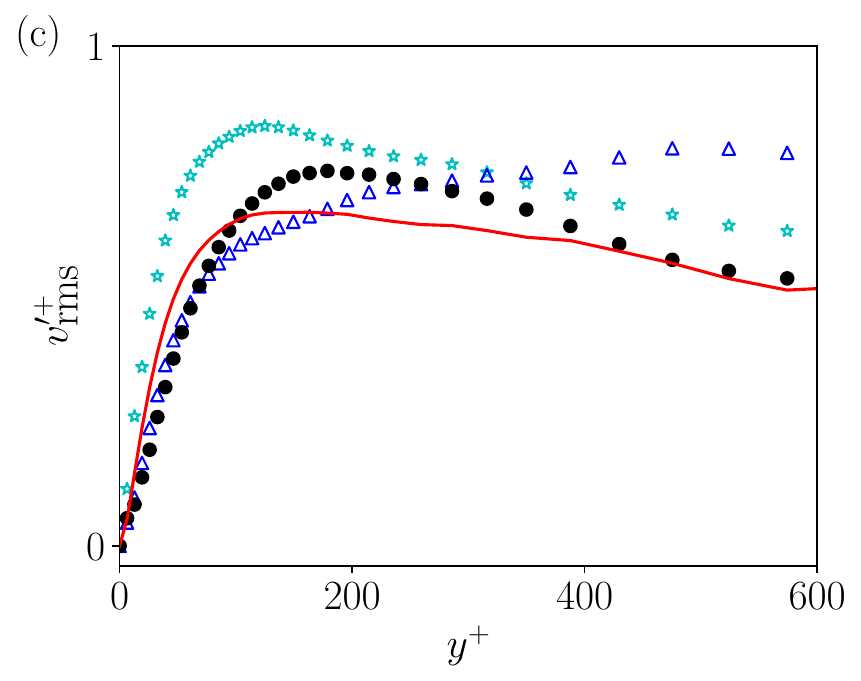}}
    \subfloat{
    \includegraphics[width=6cm,height = 4.5cm]{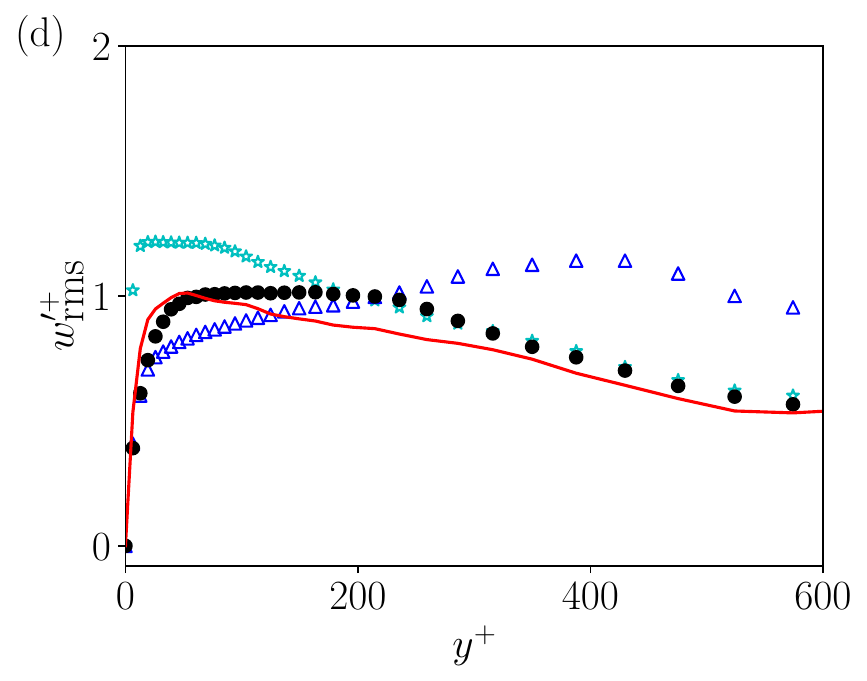}}
    \caption{\label{fig:CFvel} The mean streamwise velocity and rms fluctuating velocities at $Re_{\tau}\approx590$ at $t=400\Delta T$: (a) mean  streamwise velocity; (b) rms fluctuation of streamwise velocity; (c) rms fluctuation of transverse  velocity; (d) rms fluctuation of spanwise velocity.}
\end{figure*}

In Fig.~\ref{fig:CFvel}(a), the mean streamwise velocity predicted by various models for $Re_{\tau}\approx590$ at $t=400\Delta T$ are displayed. It can be observed that all three models are able to give reasonable mean streamwise velocity when $y^+\leq300$. However, when $y^+\geq300$, IAFNO is still able to fit the benchmark accurately while the DSM is only slightly worse, but IUFNO shows a significant deviation. Meanwhile, the rms fluctuating velocity in three directions predicted by various models for $Re_{\tau}\approx590$ at $t=400\Delta T$ are shown in Figs.~\ref{fig:CFvel}(b)(c)(d). It is observed that IAFNO has a certain shift in the streamwise direction, while in the other two directions the IAFNO model fits more closely to the benchmark especially when $ y^+\geq300$. The IUFNO model also has a shift in the streamwise direction but its amplitude is small. The DSM severely overestimates the rms fluctuating velocity near the wall, and the overall performance lies between the IAFNO model and the IUFNO model.

Obviously, the abnormal trends of IUFNO occurred at $y^+\geq 400$ for rms values of velocity flucuations in the streamwise direction, at $y^+\geq 200$ for rms fluctuations in the transverse and spanwise directions. However, the IAFNO model can predict the right trend and accurately predict the mean streamwise velocity and rms fluctuating velocities for $Re_{\tau}\approx590$ at $t=400\Delta T$.

\begin{figure*}[!t]
    \centering
    \subfloat{
    \includegraphics[width=6cm,height = 4.5cm]{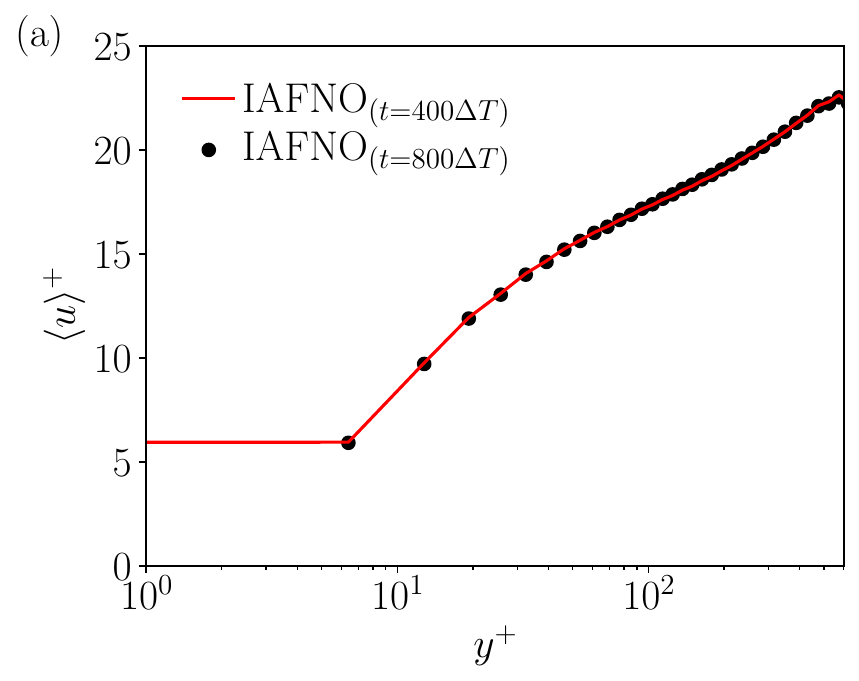}}
    \subfloat{
    \includegraphics[width=6cm,height = 4.5cm]{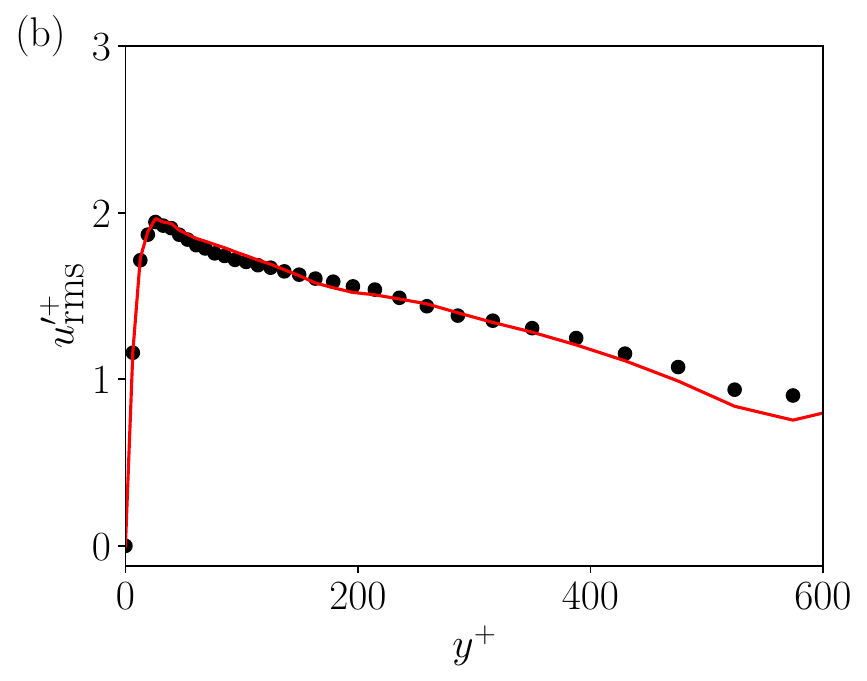}}
    \\
    \subfloat{\hspace{4mm}
    \includegraphics[width=6cm,height = 4.5cm]{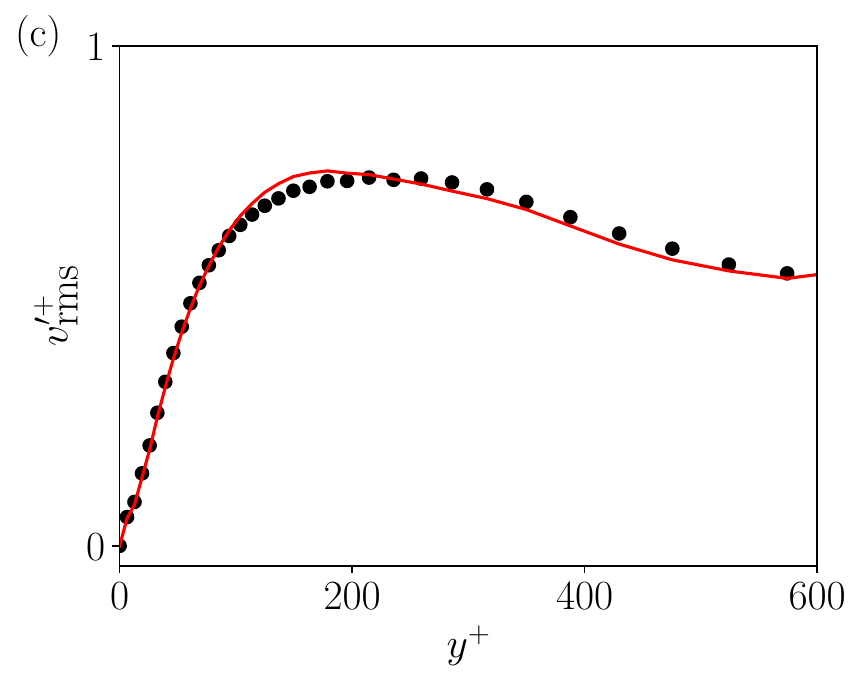}}
    \subfloat{
    \includegraphics[width=6cm,height = 4.5cm]{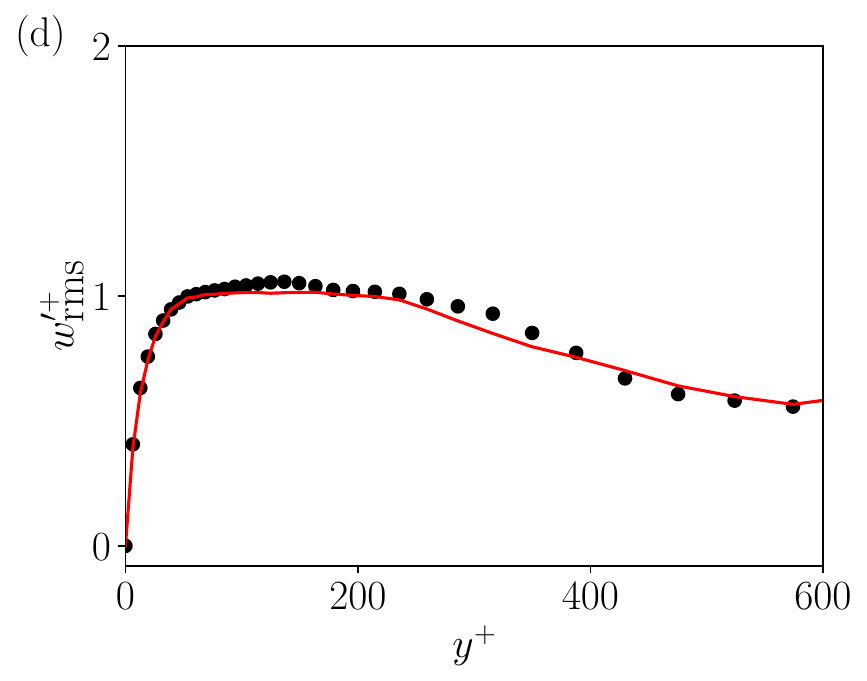}}
    \caption{\label{fig:CFvel800} The mean streamwise velocity and rms fluctuating velocities at $Re_{\tau}\approx590$ at $t=800\Delta T$: (a) mean  streamwise velocity; (b) rms fluctuation of streamwise velocity; (c) rms fluctuation of transverse  velocity; (d) rms fluctuation of spanwise velocity.}
\end{figure*}

Because the prediction result of the $800$th step is beyond the time horizon of the training data, we do not have the fDNS data as a benchmark against it. But since the channel flow already reaches a statistically steady state, we can test the ability of the IAFNO model to make stable long-term predictions by checking whether the predicted values of the IAFNO model converge or not. Fig.~\ref{fig:CFvel800} shows the mean streamwise velocity and the rms fluctuation velocities predicted by the IAFNO model at $t=400\Delta T$ and $t=800\Delta T$. The two results are seen to almost overlap, indicating that the IAFNO model converges and keeps the predicted statistics numerically stable.

\begin{figure*}[!t]
    \centering
    \subfloat{
    \includegraphics[width=6cm,height = 4.5cm]{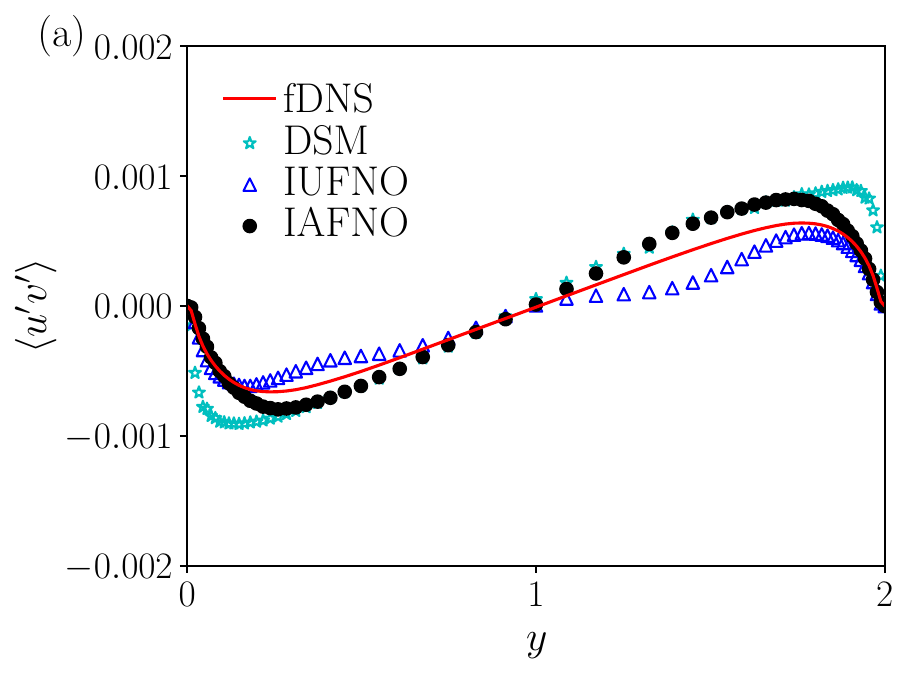}}
    \subfloat{
    \includegraphics[width=6cm,height = 4.5cm]{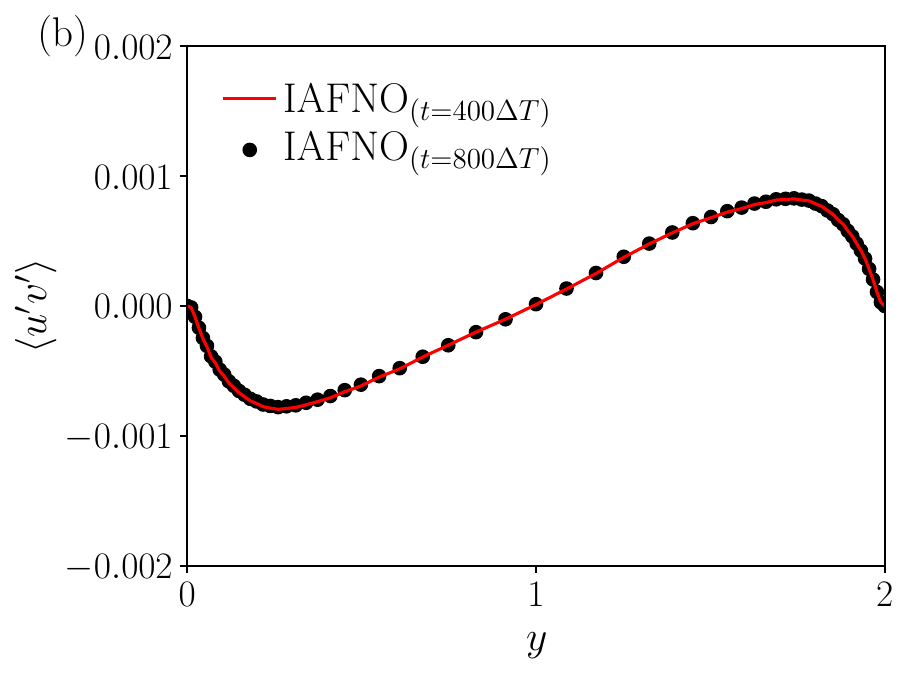}}
    \caption{\label{fig:CFrnss} The variation of Reynolds shear stress $\langle u'v' \rangle$ at $Re_{\tau}\approx590$: (a) time-averaged till $t=400\Delta T$; (b) time-averaged till $t=800\Delta T$.}
\end{figure*}

The predicted Reynolds shear stresses by the DSM and the two data-driven models are shown in Fig.~\ref{fig:CFrnss}(a). The maximum Reynolds shear stresses are seen to be located near the upper and lower walls, where both the mean shear effects and the velocity fluctuations are strong \cite{wang2024prediction}. The Reynolds shear stress between the two peaks is approximately linear with respect to the normal coordinate $y$, which is consistent with the literature \cite{kim1987turbulence}. The IAFNO model shows some deviations in the predicted values at the two peaks of the Reynolds stress, especially at the positive extreme. The DSM not only has the same deviation as the IAFNO model, but the persistence of the deviation is much wider than IAFNO. However, both IAFNO and DSM perform well in the linear part between the two peaks and maintain physical linearity, whereas the non-physical wiggles occur in the result of the IUFNO model. Therefore, the IAFNO model outperforms the IUFNO model and DSM in terms of predicting the Reynolds shear stress. In Fig.~\ref{fig:CFrnss}(b), the black dots are perfectly located on the red line, again indicating an accurate and stable prediction of the IAFNO model.

\begin{figure*}[!t]
    \centering
    \subfloat{
    \includegraphics[width=6cm,height = 4.5cm]{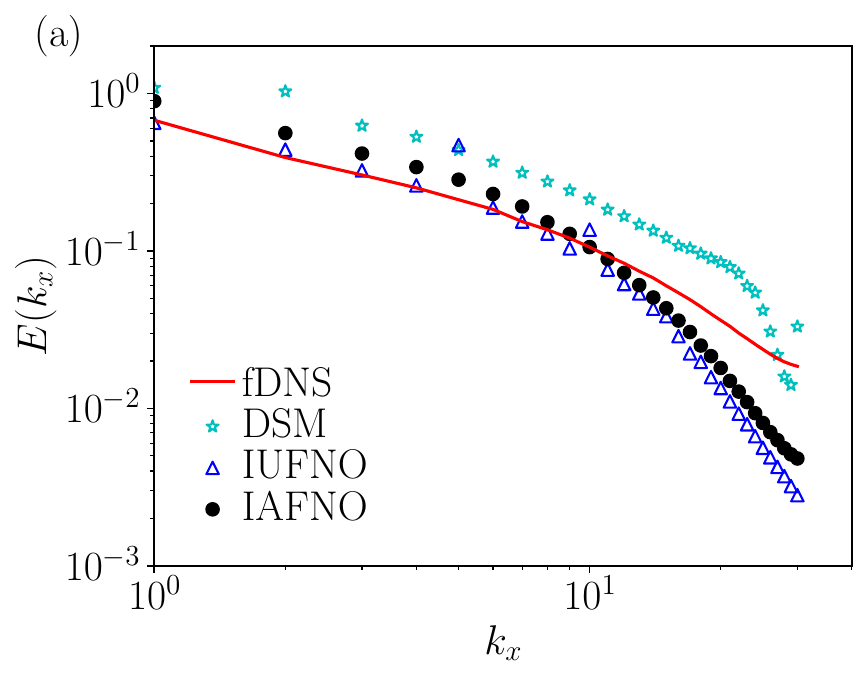}}
    \subfloat{
    \includegraphics[width=6cm,height = 4.5cm]{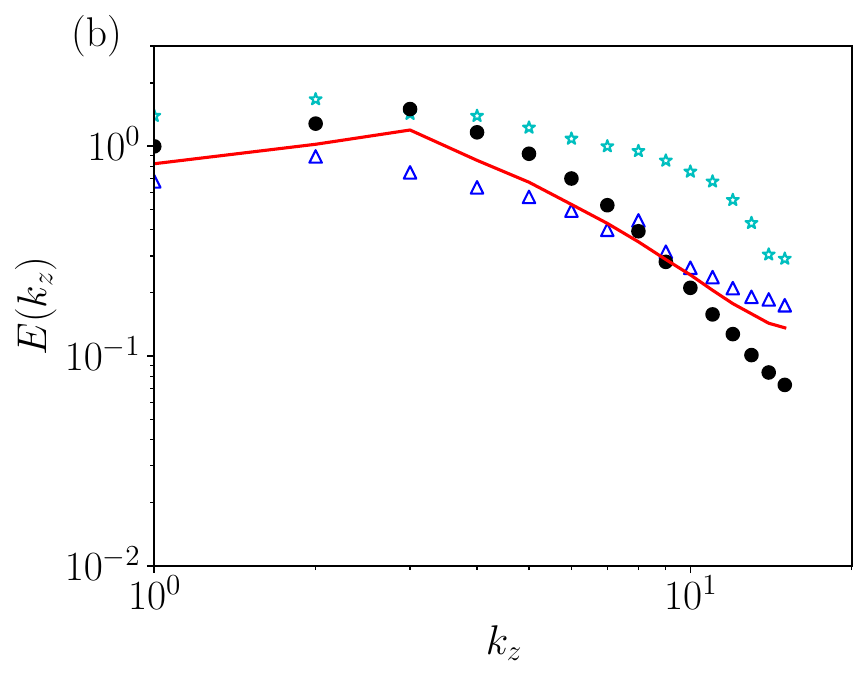}}
    \caption{\label{fig:CFspec} Energy spectrum at $Re_{\tau}\approx590$ at $t=400\Delta T$: (a) streamwise spectrum; (b) spanwise spectrum.}
\end{figure*}

To further explore the performance of the IAFNO model in terms of predicting the energy distribution, we calculate the kinetic energy spectrum in the streamwise and spanwise directions for $Re_{\tau}\approx590$ at $t=400\Delta T$ as shown in Fig.~\ref{fig:CFspec}. It is observed that the IUFNO model has nonphysical jumps in both streamwise and spanwise directions. Moreover, the difference of IUFNO model from the benchmark is larger than that of the IAFNO model. The results of DSM deviate upward at the very beginning of the energy spectrum, and the amplitude of this deviation is so large that the DSM is obviously the worst. Hence, the prediction of IAFNO model on the kinetic energy spectrum is the most accurate.

\begin{figure*}[!t]
    \centering
    \includegraphics[width=14cm,height=12.66cm]{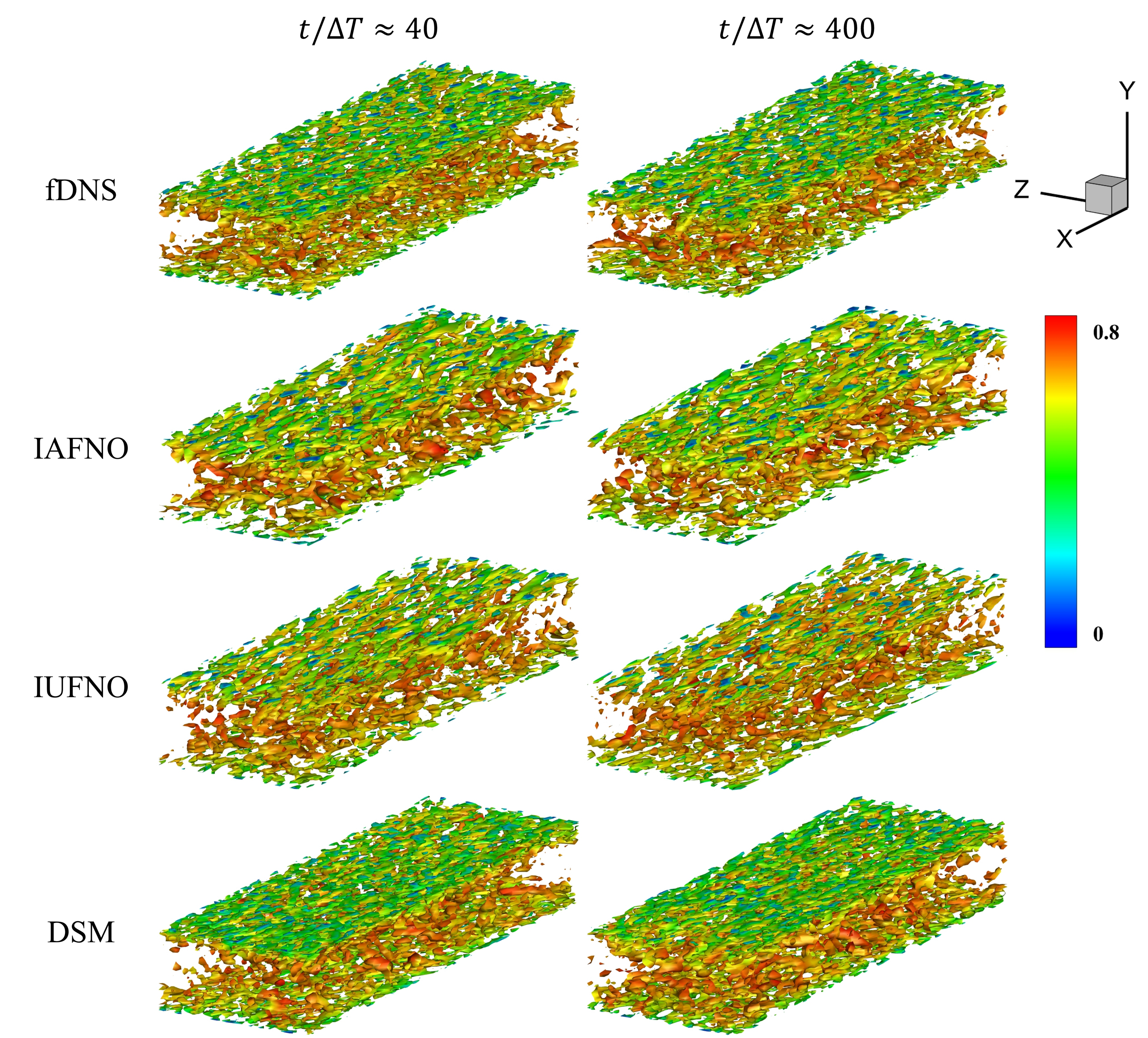}
    \caption{\label{fig:CFCoutour} The iso-surface of the Q-criterion at $Q = 0.05$ colored by the streamwise velocity at  $t/\Delta T \approx 40$ and $t/\Delta T \approx 400$ in the turbulent channel flow.}
\end{figure*}

We then investigate the reconstruction of the vortex structure in the turbulent channel flow, by comparing the Q-criterion predicted by the IAFNO, IUFNO models and DSM with fDNS data in Fig.~\ref{fig:CFCoutour}. It is seen that all three models are able to reconstruct the vortex structure well. However, the IAFNO model predicts better results than the IUFNO model at the upper and lower surfaces. Moreover, the IAFNO model is able to predict richer vortex structures, while the prediction of the IUFNO model appears to be more sparse.

\section{Computational costs and efficiency}

We present the computational costs and efficiency for the IAFNO and IUFNO models with different hyperparameters in the forced HIT in Tab.~\ref{tab:hitcost}, free-shear turbulent mixing layer in Tab.~\ref{tab:fscost} and turbulent channel flow in Tab.~\ref{tab:cfcost}. The tables include the number of parameters of those different approaches, corresponding GPU memory occupation, the time required to train one epoch and the inference cost. Here, the inference time costs with respect to the three different turbulent are: 10 prediction steps (i.e. 1000 DNS advance steps, HIT), 10 prediction steps (i.e. 2000 DNS advance steps, free-shear turbulent mixing layer), and 50 prediction steps (i.e. 10000 DNS advance steps, turbulent channel flow). All data-driven models are trained and tested on a NVIDIA A100 40G PCIe GPU, while the CPU used for loading model parameters and data is an Intel(R) Xeon(R) Gold 6248R CPU @3.00 GHz (except for IUFNO in the test of turbulent channel flow, during which we use a NVIDIA A100 80G PCIe GPU and an AMD EPYC 7763 @2.45 GHz CPU). The LES simulations are implemented on a computing cluster, where the type of CPU is Intel(R) Xeon(R) Gold 6148 with 16 cores each @2.40 GHz (for forced HIT and free-shear turbulent mixing layer) and Intel(R) Xeon(R) Gold 6148 with 64 cores (turbulent channel flow).

\begin{table*}[!t]
\centering
\caption{\label{tab:hitcost}Computing costs and computational efficiency of different approaches in forced HIT.}
\renewcommand{\arraystretch}{1.5}
\begin{tabular}{ccccc}
\hline\hline
\vspace{-2mm}
\mbox{Model}&\mbox{Training time}&\mbox{Number of parameters}&\mbox{GPU memory occupation}&\mbox{Inference}\\
\mbox{}&\mbox{(s/epoch)}&\mbox{($\times10^6$)}&\mbox{(MB)}&\mbox{(s)}\\
\hline
\mbox{DSM$_{(\times16~\mathrm{cores})}$}&\mbox{N/A}&\mbox{N/A}&\mbox{N/A}&\mbox{8.55}\\
\mbox{IUFNO$_{(L=10)}$}&\mbox{2136}&\mbox{83.02}&\mbox{15273}&\mbox{4.03}\\
\mbox{IUFNO$_{(L=20)}$}&\mbox{4072}&\mbox{83.02}&\mbox{22649}&\mbox{7.50}\\
\mbox{IUFNO$_{(L=40)}$}&\mbox{8012}&\mbox{83.02}&\mbox{37673}&\mbox{14.10}\\
\mbox{IAFNO$_{(L=10)}$}&\mbox{509}&\mbox{1.215}&\mbox{5247}&\mbox{1.75}\\
\mbox{IAFNO$_{(L=20)}$}&\mbox{1009}&\mbox{1.215}&\mbox{8451}&\mbox{2.37}\\
\mbox{IAFNO$_{(L=40)}$}&\mbox{2025}&\mbox{1.215}&\mbox{14857}&\mbox{3.98}\\
\hline\hline
\end{tabular}
\end{table*}

\begin{table*}[!t]
\centering
\caption{\label{tab:fscost}Computing costs and computational efficiency of different approaches in free-shear turbulent mixing layer.}
\renewcommand{\arraystretch}{1.5}
\begin{tabular}{ccccc}
\hline\hline
\vspace{-2mm}
\mbox{Model}&\mbox{Training time}&\mbox{Number of parameters}&\mbox{GPU memory occupation}&\mbox{Inference}\\
\mbox{($L=20$)}&\mbox{(s/epoch)}&\mbox{($\times10^6$)}&\mbox{(MB)}&\mbox{(s)}\\
\hline
\mbox{DSM$_{(\times16~\mathrm{cores})}$}&\mbox{N/A}&\mbox{N/A}&\mbox{N/A}&\mbox{50.72}\\
\mbox{IUFNO$_{(bs=2)}$}&\mbox{7763}&\mbox{83.02}&\mbox{31335}&\mbox{2.89}\\
\mbox{IAFNO$_{(bs=2)}$}&\mbox{2005}&\mbox{3.206}&\mbox{11687}&\mbox{1.02}\\
\mbox{IAFNO$_{(bs=5)}$}&\mbox{1881}&\mbox{3.206}&\mbox{26257}&\mbox{1.03}\\
\hline\hline
\end{tabular}
\end{table*}

\begin{table*}[!t]
\centering
\caption{\label{tab:cfcost}Computing costs and computational efficiency of different approaches in the turbulent channel flow.}
\renewcommand{\arraystretch}{1.5}
\begin{tabular}{ccccc}
\hline\hline
\vspace{-2mm}
\mbox{Model}&\mbox{Training time}&\mbox{Number of parameters}&\mbox{GPU memory occupation}&\mbox{Inference}\\
\mbox{}&\mbox{(s/epoch)}&\mbox{($\times10^6$)}&\mbox{(MB)}&\mbox{(s)}\\
\hline
\mbox{DSM$_{(\times64~\mathrm{cores})}$}&\mbox{N/A}&\mbox{N/A}&\mbox{N/A}&\mbox{315.3}\\
\mbox{IUFNO}&\mbox{4400$\sim$4600}&\mbox{82.05}&\mbox{79440}&\mbox{14.90}\\
\mbox{IAFNO}&\mbox{1500}&\mbox{4.212}&\mbox{33985}&\mbox{7.48}\\
\hline\hline
\end{tabular}
\end{table*}

In the first column of Tab.~\ref{tab:hitcost}, the subscript indicates the number of implicit layers. It is shown that the number of implicit layers of the model is approximately proportional to the time it takes to train per epoch as well as the GPU memory usage. Furthermore, since we are more interested in what advantages the IAFNO model has over the IUFNO model when they contain the same number of implicit layers, we now compare row 2 over row 5, row 3 over row 6, and row 4 over row 7. It is obvious that using the same number of implicit layers, the computational efficiency of the IAFNO model is $\textbf{4}$ times higher than that of the IUFNO model, the number of parameters is $\textbf{1/80}$ of the IUFNO model, and the GPU memory occupation is only $\textbf{1/3}\sim\textbf{2/5}$ of the IUFNO model. The fifth column indicates that the inference time cost of the IUFNO model will gradually exceed the DSM when the number of implicit layers increases, while the IAFNO model is always faster than the DSM and it is $\textbf{2}\sim\textbf{3}$ times faster than the IUFNO model. Here, the data-driven models are implemented on a single-core CPU, whereas the DSM model is performed on a CPU with 16 cores. Therefore, if we only use one single core to carry out the DSM calculation, the actual inference time of the DSM model is 16 times greater than the time shown in Tab.~\ref{tab:hitcost}.

In Tab.~\ref{tab:fscost}, the GPU memory usage for data-driven models is increased significantly compared to the situation of HIT due to the increase of the flow field's resolution. The efficiency of IAFNO model remains nearly \textbf{4} times higher than that of IUFNO model, and the memory usage remains around \textbf{1/3} of that of IUFNO model. The inference time cost for the IUFNO is \textbf{3/50} of that of the DSM, while the IAFNO is only \textbf{1/50} of the DSM.

Moreover, in Tab.~\ref{tab:cfcost}, it can be observed that the efficiency of IAFNO model is \textbf{3} times higher than that of IUFNO. The number of parameters and GPU memory occupation of IAFNO are only \textbf{1/20} and \textbf{1/2} of IUFNO, respectively. The inference time cost of IAFNO is only \textbf{1/42} of that of the DSM.


\section{Conclusion}
\label{conclusion}

In this work, inspired by the FourCastNet \cite{pathak2022fourcastnet}, we proposed an implicit adaptive Fourier neural operator (IAFNO) model to predict long-term large-scale dynamics of three-dimensional turbulence. The IAFNO model is verified by the comparison with the DSM and IUFNO model in the large-eddy simulations of three types of 3D turbulence, including forced homogeneous isotropic turbulence, free-shear turbulent mixing layer, and turbulent channel flow. These numerical simulations demonstrate that: 

1) Compared with the model which only use AFNO as the backbone, the IAFNO model with the same network depth is able to stably and accurately predict a variety of statistics of flow fields, and the instantaneous spatial structures of vorticity over a long period of time.

2) The IAFNO model can predict the various physical statistics of flow fields more accurately than the IUFNO model. The IAFNO model is also more accurate than the DSM, especially in the two more complex turbulence, including free-shear turbulent mixing layer and turbulent channel flow.

3) While IAFNO model has the highest accuracy, in the tests of HIT and turbulent mixing layer, compared with the IUFNO model under the same batchsize and network depth, the computational efficiency of IAFNO is $\textbf{4}$ times higher than IUFNO, the number of parameters is $\textbf{1/80}$ and $\textbf{1/30}$ of IUFNO, respectively, and the GPU memory occupation is only $\textbf{1/3}$ of IUFNO. Moreover, in the test of turbulent channel flow, the computational efficiency of IAFNO is $\textbf{3}$ times higher than IUFNO, the number of parameters and GPU memory occupation are $\textbf{1/20}$ and $\textbf{1/2}$ of IUFNO, respectively. Besides, the trained IAFNO model is $\textbf{4}$, $\textbf{50}$, and $\textbf{42}$ times faster than the DSM in terms of the inference time cost in HIT, turbulent mixing layer, and turbulent channel flow, respectively.


Therefore, the proposed IAFNO approach has the great potential to efficiently solve complex 3D nonlinear problems in engineering applications.

However, although our proposed IAFNO model is able to achieve stable, efficient and accurate long-term prediction of three 3D turbulent flows in this study, one limitation is the IAFNO's reliance on data. In recent research, the physics-informed approaches have been applied to enhance the performance of operator learning and reduce the model's dependence on data by embedding PDEs into the loss functions in a manner similar to PINNs \cite{raissi2019physics}, including physics-informed DeepONets \cite{wang2021learning,wang2023long}, physics-informed Fourier neural operator \cite{li2024physics,zanardi2023adaptive} and physics-informed transformer \cite{lorsung2024physics,zhao2023pinnsformer}. Moreover, a more recent model that embeds the large-eddy simulation equations into FNO model (LESnets) can effectively simulate turbulent flows without training data and maintain the efficiency of data-driven neural operators \cite{zhao2024lesnets}, which provides a novel idea to improve the present IAFNO model.

Furthermore, we only test the IAFNO model on simple flows, whereas engineering applications often involve different Reynolds numbers and complex geometries with irregular boundaries. Therefore, it is crucial to enhance the ability of machine learning models to handle complex flow fields with parameterized boundary conditions, varying geometries and different Reynolds numbers \cite{gao2021phygeonet,li2023fourier,wu2024transolver}.


\appendix
\section{The Fourier neural operator}
\label{app1}

The Fourier neural operator (FNO) learns a non-linear mapping between two infinite dimensional spaces from a finite collection of observed input-output pairs \cite{li2020fourier,kovachki2023neural}:

\begin{equation}
G: \mathcal{A}\times\Theta\rightarrow\mathcal{U},~~\text{or equivalently}~~G_{\theta}:\mathcal{A}\rightarrow\mathcal{U},~~\theta\in\Theta ~, \label{eq:FNO}
\end{equation} where $\mathcal{A}=\mathcal{A}(D;\mathbb{R}^{d_a})$ and $\mathcal{U}=\mathcal{U}(D;\mathbb{R}^{d_u})$ are separable Banach spaces of function taking values in $\mathbb{R}^{d_a}$ and $\mathbb{R}^{d_u}$ respectively, and $D\subset\mathbb{R}^d$ is a bounded, open set. The construction of this non-linear mapping, parameterized by $\theta\in\Theta$, allows the Fourier neural operators to learn an approximation of operator $\mathcal{A}\rightarrow\mathcal{U}$. The optimal parameters $\theta\dagger\in\Theta$ are determined through a data-driven approximation \cite{li2023long}. In order to increase the depth of the neural operator to enhance its performance, an iterative architecture is then applied as followed: $v_0\mapsto v_1\mapsto\dots\mapsto v_T$, where $v_j$ for $j=0,1,\dots,T-1$ is a sequence of functions taking values in $\mathbb{R}^{d_v}$ \cite{li2020neural}. The FNO architecture is shown in Fig.~\ref{fig:ModelStructureFNO} which consists of three main steps \cite{li2020fourier}:


\begin{figure*}[htb]
    \centering
    \includegraphics[width=14cm,height=7.875cm]{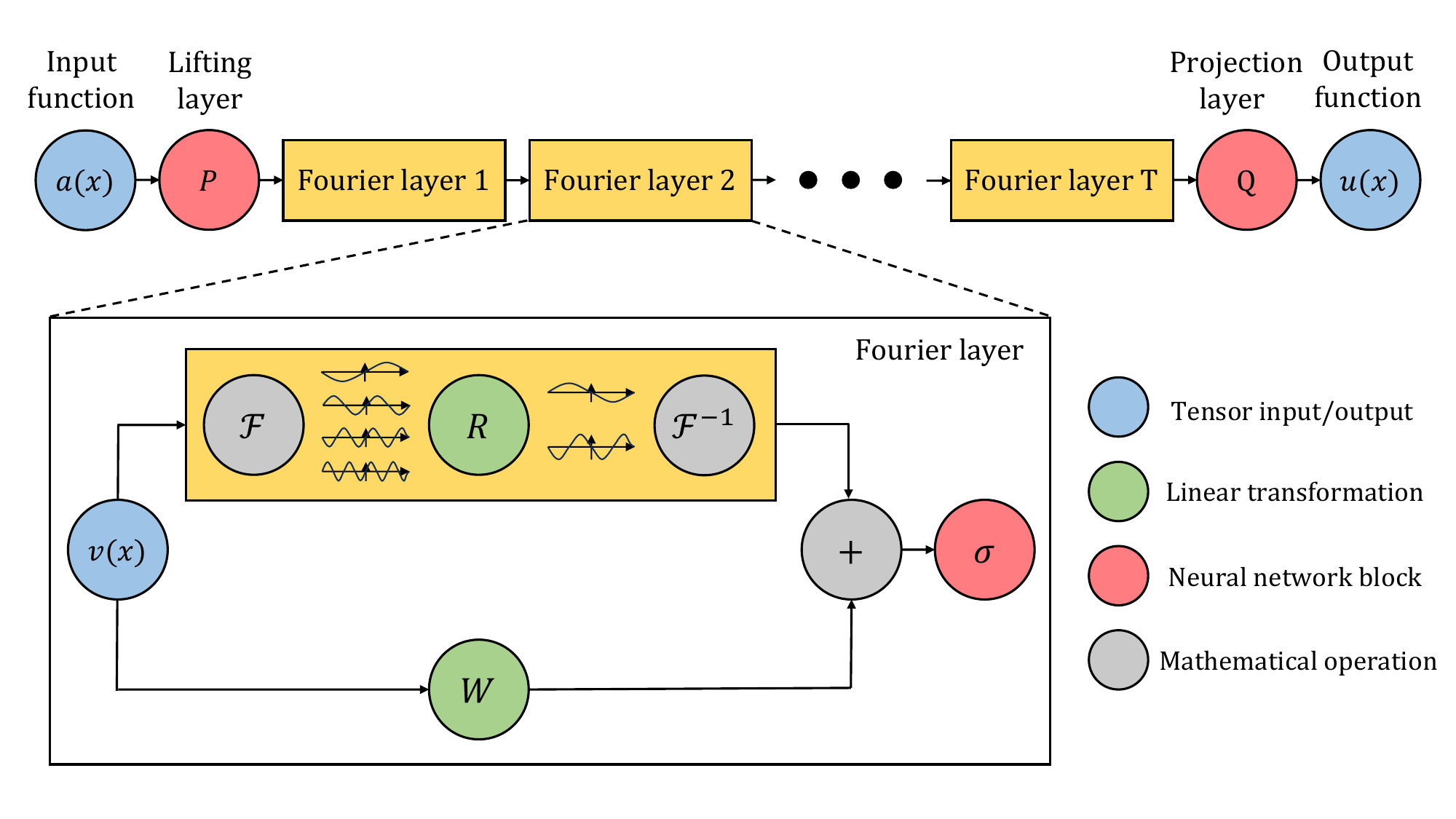}
    \caption{\label{fig:ModelStructureFNO} The architecture of FNO.}
\end{figure*}


(1) The input $a\in\mathcal{A}$ is lifted to a higher dimension channel space with a representation of $v_0(x)=P(a(x))$ by the local transformation $P$ which is commonly parameterized by a shallow fully connected neural network.

(2) The specific expression of the above mentioned iteration in higher dimension channel space is given by:

\begin{equation}
v_{t+1}:=\sigma\left( Wv_t(x)+\left( \mathcal{K}(a;\phi)v_t \right)(x) \right) ,~~\forall x\in D ~, \label{eq:iterFNO}
\end{equation} where $\mathcal{K}:\mathcal{A}\times\Theta_{\mathcal{K}}\rightarrow\mathcal{L}(\mathcal{U}(D;\mathbb{R}^{d_v}),\mathcal{U}(D;\mathbb{R}^{d_v}))$ maps to bounded linear operators on $\mathcal{U}(D;\mathbb{R}^{d_v})$ and is parameterized by $\phi \in \Theta_{\mathcal{K}}$. Here, $W:\mathbb{R}^{d_v}\rightarrow\mathbb{R}^{d_v}$ is a linear transformation, and $\sigma: \mathbb{R}\rightarrow\mathbb{R}$ is a component-wise non-linear activation function. The kernel integral operator mapping in Eq.~\ref{eq:iterFNO} is defined by: 

\begin{equation}
\left( \mathcal{K}(a;\phi)v_t \right)(x):=\int_D\kappa\left( x,y,a(x),a(y);\phi \right)v_t(y)\mathrm{d}y~, \label{eq:kernelFNO}
\end{equation}
where $\kappa_{\phi}:=\mathbb{R}^{2(d+d_a)}\rightarrow\mathbb{R}^{d_v\times d_v}$ is a neural network parameterized by $\phi \in \Theta_{\mathcal{K}}$. Here, if $\kappa_{\phi}(x,y)=\kappa_{\phi}(x-y)$ is imposed, Eq.~\ref{eq:kernelFNO} becomes a convolution operator, which can be simplified to linear transformation by Fast Fourier Transform (FFT) that can bring a significant increase of efficiency. Let $\mathcal{F}$ denotes the Fourier transform of a function $f:D\rightarrow\mathbb{R}^{d_v}$ and $\mathcal{F}^{-1}$ its inverse, we have:

\begin{equation}
\left( \mathcal{K}(a;\phi)v_t \right)(x):=\mathcal{F}^{-1}\left( 
\mathcal{F}(\kappa_{\phi})\cdot\mathcal{F}(v_t) \right)(x),~~\forall x\in D ~. \label{eq:fftFNO}
\end{equation}

Therefore, it is convenient to parameterize $\kappa_{\phi}$ in Fourier space. We can replace the term in Eq.~\ref{eq:iterFNO} with its equivalent form after FFT which is presented in Eq.~\ref{eq:fftFNO}:

\begin{equation}
v_{t+1}:=\sigma\left( Wv_t(x)+\mathcal{F}^{-1}\left( 
R_{\phi}\cdot\mathcal{F}(v_t) \right)(x) \right) ,~~\forall x\in D ~, \label{eq:fullFNO}
\end{equation} where $R_{\phi}$ is the Fourier transform of a periodic function $\kappa:\bar{D}\rightarrow\mathbb{R}^{d_v\times d_v}$ parameterized by $\phi \in \Theta_{\mathcal{K}}$. The frequency mode $k\in \mathbb{Z}^d$. The finite-dimensional parametrization is obtained by truncating the Fourier series at a maximum number of modes $k_{\mathrm{max}}=|Z_{k_{\mathrm{max}}}|=|\{k\in \mathbb{Z}^d:|k_j|\leq k_{\mathrm{max},j},\mathrm{for}j=1,\dots,d\}|$. $\mathcal{F}(v_t)\in \mathbb{C}^{n\times d_v}$ can be obtained by discretizing domain $D$ with $n\in \mathbb{N}$ points, where $v_t\in \mathbb{R}^{n\times d_v}$. By simply truncating the higher modes,  $\mathcal{F}(v_t) \in \mathbb{C}^{k_{\mathrm{max}}\times d_v}$ can be obtained, here $\mathbb{C}$ is the complex space. $R_{\phi}$ is parameterized as complex-valued-tensor $(k_{\mathrm{max}}\times d_v \times d_v)$ containing a collection of truncated Fourier modes  $R_{\phi} \in \mathbb{C}^{k_{\mathrm{max}}\times d_v\times d_v}$. Therefore, by multiplying $R_{\phi}$ and $\mathcal{F}(v_t)$, it can be derived that:

\begin{gather}
\left( 
R_{\phi}\cdot\mathcal{F}(v_t) \right)_{k,l}=\sum_{j=i}^{d_v}R_{\phi k,l,j}(\mathcal{F}v_t)_{k,j},\\ \textrm{where}~k=1,\dots,k_{\mathrm{max}},~~j=1,\dots,d_v\notag ~. \label{eq:discretFNO}
\end{gather}

(3) The output $u\in \mathcal{U}$ is obtained by $u(x) = Q(v_T(x))$, where $Q:\mathbb{R}^{d_v}\rightarrow\mathbb{R}^{d_u}$ is the projection of $v_T$ and it is parameterized by a fully connected layer.

\section{The implicit U-net enhanced Fourier neural operator}
\label{app2}

\begin{figure*}[!t]
    \centering
    \includegraphics[width=14cm,height=7.875cm]{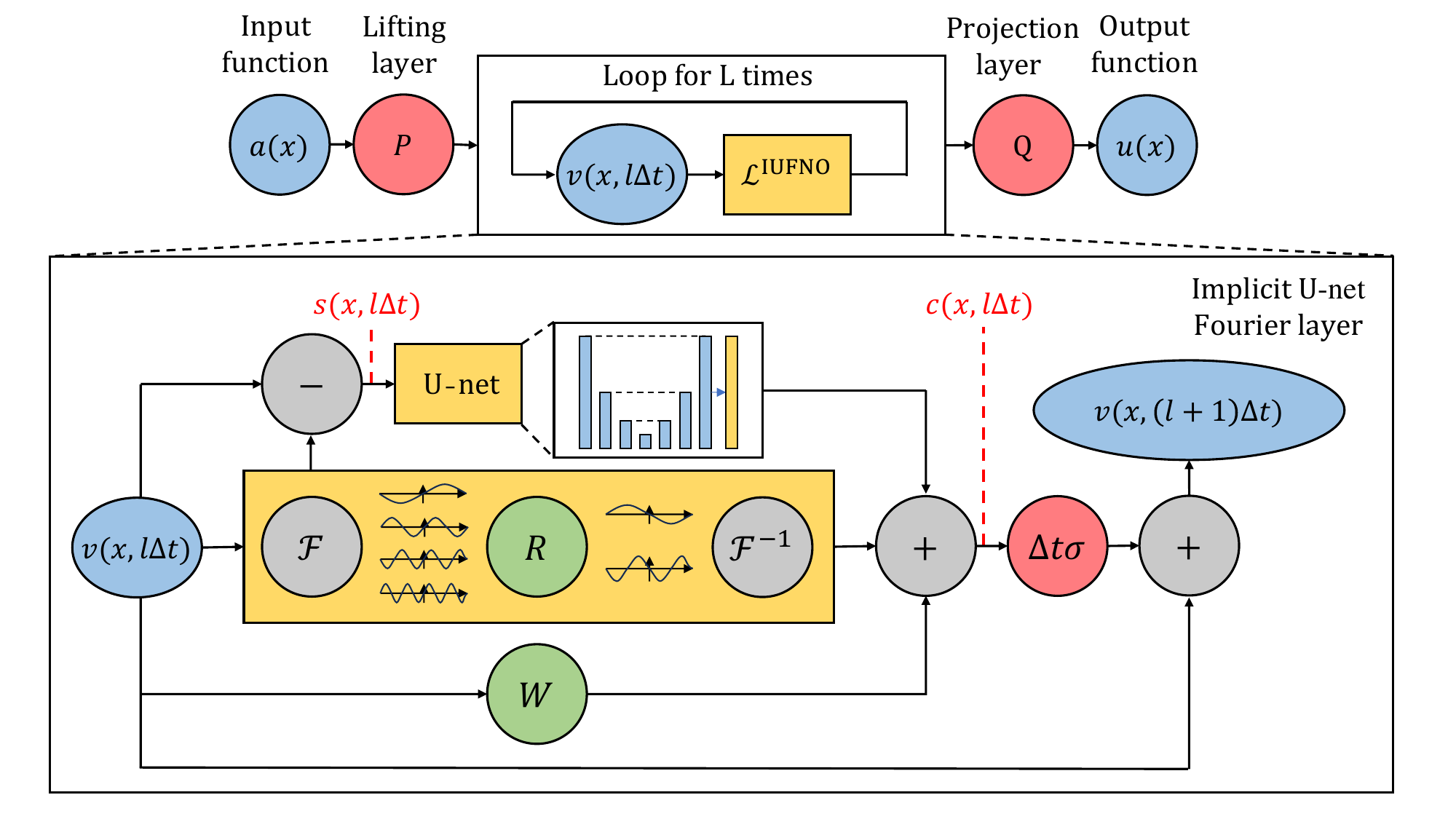}
    \caption{\label{fig:ModelStructureIUFNO} The architecture of IUFNO.}
\end{figure*}

\begin{figure*}[!t]
\begin{equation}
v(x,(l+1)\Delta t)=\mathcal{L}^{\mathrm{IUFNO}}[v(x,l\Delta t)]:=v(x,l\Delta t)+\Delta t\sigma(c(x,l\Delta t)),~~\forall x\in D ~, \label{eq:iterIUFNO}
\end{equation}
\begin{equation}
c(x,l\Delta t):=Wv(x,l\Delta t)+\mathcal{F}^{-1}(R_{\phi}\cdot(\mathcal{F}v(x,l\Delta t)))(x)+\mathcal{U}^{\ast}s(x,l\Delta t),~~\forall x\in D ~, \label{eq:unetIUFNO}
\end{equation}
\begin{equation}
s(x,l\Delta t):=v(x,l\Delta t)-\mathcal{F}^{-1}(R_{\phi}\cdot(\mathcal{F}v(x,l\Delta t)))(x),~~\forall x\in D ~. \label{eq:extIUFNO}
\end{equation}
\end{figure*}

Here, we introduce the implicit U-net enhanced Fourier neural operator (IUFNO) model \cite{li2023long}, which integrates the advantages of both IFNO and U-FNO models \cite{you2022learning,wen2022u}, and its architecture is shown in Fig.~\ref{fig:ModelStructureIUFNO}. The U-net architecture has the ability to access low-level information and high-level features simultaneously \cite{ronneberger2015u,chen2019u}. The implicit iterative approach adapted from IFNO significantly reduces the model’s parameter count, and helps alleviate the overfitting phenomenon for deep networks \cite{el2021implicit,winston2020monotone}. The IUFNO model also concludes three main steps:

(1) The tensors input to the model is converted into a high-dimensional representation via the lifting layer $P$.


(2) The generated representative is then iteratively updated through the implicit U-Fourier layers.

The fundamental difference between the IUFNO model and the FNO model focus on the different iteration method. FNO adopts a multilayer structure, where multiple Fourier layers with independent trainable parameters are connected in series. However, IUFNO adopts an implicit iterative structure, which takes the form of initializing only one layer of U-FNO layer for iterative learning. Furthermore, the IUFNO model incorporates a U-net network to effectively capture small-scale flow structures.

The formulation of iterative implicit U-Fourier layer update can be derived as shown in Eq.~\ref{eq:iterIUFNO} to Eq.~\ref{eq:extIUFNO} \cite{li2023long}.

Here, $c(l,\Delta t)\in\mathbb{R}^{d_v}$, which is specifically marked in red in Fig.~\ref{fig:ModelStructureIUFNO}, is able to capture the global-scale information of the flow field by combining large scale information learned by FFT and small-scale information $s(x,l\Delta t)$ learned by the U-net network $U^{\ast}$. In order to clearly indicate each term in the equations appear in Fig.~\ref{fig:ModelStructureIUFNO}, $s(x,l\Delta t)\in\mathbb{R}^{d_v}$ is also specifically marked in red, associated with the small-scale information obtained by subtracting the large-scale information from the complete field information $v(x,l\Delta t)$. $U^{\ast}$ is a CNN-based network which provides a symmetrical structure with an encoder and a decoder. The encoder is responsible for extracting feature representations from the input data and it is represented in Fig.~\ref{fig:ModelStructureIUFNO} as four adjacent blue bars of decreasing height. The decoder generates the output signals and it is the mirror image of the encoder. Furthermore, U-Net incorporates skip connections, enabling direct transmission of feature maps from the encoder to the decoder, thereby preserving the intricate details within the fields. The U-net architecture has a relatively small number of parameters, such that its combination with FNO has a minimal effect on the overall number of parameters. Additionally, the implicit utilization of a shared hidden layer can significantly reduce the number of network parameters, which can make the network very deep \cite{li2023long}.

(3) The output $u\in \mathcal{U}$ is obtained by $u(x) = Q(v_T(x))$, where $Q:\mathbb{R}^{d_v}\rightarrow\mathbb{R}^{d_u}$ is the projection of $v_T$ and it is parameterized by a fully connected layer.







\Acknowledgements{\textbf{Conflict of interest} \hspace{1em} The authors declare that they have no known competing financial interests or personal relationships that could have appeared to influence the work reported in this article.\\ \\
\textbf{Author contributions} \hspace{1em} 
\textbf{Yuchi Jiang}: Conceptualization, Methodology, Investigation, Coding, Validation, Writing - draft preparation, Writing - reviewing and editing. \textbf{Zhijie Li}: Conceptualization, Methodology, Investigation, Coding, Writing - reviewing and editing. \textbf{Yunpeng Wang}: Conceptualization, Investigation, Writing - reviewing and editing. \textbf{Huiyu Yang}: Conceptualization, Investigation, Writing - reviewing and editing. \textbf{Jianchun Wang}: Conceptualization, Methodology, Investigation, Supervision, Writing - reviewing and editing, Project administration, Funding acquisition.\\ \\
\textbf{Acknowlegdements} \hspace{1em} This work was supported by the National Natural Science Foundation of China (NSFC Grant Nos. 12172161, 12302283, 92052301, and 12161141017), by the NSFC Basic Science Center Program (Grant No. 11988102), by the Shenzhen Science and Technology Program (GrantNo. KQTD20180411143441009), by Department of Science and Technology of Guangdong Province (Grant No. 2019B21203001, No. 2020B1212030001, and No. 2023B1212060001), and by Innovation Capability Support Program of Shaanxi (Program No. 2023-CX-TD-30). This work was also supported by Center for Computational Science and Engineering of Southern University of Science and Technology, and by National Center for Applied Mathematics Shenzhen (NCAMS).}


\bibliographystyle{spmpsci}

\end{multicols}


\end{document}